\documentclass[prd,aps,nofootinbib,superscriptaddress,showpacs,floatfix,preprintnumbers]{revtex4}
\usepackage{hyperref,amssymb,amsmath,graphicx,xcolor}

\def \non {\nonumber}

\begin{document}

\title{Nucleon Helicity and Transversity Parton Distributions from Lattice QCD}

\author{Jiunn-Wei Chen}
\email{jwc@phys.ntu.edu.tw}
\affiliation{Department of Physics, Center for Theoretical Sciences, and Leung Center for Cosmology and Particle Astrophysics, National Taiwan University, Taipei, Taiwan 106}
\affiliation{Center for Theoretical Physics, Massachusetts Institute of Technology, Cambridge, MA 02139, USA}

\author{Saul D. Cohen}
\affiliation{Institute for Nuclear Theory, University of Washington, Seattle, WA 98195-1560}

\author{Xiangdong Ji}
\email{xji@umd.edu}
\affiliation{INPAC, Department of Physics and Astronomy, Shanghai Jiao Tong University, Shanghai, 200240, P. R. China}
\affiliation{Maryland Center for Fundamental Physics, Department of Physics, University of Maryland, College Park, Maryland 20742, USA}

\author{Huey-Wen Lin}
\email{hueywenlin@lbl.gov}
\affiliation{Physics Department, University of California, Berkeley, CA 94720}

\author{Jian-Hui Zhang}
\email{jianhui.zhang@physik.uni-regensburg.de}
\affiliation{Institut f\"ur Theoretische Physik, Universit\"at Regensburg, \\
D-93040 Regensburg, Germany}

\preprint{MIT-CTP/4776}
\preprint{INT-PUB-16-009}



\date{\today}

\begin{abstract}
We present the first lattice-QCD calculation of the isovector polarized parton distribution functions (both helicity and transversity) using the large-momentum
effective field theory (LaMET) approach for direct Bjorken-$x$ dependence. 
We first review the detailed steps of the procedure in the unpolarized case, then generalize to the helicity and transversity cases. 
We also derive a new mass-correction formulation for all three cases.
We then compare the effects of each finite-momentum correction using lattice data calculated at $M_\pi\approx 310$~MeV. Finally, we discuss the implications of these results for the poorly known antiquark structure and predict the sea-flavor asymmetry in the transversely polarized nucleon. 
\end{abstract}

\maketitle

\section{Introduction}\label{sec:intro}

Parton distribution functions (PDFs) provide a universal description of the proton's constituents (quarks, antiquarks and gluons).
They are critical inputs~\cite{Alekhin:2012ig,Gao:2013xoa,Radescu:2010zz,CooperSarkar:2011aa,Martin:2009iq,Ball:2012cx}
for the discovery of the Higgs boson, the last particle of the Standard
Model, found at the Large Hadron Collider (LHC)
through proton-proton collisions~\cite{CMS:2012nga,ATLAS:2012oga}. Despite 
this great victory, the LHC has many tasks remaining,
and the focus of the future Runs~2--5 will be to search for physics beyond the
Standard Model. In order 
to discriminate new-physics signatures from the Standard-Model background, we need to improve the precision of the latter. Unfortunately, our knowledge of
many Higgs-production cross sections remains dominated by PDF uncertainties. 
Improvement on current PDF uncertainties is important to assist LHC new-physics searches.

In addition to their applications to the energy frontier, PDFs also reveal
nontrivial structure inside the nucleon, such as the momentum and spin distributions 
of partons. Many ongoing and planned experiments at facilities around 
the world, such as Brookhaven and Jefferson Laboratory in the United States, GSI in Germany, J-PARC in
Japan, or a future electron-ion collider (EIC), are set to explore the less-known 
nucleon structures and more. 
In order to distinguish the flavor content of the PDFs, one would need to 
use nuclear data, such as neutrino scattering off heavy nuclei. However, the current understanding
of medium corrections in these cases is limited. Thus, the uncertainty in the strange
PDFs remains large. In some cases, the assumption $\overline{s}(x)=s(x)$
made in global analyses can agree with data due to the large uncertainty. 
At the LHC, strangeness can be extracted through the $W+c$ associated-production channel,
but their results are not yet well-determined. 
For example, ATLAS gets
$(s+\overline{s})/(2\overline{d})=0.96^{+0.26}_{-0.30}$ at $Q^2=1.9\mbox{ GeV}^2$ and 
$x=0.23$~\cite{Aad:2014xca}. 
CMS performs a global analysis with deep-inelastic scattering (DIS) data and the muon-charge asymmetry in 
$W$ production at the LHC to extract the ratios 
of the total integral of strange and anti-strange to the sum of the anti-up and -down, 
finding it to be $0.52^{+0.18}_{-0.15}$ at $Q^2=20\mbox{ GeV}^2$
~\cite{Chatrchyan:2013mza}.
Future high-luminosity studies may help to improve our knowledge of the strangeness. 
In the polarized case, SU(3)-flavor symmetry is often assumed due to lack of precision 
experimental data. We learn from the unpolarized case that this assumption introduces an underestimated uncertainty. 
In addition, there have been long debates concerning how big the intrinsic charm contribution is 
or whether other heavy flavors contribute. Again, the data is too inconclusive to narrow down or discriminate between the various 
proposed QCD models.

Theoretical determination of the parton distributions is complementary to
the experimental effort, especially for those not yet accessible
kinematically in experiments. On the other hand, in order to be useful for
experiments, theoretical calculations need to demonstrate that those already
measured parton distributions can be reproduced within the same approach.
This turns out to be a challenging task. It is rooted to the
nonperturbative nature of parton interactions inside the nucleon. One
hint of this nonperturbative nature can be seen in the parton distributions
extracted from experimental data. Although the net quark number of the nucleon is 3,
the quark and antiquark numbers are both infinite. This implies that there is no
hierarchy in quark-antiquark pair production through gluon emission.
The production of $N+1$ quark-antiquark pairs is as important as the
production of $N$ pairs, so truncation at a finite $N$ is impossible. This
makes the proton effectively an infinite-body system.

Lattice QCD deals with this infinite-body problem by reducing the
continuous spacetime to a discrete lattice, rendering the number of integrals in the partition function finite.
The lattice is defined in a Euclidean spacetime so that Monte Carlo
algorithms can be used to compute these integrals efficiently. The parton distributions are related to
nucleon matrix elements of quark correlators defined on a lightcone in
Minkowski space. Those lightcone correlators become local operators in
Euclidean space and lead to unphysical results (see more detailed discussion
in the next section). 
In principle, this problem can be avoided by working with moments of parton
distributions, which correspond to matrix elements of local operators,
provided all the moments can be computed to recover the whole PDF. In
practice, one can only obtain the first few (about 3) moments due to
operator mixing with lower-dimension operators with coefficients proportional
to inverse powers of the lattice spacing.
In theory, one can design more complicated operators to 
subtract the power divergence arising from the mixing of high-moment operators 
to get to even higher moments. 
However, the operator renormalization gets significantly more complicated and 
the correlators suffer from signal-to-noise problems as well. 
In recent years more and more lattice-QCD nucleon matrix elements have been directly 
calculated at the physical quark masses, a big breakthrough compared with 
a few years ago. Still, the calculations were limited to  
the first couple leading moments. Higher moments, such as
$\langle x^2 \rangle$, have not been updated using dynamical fermions for more
than a decade~\cite{Gockeler:2004wp}. 
However, there are interesting proposals to obtain higher moments by using smeared sources to overcome the power-divergent mixing problem~\cite{Davoudi:2012ya} and by using light-quark--to--heavy-quark transition currents to compute current-current correlators in Euclidean space~\cite{Detmold:2005gg}.

Recently, one of the authors proposed a new approach to calculating the full 
$x$ dependence of parton quantities, such as the parton distributions and other
parton observables~\cite{Ji:2013dva}. The method is based on the observation
that, while in the rest frame of the nucleon, parton physics corresponds to
lightcone correlations, the same physics can be obtained through
time-independent spatial correlations in the infinite-momentum frame (IMF). For
finite but large momenta feasible in lattice simulations, a large-momentum
effective field theory (LaMET) can be used to relate Euclidean
quasi-distributions to physical ones through a factorization theorem~\cite{Ji:2014gla}.
Since then, there have been many follow-up calculations to determine the
one-loop corrections needed
to connect finite-momentum quasi-distributions to IMF/lightcone distributions 
for nonsinglet leading-twist PDFs~\cite{Xiong:2013bka}, generalized parton distributions (GPDs)~\cite{Ji:2015qla} and transversity GPDs~\cite{Xiong:2015nua} in the continuum. 
Reference~\cite{Ji:2015jwa} also explores the renormalization of quasi-distributions, and establishes that the quasi-distribution is multiplicatively renormalizable at two-loop order.  
There are also proposals to improve the quark correlators to remove linear divergences in one-loop matching~\cite{Li:2016amo} and to improve the nucleon source to get higher nucleon momenta on the lattice~\cite{Bali:2016lva}.
 
This new approach was immediately implemented on the lattice,
and first results of the technique were reported at various conferences in the summer of 2013~\cite{Lin:2014gaa,Lin:2014yra}.
Preliminary studies using the LaMET approach of the Bjorken-$x$ dependence of quark, helicity and transversity
distributions, along with the pion distribution amplitude, 
show reasonable signals for the quasi-distributions. 
In 2014, we reported the first attempt to make a lattice
calculation of the unpolarized isovector quark distributions using the LaMET
formalism~\cite{Lin:2014zya}. We use lattice gauge ensembles with $N_f=2+1+1$ highly improved staggered
quarks (HISQ) (generated by the MILC Collaboration) and
clover valence fermions with pion mass $310$~MeV. We establish the
convergence of the result within the
uncertainty of the calculation as the nucleon momentum increases. Although the lattice systematics are not yet
fully under control, we obtain some qualitative features of the flavor
structure of the nucleon sea: $\bar{d}>\bar{u}$.
In an independent follow-up lattice work, 
our result was confirmed by the ETMC Collaboration in 2015~\cite{Alexandrou:2015rja} using 
the twisted-mass fermion action. 
Reference~\cite{Lin:2014zya} also reports the total polarized sea asymmetry
$\Delta \bar{u}>\Delta \bar{d}$,
which was later confirmed in updated measurements by the STAR~\cite{Adamczyk:2014xyw} and PHENIX~\cite{Adare:2015gsd}
collaborations. 

In this work, we present the first lattice-QCD results for the helicity and transversity PDFs using the LaMET approach.
We will start by briefly reviewing the LaMET approach in Sec.~\ref{sec:LaMET}, 
and then discuss the finite-momentum corrections for quasi-distributions computed on the lattice in Sec.~\ref{sec:corrections}, using the unpolarized PDF as an example.
In Sec.~\ref{sec:PolPDFs}, we generalize the results to the spin-polarized PDFs, including helicity and transversity PDFs. Finally, we present the lattice results in Sec.~\ref{sec:num}, and discuss the
implications of these results~in Sec.~\ref{sec:discussion}, focusing on the
less-known antiquark distribution. The details of the finite-momentum corrections are given in the Appendices.

\section{Review of the LaMET Approach}\label{sec:LaMET}

The original definition of the parton distribution function (PDF) of a
hadron with momentum $P=(P_0,0,0,P_z)$ depends on the correlator
of the quark bilinear operator defined on a lightcone:
\begin{equation}
h(\xi \lambda \cdot P) \equiv
  \frac{1}{2\lambda \cdot P}
  \left\langle P \left\vert
    \bar{\psi}(0) \lambda \cdot \gamma
    \Gamma \left(0,\xi \lambda \right) \psi(\xi \lambda)
  \right\vert P \right\rangle,  \label{eq:h-definition}
\end{equation}
where $\psi$ is the quark field operator, $\lambda=(1,0,0,-1)/\sqrt{2}$
is a lightlike vector with $\lambda^2=0$, $\gamma$ is the Dirac
matrix and the gauge link is
\begin{equation}
\Gamma \left( \zeta \lambda , \eta \lambda \right) \equiv
  \exp \left( ig \int_\eta^\zeta d\rho\,\lambda \cdot A(\rho \lambda) \right)
\end{equation}
with $g$ the strong coupling constant and $A$ the gauge field.
$h\left(0\right)$ is normalized to the total quark number of the hadron.
The physical PDF $q(x)$ of the hadron is the Fourier transform of $h$:
\begin{equation}
q(x,\mu) \equiv
  \int_{-\infty}^\infty \frac{d\xi \,\lambda \cdot P}{2\pi}
    e^{+ix \xi \lambda \cdot P} h(\xi \lambda \cdot P),  \label{one}
\end{equation}
where $\mu$ is the renormalization scale. This definition is invariant
under a boost along the $z$-direction. In particular, it is valid
in the rest frame where $P_z=0$.

Under the operator product expansion,
\begin{equation}
h \simeq
  \frac{1}{2\lambda \cdot P}
  \sum_{n=1}^\infty
    \frac{\left( -i\xi \right)^{n-1}}{\left( n-1 \right)!}
    \left\langle P \left\vert
    \bar{\psi}(0) \lambda \cdot \gamma \left( i\lambda \cdot D\right)^{n-1} \psi(0)
    \right\vert P\right\rangle,
\end{equation}
with $D$ the covariant derivative and higher-twist terms neglected.
The operator can be written as
\begin{eqnarray}
\bar{\psi} \lambda \cdot \gamma \left( i\lambda \cdot D\right)^{n-1} \psi &=&
  \lambda_{\mu_1} \cdots \lambda_{\mu_n} O^{\{\mu_1\cdots \mu_n\}},  \label{O1} \\
O^{\{\mu_1\cdots \mu_n\}} &=&
  \bar{\psi} \gamma^{\{\mu_1}iD^{\mu_2}\cdots iD^{\mu_n\}} \psi ,  \label{O2}
\end{eqnarray}
where $\{...\}$ indicates symmetrization of the enclosed indices.
The tensor $\lambda_{\mu_1}\cdots \lambda_{\mu_n}$ is symmetric and
hence only the symmetric part of $O$ will contribute to the PDF.

Furthermore, since $\lambda^2=0$, the tensor $\lambda_{\mu_1}\cdots
\lambda_{\mu_n}$ is automatically traceless. For example, for $n=2$ we can write
$\lambda_{\mu_1}\lambda_{\mu_2}=\left( \lambda_{\mu_1}\lambda_{\mu_2}-g_{\mu_1\mu_2}\lambda^2/4\right) +g_{\mu_1\mu_2}\lambda^2/4$, where the first term is symmetric and traceless while the second
term is the trace term. It is clear that the trace part vanishes when $\lambda^2=0$.
Therefore, only the symmetric and traceless part of $O^{\{\mu_1\cdots \mu_n\}}$ contributes in Eq.~\ref{O1}. This symmetric and traceless operator
is a twist-2 operator whose matrix element is related to moment of the parton
distribution $a_n = \int_{-1}^1 dx\,x^{n-1} q(x)$ and $q(-x) = -\bar{q}(x)$ with
\begin{equation}
\left\langle P \left\vert O^{\{\mu_1\cdots \mu_n\}}-\text{traces}%
\right\vert P\right\rangle =2a_n\left( P^{\mu_1}\cdots P^{\mu_n}-%
\text{traces}\right) .  \label{O3}
\end{equation}
One can check easily that with Eqs.~\ref{O1}--\ref{O3}, Eq.~\ref{one} is
indeed satisfied.

The definition in Eq.~\ref{one} cannot be used to compute the PDF in
Euclidean space. Spacetime points on a lightcone in Minkowski space
satisfy the equation $t^2-\mathbf{r}^2=0$. This equation becomes
$-t_E^2-\mathbf{r}^2=0$ in Euclidean space, which is only satisfied by
the point at the origin. Therefore, the quark bilinear operator
defined on a lightcone in Minkowski space becomes a local operator in
Euclidean space, which is not desirable; Eq.~\ref{one} yields
$q(x) \propto \delta(x)$ in this case.

In principle, one can use the twist-2 operators in Eq.~\ref{O3} to recover
the PDF, provided all the moments of the PDF can be computed. However,
in practice one can only obtain the first three moments because of the
difficulty to reliably subtract the power divergence arising from
mixing of higher moments with lower ones.

If we change $\lambda$ slightly away from the lightcone to make
it spacelike ($\lambda^2<0$) while proton is still at rest, we can boost the system such that $\lambda$ is equal-time.
In other words, the following two descriptions are
identical configurations seen in different Lorentz frames: 
\begin{equation}
P_z=0,\,\lambda=(\beta\gamma,0,0,-\gamma) \ \text{with}\  \beta\to 1 \quad \Leftrightarrow \quad
P_z\rightarrow \infty,\,\lambda = (0,0,0,-1)
\end{equation}
with $\gamma = 1/\sqrt{1-\beta^2}$.  
$P_z$ characterizes the hadronic state while $\lambda$ characterizes the quark bilinear operator.

\begin{figure}
\begin{center}
\includegraphics[width=0.24\textwidth]{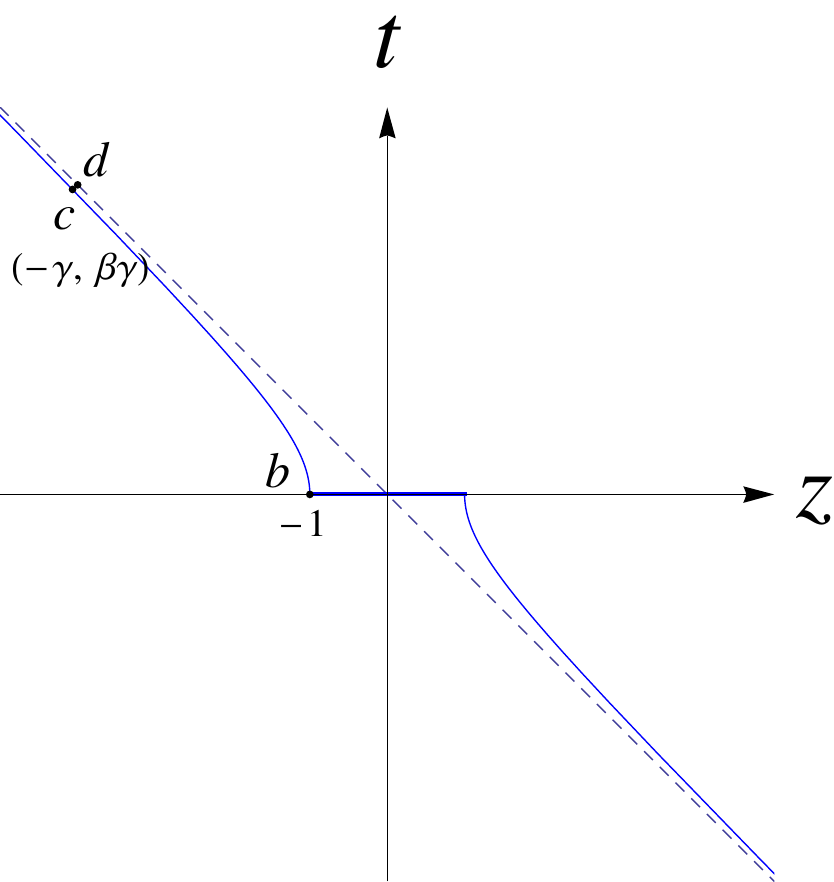} 
\end{center}
\caption{Illustration of the stepwise procedure in Eq.~\ref{eq:steps}.}
\label{steps}
\end{figure}

Ref.~\cite{Ji:2013dva} exploited the above relation and proposed a method to
compute PDFs on a Euclidean lattice in the following steps (illustrated in Fig.~\ref{steps}):
\begin{align}
\label{eq:steps}
\hspace{2in}\text{(a)}\ P_z &=\text{finite},&   \lambda&=(0,0,0,-1)  \hspace{2in}\nonumber\\
\hspace{2in}\text{(b)}\ P_z &\rightarrow \infty,&   \lambda&=(0,0,0,-1) \hspace{2in}\nonumber\\
\hspace{2in}\text{(c)}\ P_z &=0,&   \lambda&=(\beta\gamma,0,0,-\gamma)  \ \text{with}\  \beta\to 1\nonumber\hspace{2in}\\
\hspace{2in}\text{(d)}\ P_z &=0,&  \lambda^2&=0 \hspace{2in}\nonumber\\
\hspace{2in}\mbox{ or }P_z &\rightarrow\infty,& \lambda^2&=0. \hspace{2in}
\end{align}

The first step (a) is to start from the computation of Eq.~\ref{one}
with $\lambda=(0,0,0,-1)$ and a nonzero but finite $P_z$. Note
that now the quark bilinear is equal-time, and this quantity, referred to as a
``quasi-distribution'' $\tilde{q}(x,\Lambda ,P_z)$, 
can be computed on a Euclidean lattice with an ultraviolet (UV) cutoff $\Lambda$.
With $\lambda^2 \neq 0$, $O^{\{\mu_1\cdots \mu_n\}}$ in Eq.~\ref{O1} is symmetric
but not traceless. However, its structure only differs from a twist-2 operator
by trace terms. Therefore, by Eq.~\ref{O3}, its matrix element
$\left\langle P\left\vert O^{\{\mu_1\cdots \mu_n\}}\right\vert P\right\rangle$
is still related to the moment of the PDF $a_n$ plus trace terms. As discussed in
Ref.~\cite{Ji:2013dva}, the quark-level trace contribution on the left-hand
side of Eq.~\ref{O3} is a twist-4 effect and is an
$\mathcal{O}(\Lambda_\text{QCD}^2/P_z^2)$ correction, while the trace
contribution on the right-hand side is an $\mathcal{O}(M^2/P_z^2)$ correction
with $M$ the nucleon mass. There is also an $\mathcal{O}(\alpha_s\equiv g^2/4\pi)$
quantum correction to the operator $O^{\{\mu_1\cdots \mu_n\}}$ which
could depend on $P_z$ as well. Taking into account these $P_z$-dependent
corrections, one can take $P_z\rightarrow \infty$ and go from step
(a) to step (b). From step (b) to (c), nothing needs to be done, since, as explained above, they are the same system viewed in different Lorentz frames.

Going from step (c) to step (d) is nontrivial, but one can instead go from
(b), which is identical to (c), to (d) and use the boost invariance of  $\lambda^2=0$
to bring (d) to the $P_z\rightarrow \infty$ frame. Now both (b) and (d)
have the same hadronic state with $P_z\rightarrow \infty$ but with
different quark bilinear operators: the one in (b) with $\lambda^2 < 0$
while the one in (d) with $\lambda^2=0$. From the discussion above, this
difference yields $\mathcal{O}(\Lambda_\text{QCD}^2/P_z^2)$ and
$\mathcal{O}(M^2/P_z^2)$ corrections which vanish as
$P_z\rightarrow \infty$. The remaining correction is the
$\mathcal{O}(\alpha_s)$ Wilson coefficient renormalization in the operator
product expansion of the quark bilinear. It is governed by short-distance physics
and is independent of the hadronic state. 

The lightlike condition $\lambda^2=0$ in (d) implies that the quark bilinear
operator is boost
invariant, and hence, the Wilson coefficient does not depend on $P_z$, but
this is not the case for (b). This is because in the former case $P_z$ is
taken above the UV cutoff and is no longer dynamical, while in the latter
case $P_z$ is below the UV cutoff and is still dynamical. The former can
be considered as an effective field theory of the latter which is analogous
to the relation between the heavy-quark effective theory (HQET) and its full
theory \cite{Ji:2014gla}. In HQET, the heavy-quark mass is taken above the
UV cutoff and is no longer a dynamical quantity while in the full theory the
heavy-quark mass is below the UV cutoff and is dynamical. The difference
between the two theories are compensated by higher-dimensional operators or
counterterms in the effective field theory which encode the short-distance
physics of the full theory that is integrated out in the effective theory.
This ``matching'' procedure can be
implemented order by order in powers of $\alpha_s$ in perturbation theory.

Summarizing the above discussion, the quasi-distribution $\tilde{q}(x,\Lambda,P_z)$,
which can be computed in Euclidean space with $\lambda=(0,0,0,-1)$
and nucleon momentum $P_z$, can be related to the $P_z$-independent
physical distribution $q(y,\mu)$ with $\lambda=(1,0,0,-1)/\sqrt{2}$ through~\cite{Ji:2013dva}
\begin{equation}
\tilde{q}(x,\Lambda ,P_z) = 
  \int \frac{dy}{\left\vert y\right\vert} 
    Z\left( \frac{x}{y}, \frac{\mu}{P_z}, \frac{\Lambda}{P_z}\right) q(y,\mu) +
  \mathcal{O}\left( \frac{\Lambda_\text{QCD}^2}{P_z^2},\frac{M^2}{P_z^2}\right) +\ldots .  \label{Z}
\end{equation}
where $\mu$ is the renormalization scale, usually in the $\overline{\text{MS}}$ scheme, $\Lambda$
will be set by the lattice spacing, and $Z$ is the kernel from the matching.
Here, we have concentrated on the flavor nonsinglet case such that the
mixing with the gluon PDF is not needed.

\section{Finite-$P_z$ Corrections for Unpolarized PDFs}\label{sec:corrections}

In this section, we detail the procedure to implement the $P_z$ corrections
needed to extract the physical $q(x,\mu)$ from the quasi-distribution
$\tilde{q}(x,\Lambda,P_z)$ computed from the lattice. We first explain each of the
corrections and then summarize the procedure at the end of the section.

\subsection{One-Loop Matching}

In the limit $P_z \rightarrow \infty$, the matching becomes the most
important $P_z$ correction. The factor
$Z\left(\xi=\frac{x}{y},\frac{\mu}{P_z},\frac{\Lambda}{P_z}\right)$
has been computed up to one loop in
Ref.~\cite{Xiong:2013bka} using a momentum-cutoff regulator instead of a
lattice regulator. Therefore, this $Z$ factor is accurate up to the
leading logarithm but not for the numerical constant. To determine this
constant, a lattice perturbation theory calculation using the same lattice
action is required.

At tree level, the $Z$ factor is just a delta function,
\begin{equation}
Z(\xi) = \delta (\xi-1) + \frac{\alpha_s}{2\pi} \overline{Z}(\xi)
  + \mathcal{O}\left(\alpha_s^2 \right),
\end{equation}
such that
\begin{equation}
\widetilde{q}(x) \simeq q(x)
  + \frac{\alpha_s}{2\pi} \int \frac{dy}{\left\vert y\right\vert}\,
    \overline{Z}\!\left(\frac{x}{y}\right)q(y).
\end{equation}
Since the difference between $\widetilde{q}(x)$ and $q(x)$ starts at
the loop level, we can rewrite the above equation as
\begin{equation}
q(x) \simeq \widetilde{q}(x)
  - \frac{\alpha_s}{2\pi} \int \frac{dy}{\left\vert y\right\vert}\,
    \overline{Z}\!\left(\frac{x}{y}\right)\widetilde{q}(y)
\label{q}
\end{equation}
with an error of $\mathcal{O}\left(\alpha_s^2\right)$~\cite{Ma:2014jla}.

$\overline{Z}(\xi)$ is a singular function of $\xi$ with terms like
$a/\left(1-\xi\right)^2$ and
$\left(b\ln \left\vert 1-\xi \right\vert + c\right)/(1-\xi)$~\cite{Xiong:2013bka}. The value of $\left\vert \xi \right\vert$ is not bounded by unity when $P_z$ is
finite. This is because $\overline{Z}$ describes the evolution of partons, which
includes the gluon emission process where a mother quark splits into a daughter
quark and a gluon. When $P_z$ is finite, the gluon can travel in the opposite
direction from the mother quark, which makes the momentum fraction of the
daughter quark bigger than that of the mother quark; hence, $\xi$ can be bigger
than one. To show that these singular terms are not harmful, we use the fact
that $\overline{Z}$ has the structure
\begin{equation}
\overline{Z}(\xi) = \left(Z^{(1)}(\xi) - C\delta(\xi-1)\right),
\end{equation}
with the first term coming from gluon emission and the second term from the quark
self-energy diagram. $C=\int_{-\infty}^{\infty}\!d\xi^{\prime}\,Z^{(1)}(\xi
^{\prime})$ such that $\int\!d\xi\,\overline{Z}(\xi)=0$ and particle-number
conservation is satisfied. Using this, Eq.~\ref{q} becomes
\begin{equation}
q(x) \simeq \widetilde{q}(x)
  - \frac{\alpha_s}{2\pi} \int_{-\infty}^{\infty}\!dy\,
  \left[ Z^{(1)}\!\left(\frac{x}{y}\right)
    \frac{\widetilde{q}(y)}{\left\vert y\right\vert}
  - Z^{(1)}\!\left(\frac{y}{x}\right)
    \frac{\widetilde{q}(x)}{\left\vert x\right\vert}\right] .  \label{cov}
\end{equation}
Then, when $\xi=x/y$ is close to one, the double poles
$1/\left(1-\xi\right)^2$ in $Z^{(1)}(\xi)$ cancel out. The single pole
$\left(b\ln \left\vert 1-\xi \right\vert + c\right)/(1-\xi)$ is odd in $(y-x)$,
which is not an endpoint singularity in $y$ because $\xi$ is not bounded
by unity. Therefore, the integral and $q(x)$ are finite.

Another observation is that the double-pole structure in $Z^{(1)}$ makes
$\widetilde{q}(x)$ behave like $\alpha_s/x^2$ at large $x$. This can be
seen by approximating $\widetilde{q}(y)$ in the integral by $q(y)$ with $y$
bounded by unity. So even though $\widetilde{q}(x)$ is finite, its higher
moments are not. This is consistent with the fact that moments of
$\tilde{q}(x)$, $q(x)$ and $Z(x)$ satisfy
\begin{equation}\label{momentrelation}
\left\langle x^n \right\rangle_{\tilde{q}} =
  \left\langle x^n \right\rangle_Z \left\langle x^n \right\rangle_q,
\end{equation}
where
\begin{equation}
\langle x^{n}\rangle_{f}=\int_{-\infty}^{\infty}dx\,x^{n}f(x).
\end{equation}
Note that for $f=q$ this is identical to the usual moments since $q(x)$ vanishes outside $[-1,1]$. Eq.~\ref{momentrelation} can be obtained by taking moments of Eq.~\ref{Z}, and the divergence
of higher moments of $Z$ implies the divergences of higher moments of $\tilde{q}$.

\subsection{$M^{2n}/P_z^{2n}$ Correction}

In this subsection, we show how the $M^2/P_z^2$ correction to all
orders (denoted as $M^{2n}/P_z^{2n}$) can be computed exactly. We will
make use of the property
\begin{equation}
\lambda_{\mu_1}\cdots \lambda_{\mu_n} P^{(\mu_1}\cdots P^{\mu_n)} =
  \lambda_{(\mu_1}\cdots \lambda_{\mu_n)} P^{\mu_1}\cdots P^{\mu_n},
\end{equation}
where $(...)$ means the indices enclosed are symmetric and traceless.

A useful identity is
\begin{equation}
\lambda_{(\mu_1}\cdots \lambda_{\mu_n)} =
  \sum_{i=0}^{i_\text{max}} B_{n,i} \left(\lambda^2\right)^i
    \left(\frac{\partial^2}{\partial\lambda_\alpha \partial\lambda^\alpha}\right)^i
    \lambda_{\mu_1}\cdots \lambda_{\mu_n},  \label{a1}
\end{equation}
where $i_\text{max}=\frac{n-\text{Mod}[n,2]}{2}$ and $B_{n,0}=1$.
The $B$ coefficients can be determined by the tracelessness of
$\lambda_{(\mu_1}\cdots \lambda_{\mu_n)}$ which implies
\begin{equation}
g^{\mu_1 \mu_2}P^{\mu_3}\cdots P^{\mu_n}
  \lambda_{(\mu_1}\cdots \lambda_{\mu_n)} = 0,
\end{equation}
or
\begin{equation}
\sum_{i=0}^{i_\text{max}} B_{n,i} \left(\lambda^2\right)^i
  \left(\frac{\partial^2}{\partial\lambda_\alpha \partial\lambda^\alpha}\right)^i
  \lambda^2 \left(\lambda \cdot P\right)^{n-2} = 0.  \label{C}
\end{equation}
The left-hand side of this equation is a polynomial of powers of $\lambda^2$
with each term involving at most two $B$ coefficients. Then, the identity of
Eq.~\ref{C} yields the following recurrence relation:
\begin{equation}
B_{n,i} = -\frac{B_{n,i-1}}{4i(n-i+1)}.  \label{a3}
\end{equation}

To implement the $M^{2n}/P_z^{2n}$ correction, we first compute the ratio of the moments 
\begin{align}
K_n  &\equiv\frac{\left\langle x^{n-1}\right\rangle_{\tilde q}}{\left\langle x^{n-1}\right\rangle_q}=\frac{\lambda_{(\mu_1}\cdots \lambda_{\mu_n)}
  P^{\mu_1}\cdots P^{\mu_n}}{\lambda_{\mu_1}\cdots \lambda_{\mu_n}
  P^{\mu_1}\cdots P^{\mu_n}}  \notag \\
&= \sum_{i=0}^{i_\text{max}} C_{n-i}^i c^i   \label{Kn},
\end{align}
where $C$ is the binomial function and
$c=-\lambda^2 M^2/4\left(\lambda\cdot P\right)^2 = M^2/4 P_z^2$
with $\lambda^\mu = (0,0,0,-1)$ and $\lambda \cdot P=P_z$. 

As shown in Appendix B, the above factors can be converted to the following relation between PDFs
\begin{align}
q(x)&=\sqrt{1+4c}\sum_{n=0}^\infty \frac{f_-^n}{f_+^{n+1}}\Big[(1+(-1)^n)\tilde q\Big(\frac{f_+^{n+1}x}{2f_-^n}\Big)+(1-(-1)^n)\tilde q\Big(\frac{-f_+^{n+1}x}{2f_-^n}\Big)\Big]\non\\
&=\sqrt{1+4c}\sum_{n=0}^\infty \frac{(4c)^n}{f_+^{2n+1}}\Big[(1+(-1)^n)\tilde q\Big(\frac{f_+^{2n+1}x}{2(4c)^n}\Big)+(1-(-1)^n)\tilde q\Big(\frac{-f_+^{2n+1}x}{2(4c)^n}\Big)\Big],
\end{align}
where $f_{\pm}=\sqrt{1+4c}\pm 1$. Unlike the mass-correction expression of Ref.~\cite{Alexandrou:2015rja}, particle number is conserved in this expression.

\subsection{$\Lambda_\text{QCD}^2/P_z^2$ Correction}

This correction comes from the trace part on the left-hand side of Eq.~\ref{O3},
which is a twist-4 effect and can be implemented by adding a
$\tilde{q}_\text{twist-4}$ contribution to $\tilde{q}$, such that
\begin{equation}
\tilde{q}(x,\Lambda,P_z) \rightarrow
  \tilde{q}(x,\Lambda,P_z) + \tilde{q}_\text{twist-4}(x,\Lambda,P_z).
\end{equation}
As derived in Appendix~C,
\begin{equation}
\tilde{q}_\text{twist-4}(x,\Lambda,P_z) =
  \frac{1}{8\pi}\int_{-\infty}^\infty\!dz\,\Gamma_0 \left(-ixzP_z\right)
  \left\langle P\left\vert \mathcal{O}_\text{tr}(z)\right\vert P\right\rangle,
  \label{t4-1}
\end{equation}
where $\Gamma_0$ is the incomplete Gamma function and
\begin{align}
\mathcal{O}_\text{tr}(z) &=
  \int_0^z\!dz_1\,\bar{\psi}(0) \Big[ \gamma^\nu \Gamma\left(0,z_1\right)
    D_\nu\Gamma \left(z_1,z\right) \notag \\
&{} + \int_0^{z_1}\!dz_2\, \lambda \cdot \gamma
  \Gamma\left(0,z_2 \right) D^\nu \Gamma \left(z_2,z_1\right)
  D_\nu \Gamma\left(z_1,z\right) \Big] \psi (z\lambda).  \label{t4-2}
\end{align}
Instead of computing these corrections directly on the lattice, we only
parametrize and fit them as a $1/P_z^2$ correction after we have
removed other leading-$P_z$ corrections.

\subsection{Summary of the Finite $P_z$ Corrections}

Here we summarize the procedure needed to implement finite-$P_z$ corrections,
and use the unpolarized $u(x)-d(x)$ PDF as an example. We focus on the flavor-nonsinglet PDF such that there is no mixing with the gluon PDF.
The generalization to the polarized case will be shown in the next section.

We start with the computation of the equal-time correlator on a Euclidean
lattice:
\begin{equation}
h_\text{lat}\left(z,P_z,\Lambda\right) =
  \frac{1}{2P_z} \left\langle P\left\vert \bar{\psi}(0) \gamma_z
  \left(\prod_n U_z (n\hat{z})\right) \psi(z) \right\vert P\right\rangle,
  \label{hlat}
\end{equation}
which is the lattice version of Eq.~\ref{eq:h-definition} with
$\lambda^\mu=(0,0,0,-1)$ and $U_\mu$ is a discrete gauge link in the $\mu$
direction. $h_\text{lat}\left(0\right)$ is the total quark number of
the hadronic state. The quasi-PDF is the Fourier transform of $h_\text{lat}$:
\begin{equation}
\widetilde{q}(x,P_z,\Lambda) \equiv
  \int_{-\infty}^\infty \frac{dz\,P_z}{2\pi} e^{ixzP_z} h_\text{lat}.
\end{equation}

The second step is to implement the one-loop matching to convert
$\widetilde{q}(x,P_z,\Lambda)$ in the lattice scheme to $q_I(x,P_z,\mu)$
in the $\overline{\text{MS}}$ scheme:

\begin{equation}
q_I(x) \simeq \widetilde{q}(x) -
  \frac{\alpha_s}{2\pi} \int_{-\infty}^\infty\!dy\,
  \left[ Z^{(1)}\!\left(\frac{x}{y}\right)
    \frac{\widetilde{q}(y)}{\left\vert y\right\vert} - 
    Z^{(1)}\!\left(\frac{y}{x}\right)
    \frac{\widetilde{q}(x)}{\left\vert x\right\vert} \right] .  \label{Z(1)}
\end{equation}
We will use the $Z^{(1)}$ factor derived in Ref.~\cite{Xiong:2013bka} (also listed in Appendix~A for completeness),
which matches the momentum cutoff scheme with the $\overline{\text{MS}}$
scheme. This $Z^{(1)}$ factor is accurate up to the leading logarithm
but not for the numerical constant. One should replace the momentum cutoff
calculation with a lattice perturbation theory calculation to get the
correct numerical constant in the future.

The third step is to remove the $\mathcal{O}(M^{2n}/P_z^{2n})$
correction from $q_I (x,P_z,\mu)$:
\begin{align}
q_{II}(x)&=\sqrt{1+4c}\sum_{n=0}^\infty \frac{f_-^n}{f_+^{n+1}}\Big[(1+(-1)^n) q_I\Big(\frac{f_+^{n+1}x}{2f_-^n}\Big)+(1-(-1)^n)q_{I}\Big(\frac{-f_+^{n+1}x}{2f_-^n}\Big)\Big]\non\\
&=\sqrt{1+4c}\sum_{n=0}^\infty \frac{(4c)^n}{f_+^{2n+1}}\Big[(1+(-1)^n) q_I\Big(\frac{f_+^{2n+1}x}{2(4c)^n}\Big)+(1-(-1)^n) q_I\Big(\frac{-f_+^{2n+1}x}{2(4c)^n}\Big)\Big].
\end{align}
This gives rise to $q_{II}(x,P_z,\mu)$ whose remaining
$\mathcal{O}\left(\Lambda_\text{QCD}^2/P_z^2\right)$ correction can be removed by
adding the twist-4 contribution $\tilde{q}_\text{twist-4}$ defined in
Eqs.~\ref{t4-1}--\ref{t4-2}
\begin{equation}
q(x,\mu) = q_{II}(x,P_z,\mu) + \tilde{q}_\text{twist-4}(x,P_z,\mu).
\label{final step}
\end{equation}
$\tilde{q}_\text{twist-4}(x,P_z,\Lambda)$ can be computed on the lattice and in
principle another matching to the $\overline{\text{MS}}$ scheme is
required to obtain $\tilde{q}_\text{twist-4}(x,P_z,\mu)$. However, the difference
is $\mathcal{O}\left(\alpha_s\,\tilde{q}_\text{twist-4}\right)$ and hence negligible. In this work, the effect of $\tilde{q}_\text{twist-4}$ will be
parametrized, fitted to data and extrapolated $P_z \to \infty$ to obtain the $P_z$-independent left-hand side of
Eq.~\ref{final step}.

\section{Spin-Polarized PDFs}\label{sec:PolPDFs}

In this section, the finite-$P_z$ corrections for the longitudinally
polarized PDF (helicity) and the transversely polarized PDF
(transversity) is documented.

\subsection{Helicity}

The lattice definition of the helicity distribution is
\begin{equation}
\Delta \widetilde{q}(x,P_z,\Lambda) =
  \int \frac{dz\,P_z}{2\pi} e^{ixzP_z} \Delta h_\text{lat},
\end{equation}
with
\begin{align}
\Delta h_\text{lat}\left(z,P_z,\Lambda \right) &=
  \frac{1}{2M S_z} \left\langle P,S_z\left\vert
    \bar{\psi}(0)\gamma^z\gamma^5 \left(\prod_n U_z(n\hat{z})\right) \psi(z)
    \right\vert P,S_z\right\rangle  \notag \\
&\simeq \frac{1}{2M S_z} \sum_{n=1}^\infty
  \frac{\left(-iz\right)^{n-1}}{\left(n-1\right)!}
  \left\langle P,S_z\left\vert
  \bar{\psi}(0)\gamma^z\gamma^5\left(iD^z\right)^{n-1}\psi (0)
  \right\vert P,S_z\right\rangle ,  \label{Deltah}
\end{align}
where $M S_z = \sqrt{P_z^2+M^2}$.  The one-loop matching is
\begin{equation}
\Delta q_I(x) \simeq \Delta \widetilde{q}(x) -
  \frac{\alpha_s}{2\pi} \int_{-\infty}^\infty\!dy
    \left[ \Delta Z^{(1)}\!\!\left(\frac{x}{y}\right)
      \frac{\Delta \widetilde{q}(y)}{\left\vert y\right\vert}
      - Z^{(1)}\!\!\left(\frac{y}{x}\right)
      \frac{\Delta \widetilde{q}(x)}{\left\vert x\right\vert}\right] ,
\end{equation}
where $\Delta Z^{(1)}$ from the vertex correction is given in Appendix~A, while the factor $Z^{(1)}$ from wavefunction renormalization of Eq.~\ref{Z(1)} is the same as in the unpolarized case.

The symmetric operator in Eq.~\ref{Deltah} is
\begin{equation}
\Delta O^{\{\mu_1\cdots \mu_n\}} = 
  \bar{\psi} \gamma^{\{\mu_1}\gamma^5iD^{\mu_2}\cdots iD^{\mu_n\}} \psi ,
\end{equation}
whose symmetric traceless version is a twist-2 operator with matrix element 
\begin{equation}
\left\langle P,S_z\left\vert \Delta O^{(\mu_1\cdots \mu_n)}
  \right\vert P,S_z\right\rangle =
  2 \Delta a_n MS^{(\mu_1}P^{\mu_2}\cdots P^{\mu_n)},
\end{equation}
where $\Delta a_n=\int\!dx\,x^{n-1}\Delta q(x)$ and $S$ is the polarization
vector with $S^2=-1$. Using Eq.~\ref{a1}, we have
\begin{align}
\bar{K}_n &\equiv\frac{\left\langle x^{n-1}\right\rangle_{\Delta \tilde q}}{\left\langle x^{n-1}\right\rangle_{\Delta q}}=
  \frac{\lambda_{(\mu_1} \cdots \lambda_{\mu_n)} S^{\mu_1}P^{\mu_2}\cdots P^{\mu_n}}
       {\left(\lambda \cdot S\right) \left(\lambda \cdot P\right)^{n-1}} \notag \\
&= \left[ 1+B_{n,1} \left(n-1\right) \left(n-2\right) \tilde{c}
  + B_{n,2}\left(n-1\right) \left(n-2\right) \left(n-3\right) \left(n-4\right)
    \tilde{c}^2
  + \cdots \right]  \notag \\
&= \sum_{i=0}^{i_{\text{max}}} \left(\frac{n-i}{n}\right) C_{n-i-1}^i
  c^i,  \label{Knbar}
\end{align}
where $\tilde{c}=-4c$ and we have used $S\cdot P=0$. The $\mathcal{O}(M^{2n}/P_z^{2n})$
correction can be removed by (see Appendix~B for the detailed derivation)
\begin{align}
\Delta q_{II}(x)&=\frac{2a}{f_+}\Big[\Delta q_I\left(\frac{f_+}{2}x\right)-r\Big(\Delta q_I\left(-\frac{f_+}{2}\frac{x}{r}\right)-\int_{-\infty}^x \frac{dy}{y} \,b\, \Delta(y)\Big)\non\\
&+r^2\Big(\Delta q_I\left(\frac{f_+}{2}\frac{x}{r^2}\right)-\int_{-\infty}^x \frac{dy}{y} \, b\,\Delta\Big(-\frac{y}{r}\Big)+\int_{-\infty}^x \frac{dy}{y}\int_{-\infty}^y \frac{dz}{z}\, b^2 \Delta(z)-\int_{-\infty}^{-\frac{x}{r}}\frac{dy}{y}\, b\,\Delta(y)\non\\
&+\int_{-\infty}^x \frac{dy}{y}\int_{-\infty}^{-\frac{y}{r}}\frac{dz}{z}\, b^2\Delta(z)\Big)\Big]+\mathcal O(r^3),
\label{h_mass}
\end{align}
where
\begin{equation}
a=1+4c, \hspace{2em} b=\frac{f_+}{\sqrt{1+4c}}, \hspace{2em} r=\frac{f_-}{f_+}, \hspace{2em} \Delta(y)=\Delta q_I\left(\frac{f_+}{2}y\right)+\Delta q_I\left(-\frac{f_+}{2}\frac{y}{r}\right).
\end{equation}

The remaining $\mathcal{O}\left(\Lambda_\text{QCD}^2/P_z^2\right)$ correction can be removed by adding the twist-4 contribution $\Delta \tilde{q}_\text{twist-4}$
\begin{equation}
\Delta q(x,\mu) = 
  \Delta q_{II}(x,P_z,\mu) + \Delta\tilde{q}_\text{twist-4}(x,P_z,\mu).
\end{equation}
Again, in principle $\Delta \tilde{q}_\text{twist-4}$ (see Eq.~\ref{tildeq-twist4}) might be computed on the lattice
directly, but here we just parametrize and fit it.

\subsection{Transversity}

The lattice definition of the transversity distribution is
\begin{equation}
\delta \widetilde{q}(x,P_z,\Lambda) =
  \int \frac{dz\,P_z}{2\pi} e^{ixzP_z} \delta h_\text{lat},
\end{equation}
with 
\begin{align}
\delta h_\text{lat}\left(z,P_z,\Lambda \right) &= \frac{1}{2P_z}
  \left\langle P,S^x\left\vert
  \bar{\psi}(0) i\gamma^x\gamma^z\gamma^5\left(\prod_n U_z(n\hat{z})\right) \psi(z)
  \right\vert P,S^x\right\rangle   \notag \\
&\simeq \frac{1}{2P_z} \sum_{n=1}^\infty
  \frac{\left(-iz\right) ^{n-1}}{\left(n-1\right)!}
  \left\langle P,S^x\left\vert
  \bar{\psi}(0)i\gamma^x\gamma^z\gamma^5\left(iD^z\right)^{n-1}\psi(0)
  \right\vert P,S^x\right\rangle .  \label{deltah}
\end{align}
The one-loop matching is
\begin{equation}
\delta q_I(x) \simeq \delta \widetilde{q}(x) -
  \frac{\alpha_s}{2\pi} \int_{-\infty}^\infty\!dy
    \left[ \delta Z^{(1)}\!\!\left(\frac{x}{y}\right)
      \frac{\delta \widetilde{q}(y)}{\left\vert y\right\vert}
      - Z^{(1)}\!\!\left(\frac{y}{x}\right)
      \frac{\delta \widetilde{q}(x)}{\left\vert x\right\vert}\right] ,
\end{equation}
where $\delta Z^{(1)}$ is given in Appendix~A, while the factor $Z^{(1)}$ is again the same as in the unpolarized case.

For the mass correction of the transversely polarized case, we need to
compute
$t_\alpha \lambda_{(\mu_1}\cdots \lambda_{\mu_n)}S^{[\alpha}P^{\mu_1]}P^{\mu_2}\cdots P^{\mu_n}$ where the vector $t$
 and $S$ are transverse to $P$ ($t\cdot P=S\cdot P=0$) 
and $S^{[\alpha}P^{\mu_1]}=\left(S^\alpha P^{\mu_1}-S^{\mu_1}P^\alpha \right)/2$. The ratio factor between moments is given by
\begin{align}
& t_\alpha \lambda_{(\mu_1}\cdots \lambda_{\mu_n)}
  S^{[\alpha}P^{\mu_1]}P^{\mu_2}\cdots P^{\mu_n}  \notag \\
&= \frac{1}{2} \left(1+\sum_{i=1}^{i_\text{max}}
  B_{n,i} \left(\lambda^2\right)^i
    \left(\frac{\partial^2}{\partial \lambda_\alpha \partial \lambda^\alpha}\right)^i\right)
  \left[ \left(t\cdot S\right) \left(\lambda \cdot P\right)^n\right]  \notag \\
&= \frac{1}{2}\left(t\cdot S\right) \left(\lambda \cdot P\right)^n K_n,
\end{align}
where we have used $t\cdot P=0$ in the first equality and $S\cdot P=0$ in
the second one and then taken $t^\mu =(0,1,0,0)$. The ratio factor $K_n$ is the same as the unpolarized case in Eq.~\ref{Kn}. We therefore have
\begin{align}
\delta q_{II}(x)&=\sqrt{1+4c}\sum_{n=0}^\infty \frac{f_-^n}{f_+^{n+1}}\Big[(1+(-1)^n)\delta q_I\Big(\frac{f_+^{n+1}x}{2f_-^n}\Big)+(1-(-1)^n)\delta q_I\Big(\frac{-f_+^{n+1}x}{2f_-^n}\Big)\Big]\non\\
&=\sqrt{1+4c}\sum_{n=0}^\infty \frac{(4c)^n}{f_+^{2n+1}}\Big[(1+(-1)^n)\delta q_I\Big(\frac{f_+^{2n+1}x}{2(4c)^n}\Big)+(1-(-1)^n)\delta q_I\Big(\frac{-f_+^{2n+1}x}{2(4c)^n}\Big)\Big].
\end{align}

The remaining $\mathcal{O}\left(\Lambda_\text{QCD}^2/P_z^2\right)$ correction
can be removed by adding the twist-4 contribution
$\delta \tilde{q}_\text{twist-4}$
\begin{equation}
\delta q(x,\mu) = \delta q_{II}(x,P_z,\mu)
  + \delta \tilde{q}_\text{twist-4}(x,P_z,\mu).
\end{equation}
Again, in principle $\delta \tilde{q}_\text{twist-4}$ can be computed on the lattice
directly, but here we just parametrize and fit it.

\section{Numerical Results}\label{sec:num}

\begin{figure}
\begin{center}
\includegraphics[width=0.48\textwidth]{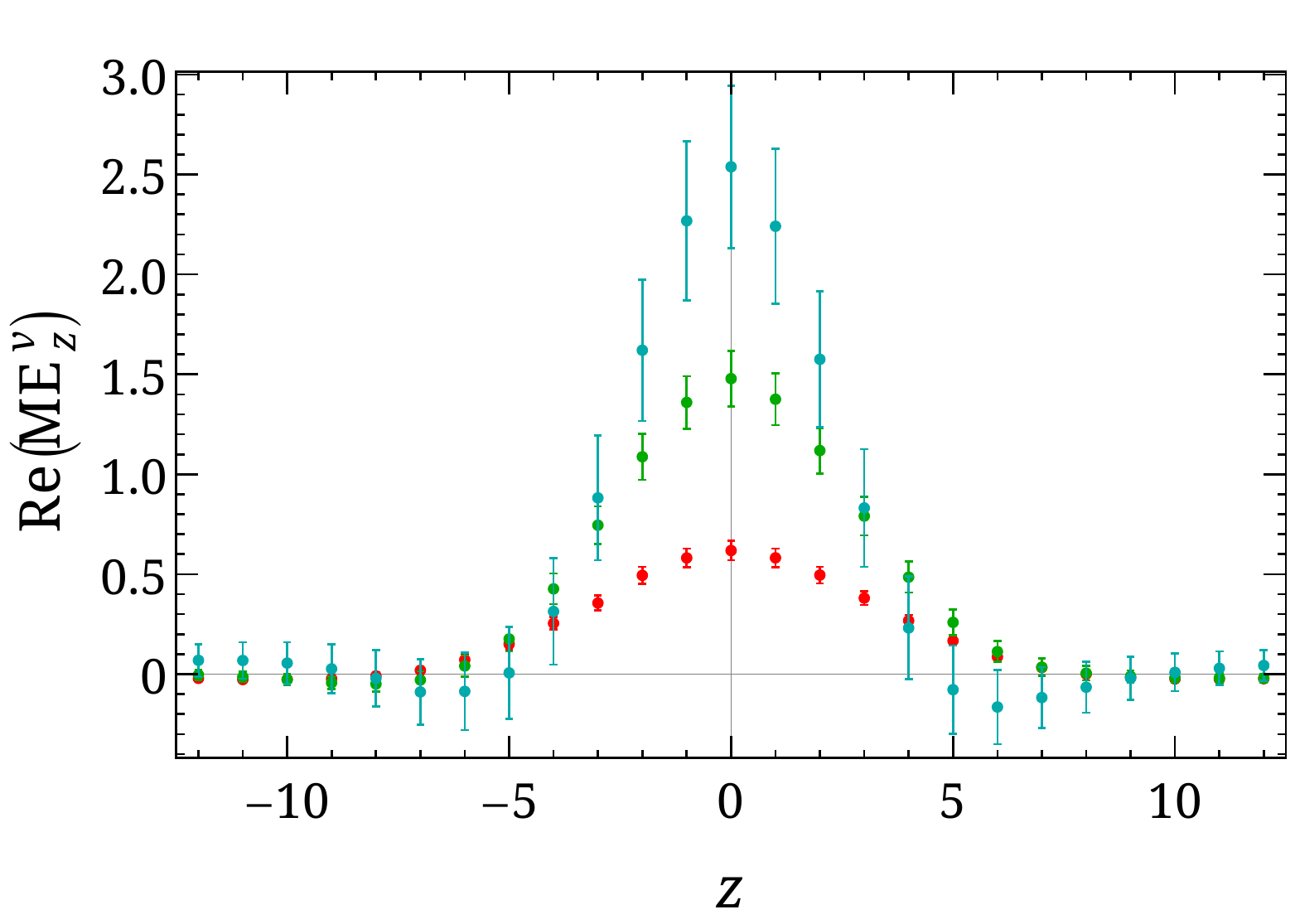} 
\includegraphics[width=0.48\textwidth]{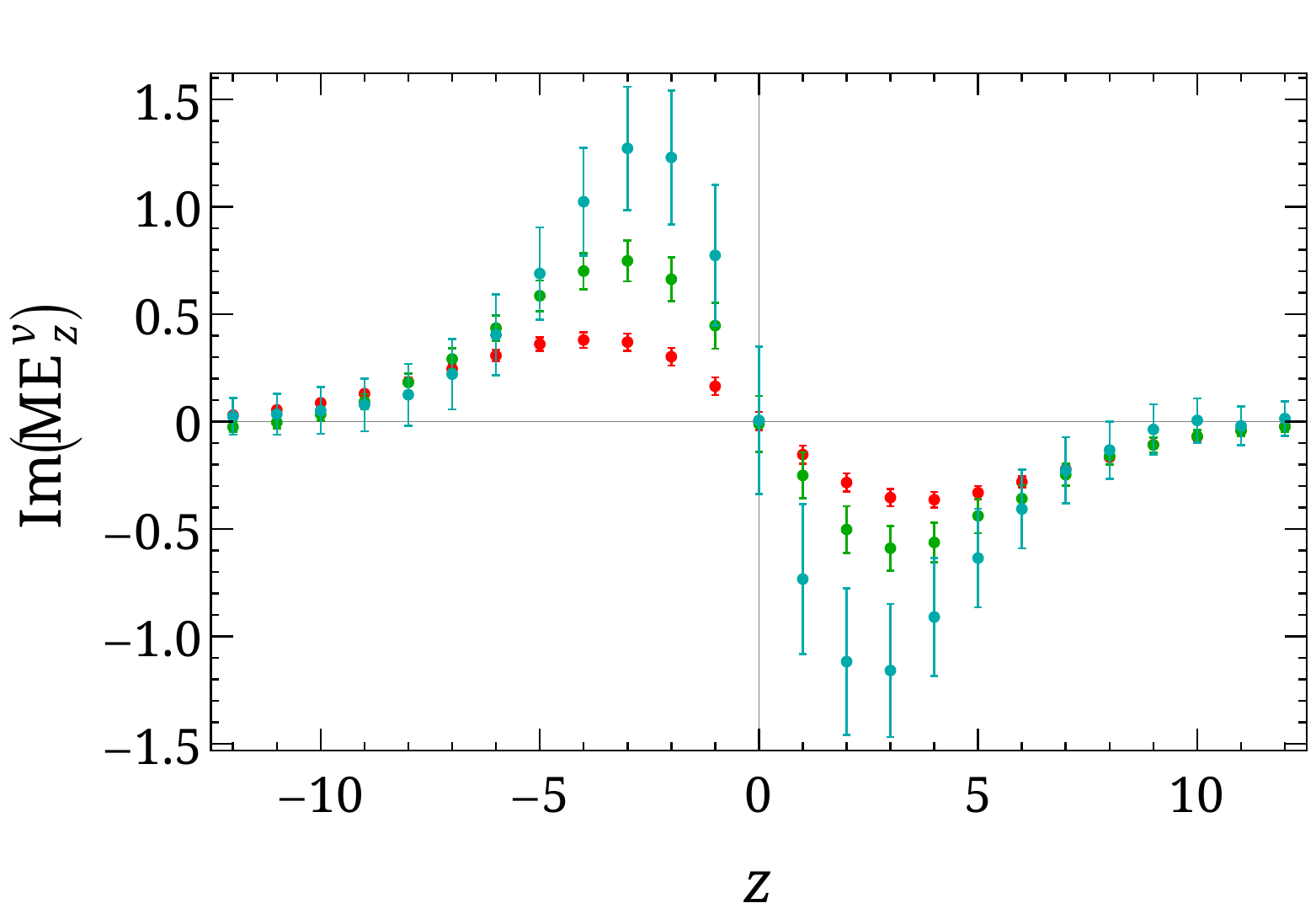}
\includegraphics[width=0.48\textwidth]{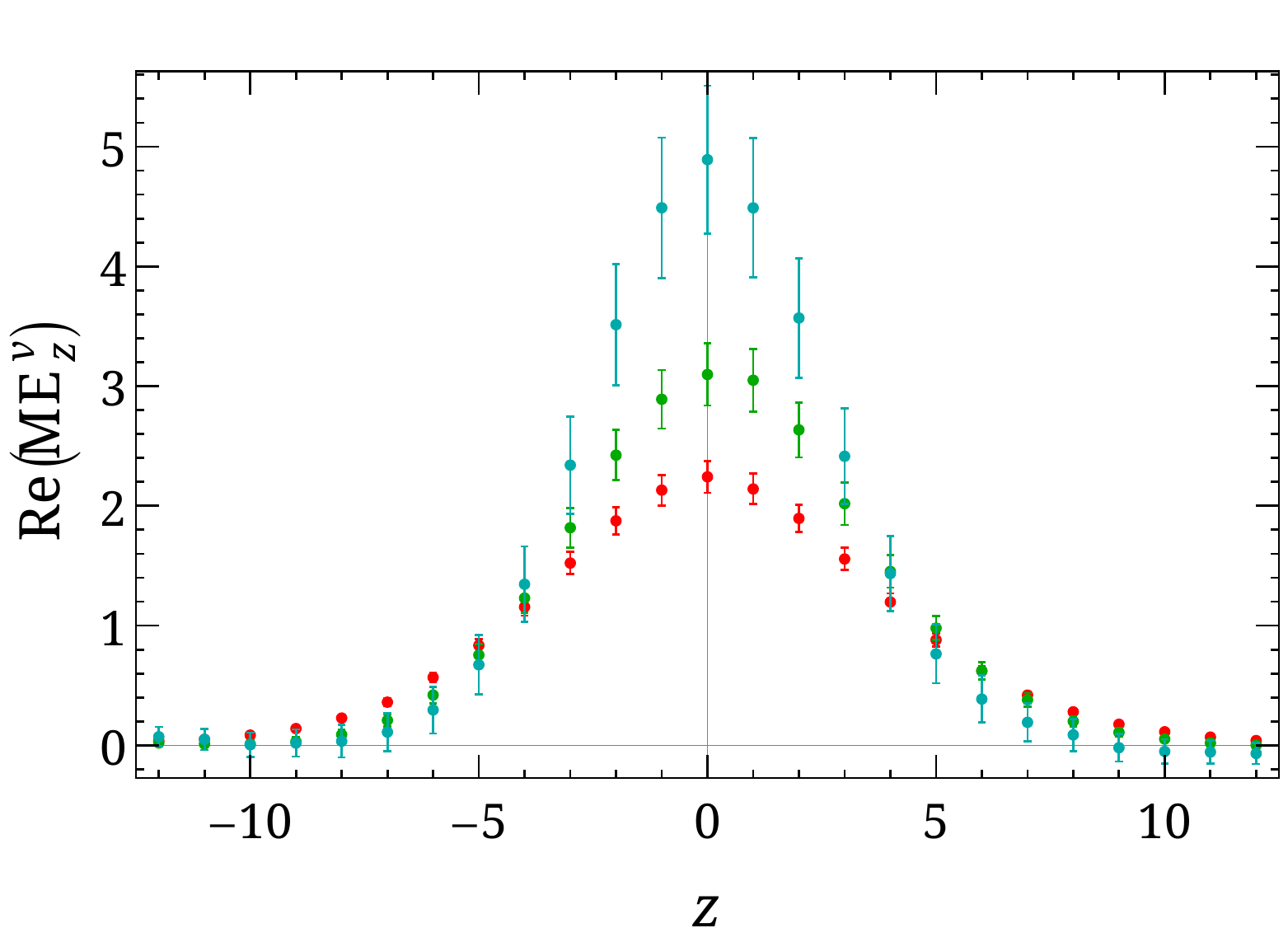} 
\includegraphics[width=0.48\textwidth]{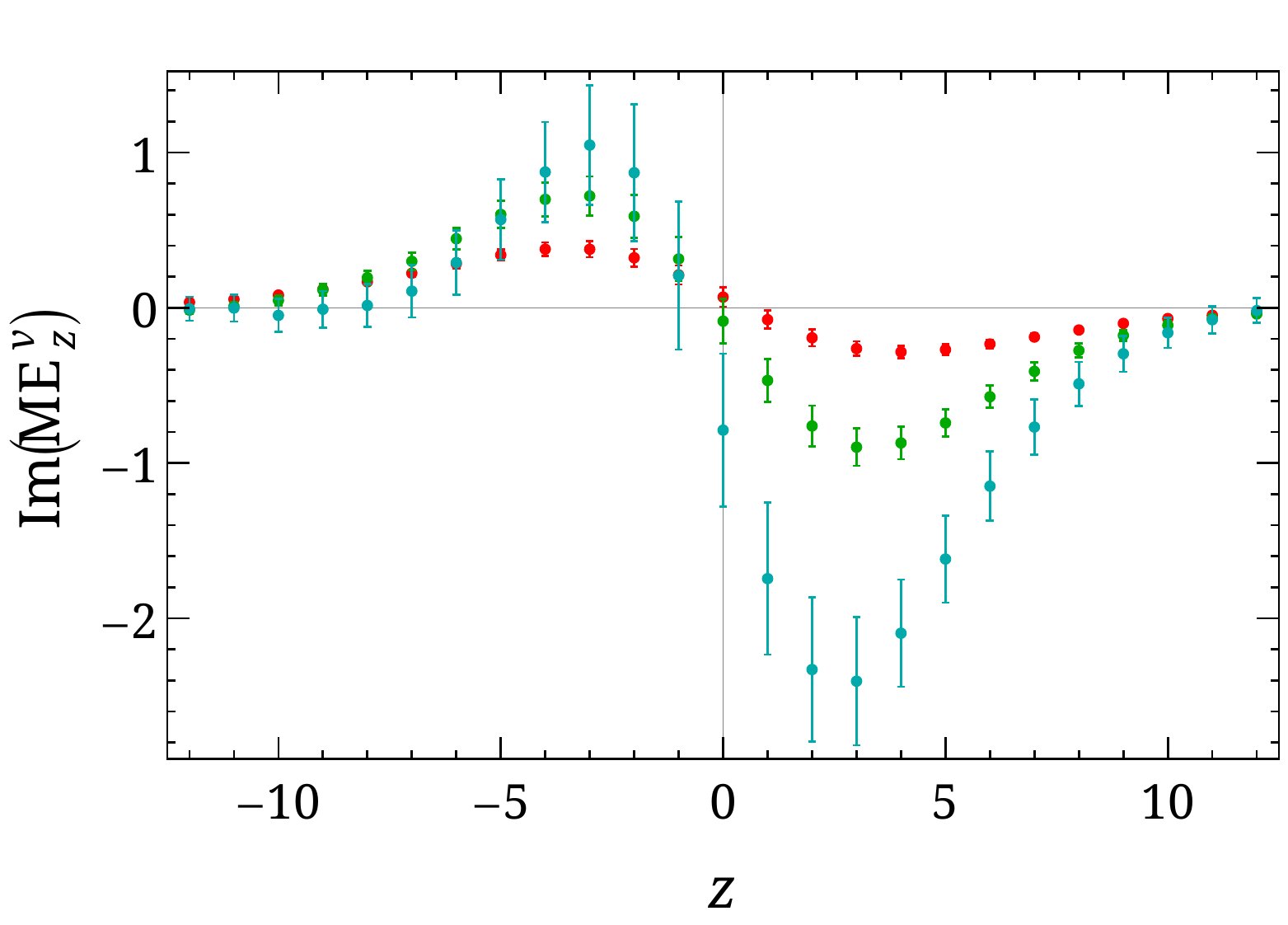}
\includegraphics[width=0.48\textwidth]{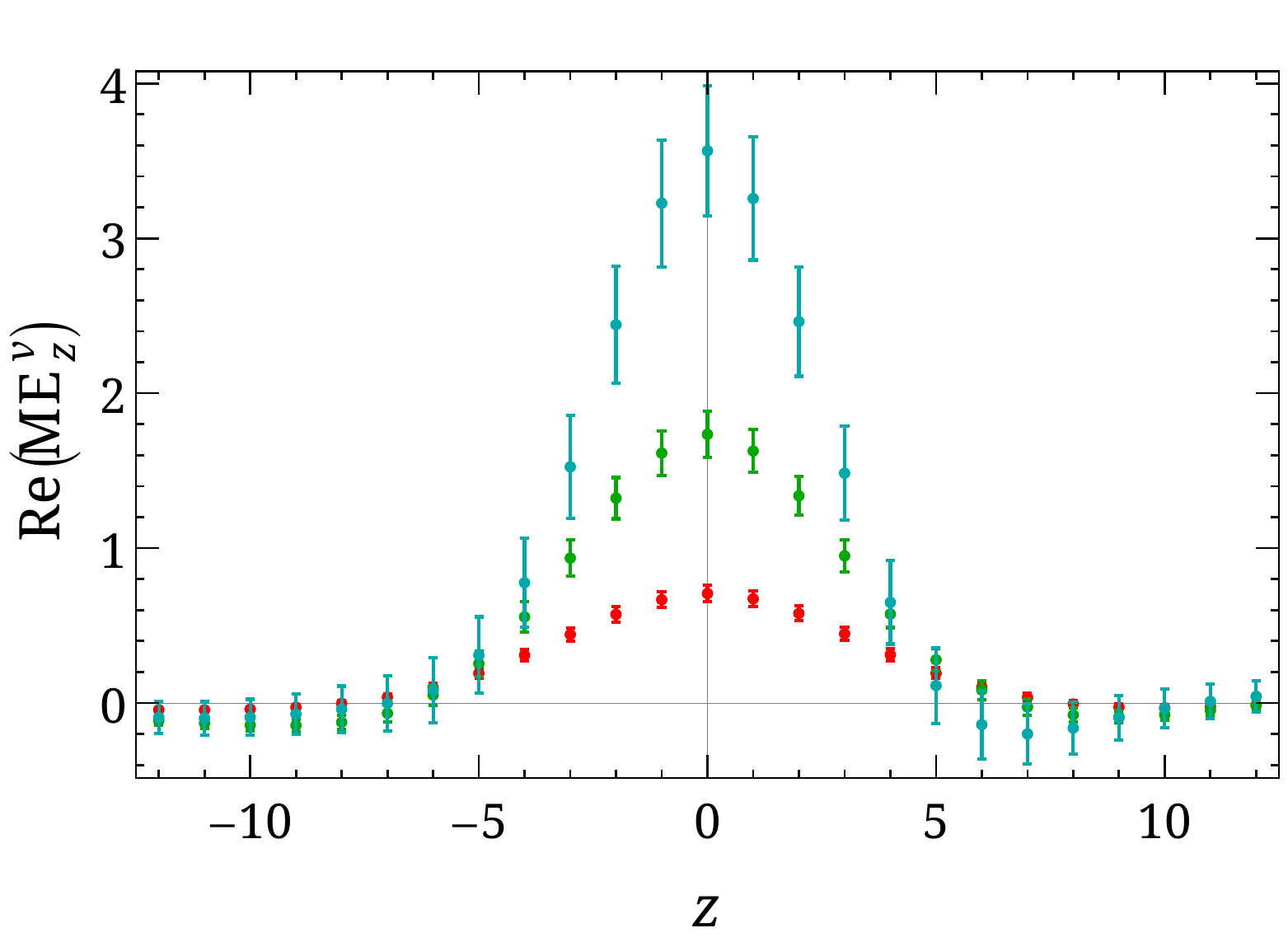} 
\includegraphics[width=0.48\textwidth]{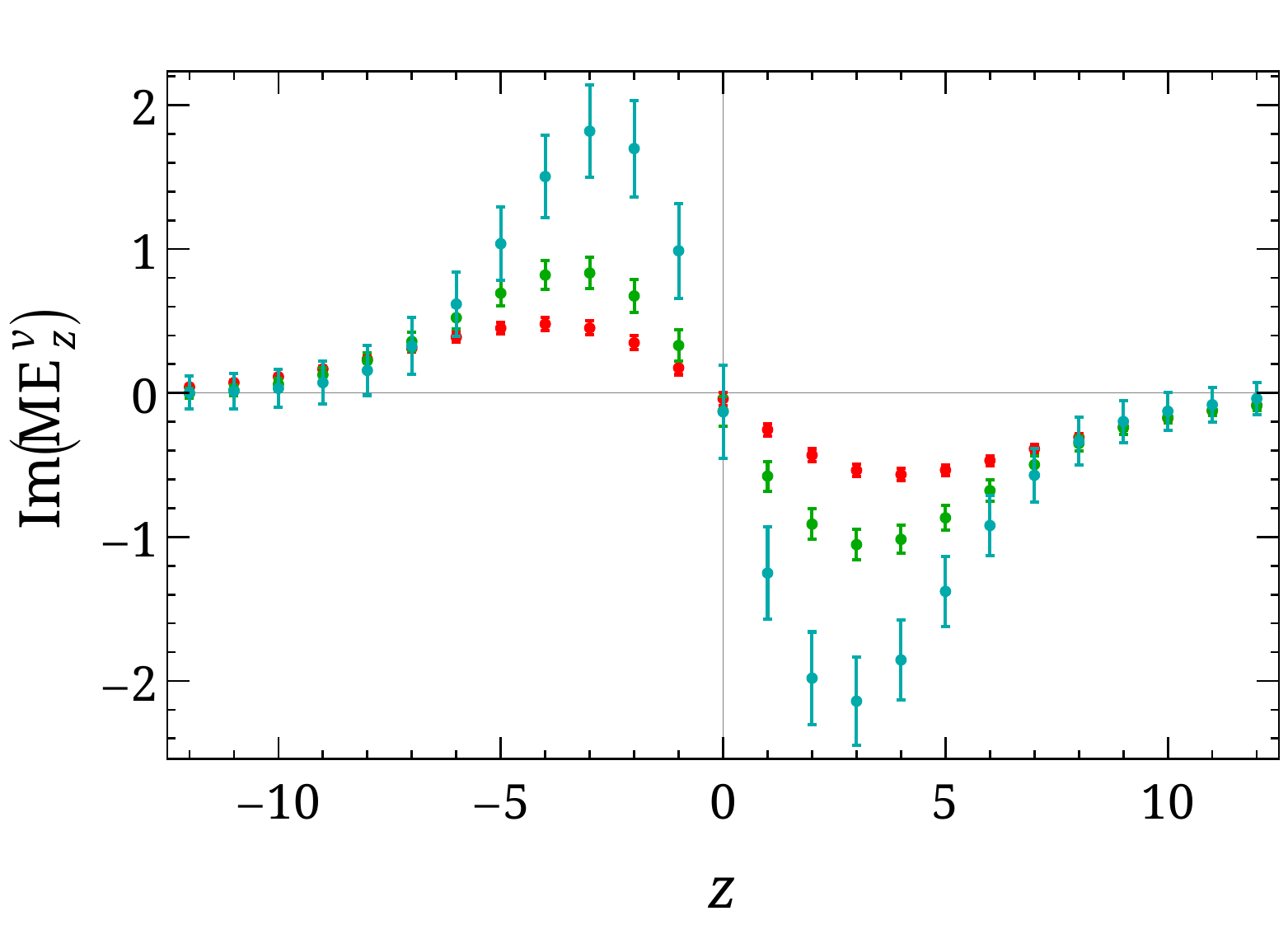}
\end{center}
\caption{The real (left) and imaginary (right) parts of the nonlocal isovector matrix elements (defined in Eq.~\ref{eq:qlat}) needed to determine the quark density (top), helicity (middle) and transverse (bottom) PDFs as functions of the length $z$ of the gauge connection between the quark and antiquark fields in the current insertion. The different colors from bottom to top indicate 
boosted momentum $P_z$ (in units of $2\pi/L$) of 1 (red), 2 (green), 3 (cyan)
.}
\label{fig:ME}
\end{figure}

In this paper, we report the results of a lattice-QCD calculation using clover valence fermions on an ensemble of gauge configurations with lattice spacing $a=0.12$~fm, box size $L\approx 3$~fm and pion mass $M_\pi \approx 310$~MeV with $N_f=2+1+1$ (degenerate up/down, strange and charm) flavors of highly improved staggered quarks (HISQ)~\cite{Follana:2006rc} generated by MILC Collaboration~\cite{Bazavov:2012xda}. The gauge links are hypercubic (HYP)-smeared~\cite{Hasenfratz:2001hp} and then clover parameters are tuned to recover the lowest pion mass of the staggered quarks in the sea\footnote{This setup is the same as the one used in works done by PNDME~\cite{Bhattacharya:2015wna,Bhattacharya:2015esa,Bhattacharya:2013ehc}.}.
HYP smearing has been shown to significantly improve the discretization effects on operators and shift their corresponding renormalizations toward their tree-level values (near 1 for quark bilinear operators). 
The volume of this ensemble is large enough, $M_\pi L \approx 4.5$, that there is no visible finite-volume correction in current lattice-QCD calculations of nucleon matrix elements. 
The results shown in this work are done using correlators calculated from 3 source locations on 449 configurations.

On the lattice, we first calculate the time-independent, nonlocal (in space, chosen to be the $z$ direction) correlators of a nucleon with finite-$P_z$ boost
\begin{align}
\label{eq:qlat}
\tilde{h}_\text{lat}(z,\mu,P_z) =  
  \left\langle \vec{P} \right|
    \bar{\psi}(z) \Gamma \left( \prod_n U_z(n\hat{z})\right) \psi(0)
  \left| \vec{P} \right\rangle,
\end{align}
where $U_z$ is a discrete gauge link in the $z$ direction and $\vec{P}=\{0,0,P_z\}$ is the momentum of the nucleon. $\Gamma=\gamma_z$, $\gamma_z\gamma_5$ and $\sigma_{xz}\gamma_5$ for the unpolarized, helicity, and transversity distributions, respectively. The spin direction is along the $z$-direction for helicity and  $x$-direction for transversity.
In this work, we are only studying 
isovector quantities (such as the up-down flavor asymmetry).
To control the systematics due to contamination by nearby excited-state quantities, 
we make a simultaneous fit of the nucleon matrix element correlators,
using two source-sink nucleon separations, 0.96 and 1.2~fm; 
the detailed procedure is described in Ref.~\cite{Bhattacharya:2013ehc} for the nucleon charges.
Examining the individual fits to each source-sink nucleon separation, we do not see noticeable excited-state contamination for either 
separation at the current statistics. 
Figure~\ref{fig:ME} shows the bare lattice nucleon matrix elements at the three boost momenta used here: $\{1,2,3\} 2\pi/L$, which correspond to nucleon momenta of 0.43, 0.86 and 1.29~GeV, respectively. 
We note that in all three cases, the matrix elements vanish when the link length reaches 10--12.
The signal-to-noise ratios worsen as the nucleon is increasingly boosted, so to push this method forward, future studies should investigate methods for improving nucleon momentum sources.

We then take the integrals to transform the lattice matrix elements as functions of spatial link length $z$ into the quasi-distributions as functions of parton momentum fraction $x$:
\begin{align}
\label{eq:qlat}
\tilde{q}(x,\Lambda,P_z) = \int \frac{dz}{4\pi} e^{-i z k} C_\Gamma \tilde{h}_\text{lat}(z,\Lambda,P_z),
\end{align}
where $x=k/P_z$, $\Lambda$ is the renormalization scale set by the lattice spacing $a$ and $C_\Gamma=(P_z/M S_z)$ for helicity and $C_\Gamma=1$
for unpolarized and transversity PDFs.
We have sampled $\delta k$ as finely as 0.002 but have
not observed any dependence in downstream results on the choice of interval used here.
Since the matrix elements go to zero beyond about 12, the integral does not depend sensitively on the 
choice of maximum $z$. 
The normalization of the long-link operators is currently 
estimated through zeroth moment of the quark distribution,
\begin{equation}
\label{eq:49}
\tilde{q}(x,\mu,P_z) \rightarrow \frac{\tilde{q}(x,\mu,P_z)}{\int\!dx\,\tilde{q}(x,\mu,P_z)} \times
  g_V^\text{local}(\mu=2\mbox{ GeV})_{\overline{\text{MS}}}.
\end{equation}
This choice reduces the systematic uncertainty arising from the matching
and other systematics such as finite-volume effects and lattice discretization. 
Given that the lattice renormalization constants for most observables are close to 1 on this ensemble,
we will get reasonable cancellation of the remaining factors. 
Similar normalizations apply to the helicity and transversity. 
The normalization of each distribution is then set by multiplying in
the corresponding vector, axial or tensor charge, as obtained on the
same lattices by Ref.~\cite{Bhattacharya:2013ehc} using standard techniques.

\begin{figure}
\includegraphics[width=0.32\textwidth]{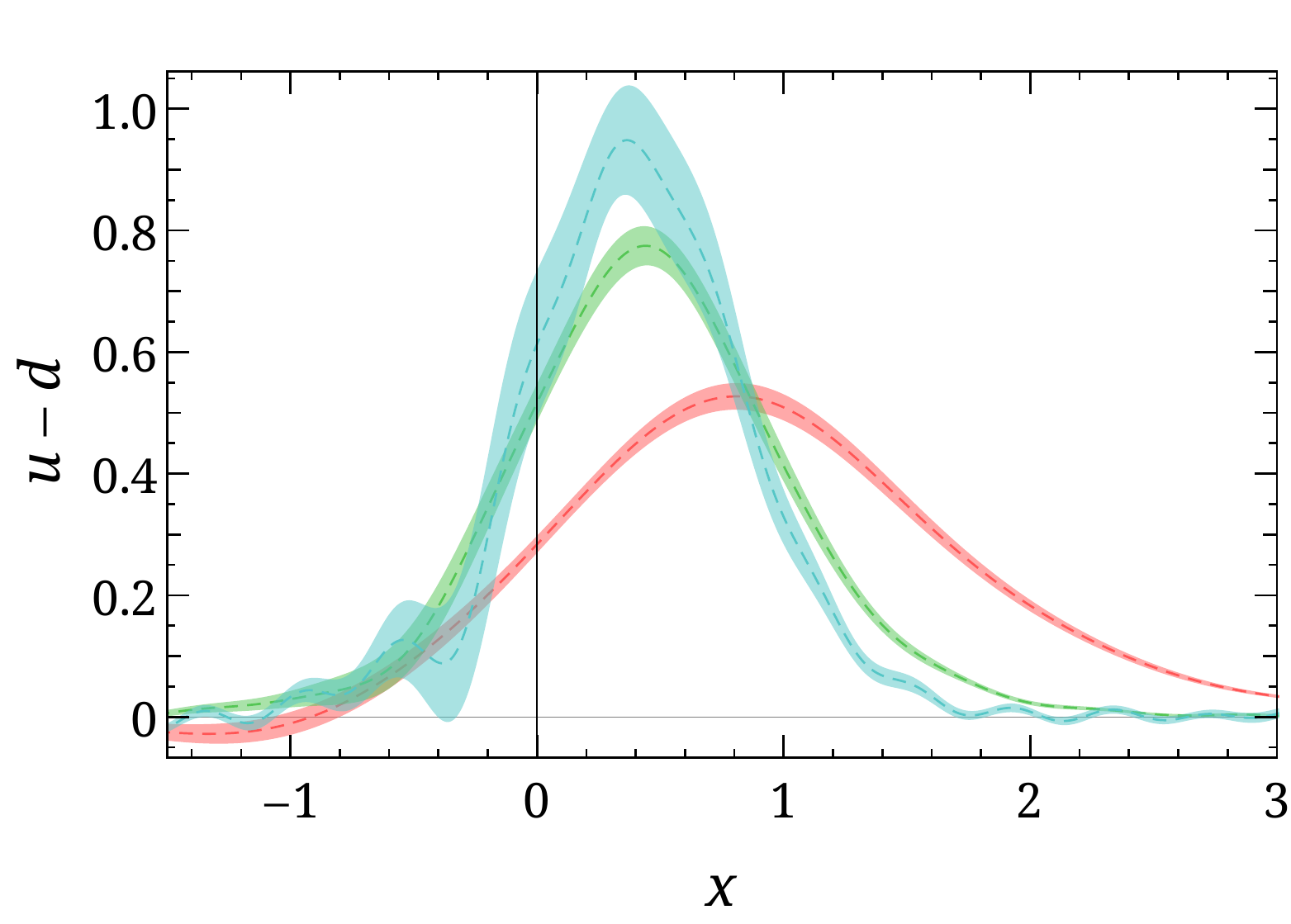}
\includegraphics[width=0.32\textwidth]{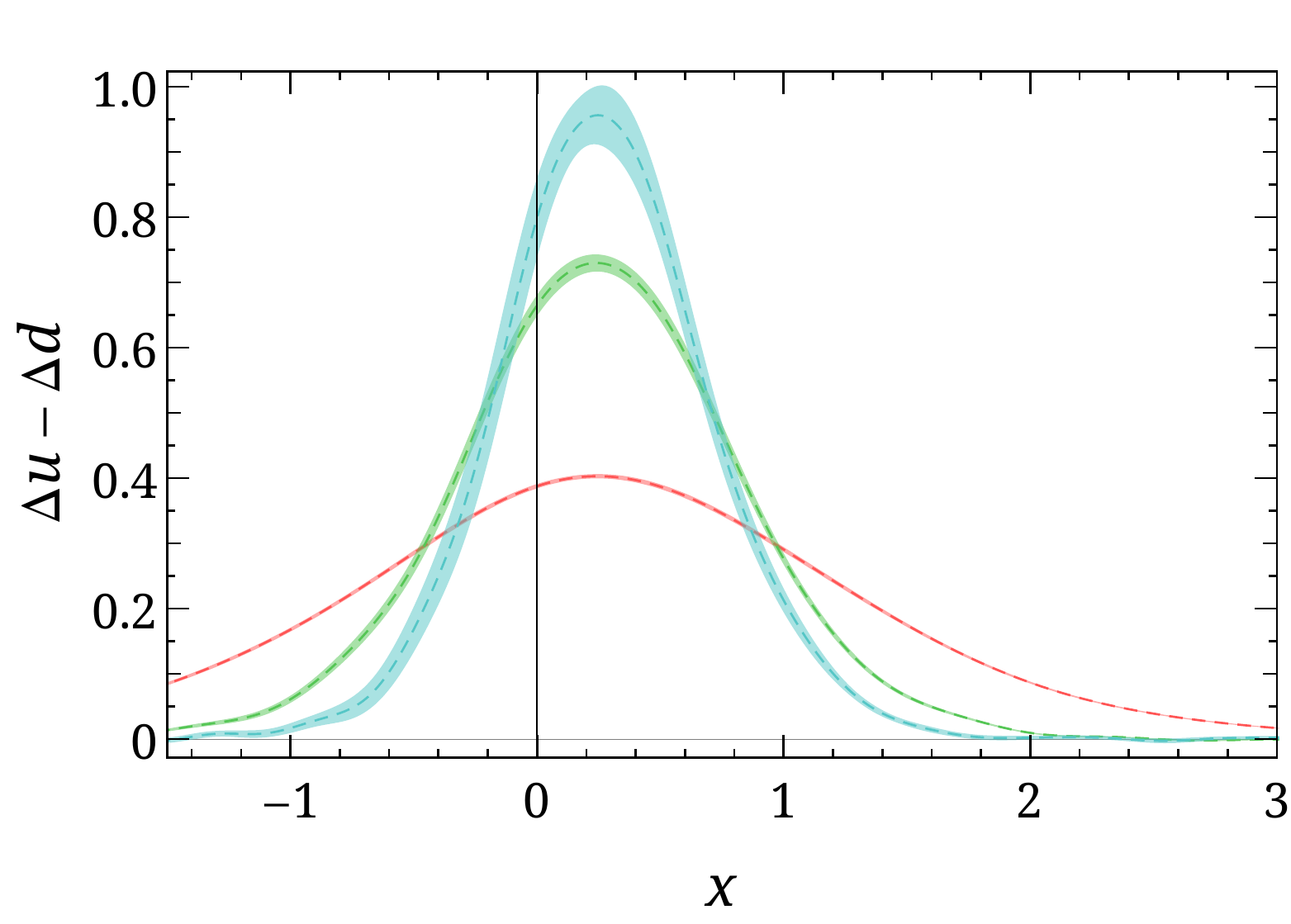}
\includegraphics[width=0.32\textwidth]{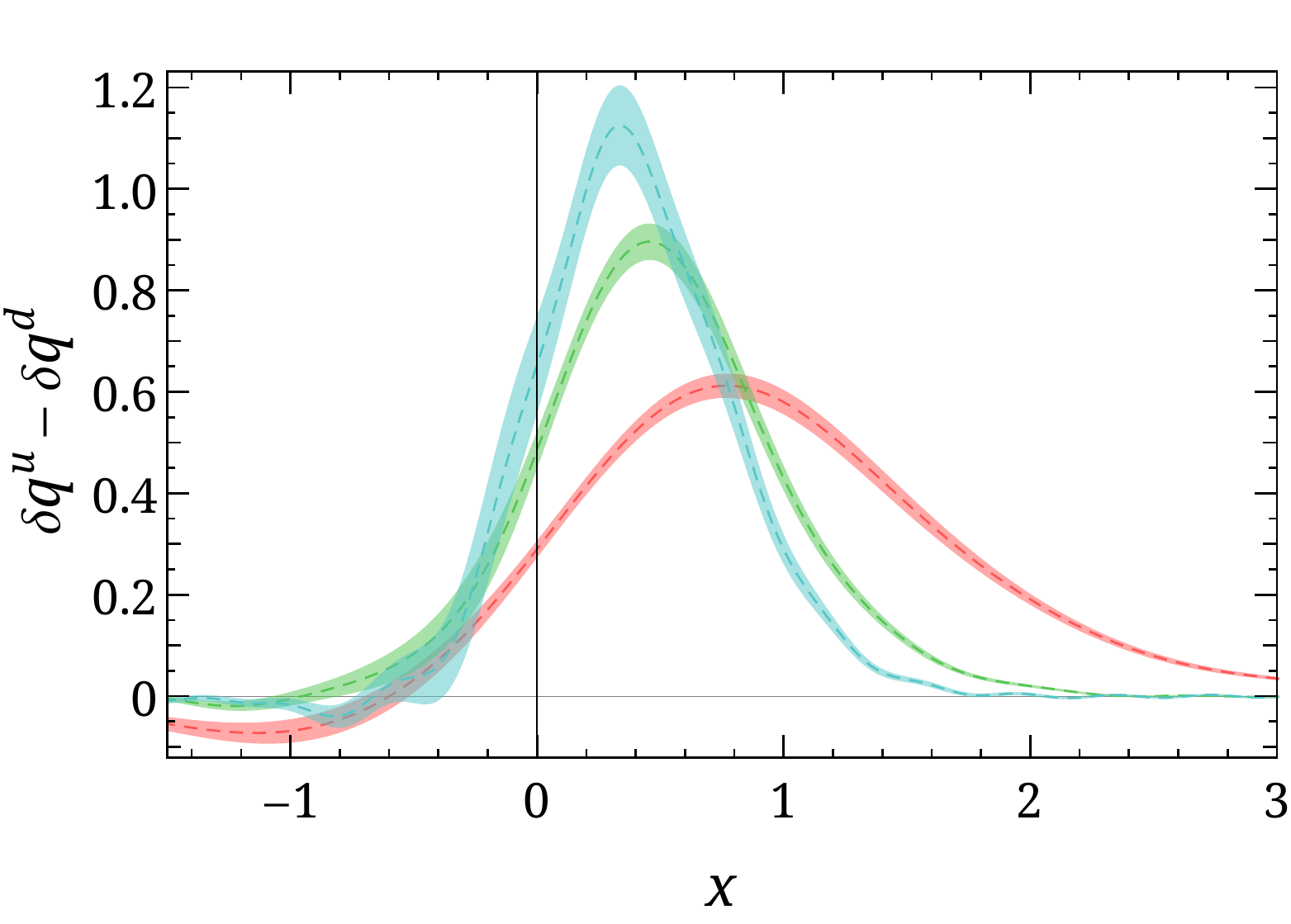}
\caption{The nucleon isovector quasi-PDFs of Eq.~\ref{eq:49}
for the quark density (left), helicity (middle) and transversity (right) as functions of $x$. The different colors from $P_z$ (in units of $2\pi/L$) 1 (red), 2 (green), 3 (cyan).
We see the data converging at large $P_z$.}
\label{fig:intx}
\end{figure}

The isovector nucleon quark, helicity and transversity quasi-distributions are shown in 
Fig.~\ref{fig:intx}, using in the same color scheme to indicate different boosted momenta.
We see that our lattice-QCD result has nonzero values for $q(x)$, $\Delta q(x)$ and $\delta q(x)$ at $x \ge 1$ and that it does not vanish until $x \approx 1.5$. %
In all three cases, the smallest momentum has the widest distribution, spreading out to 
large positive and negative $x$, beyond $|x|=1$. 
As we discussed after Eq.~\ref{q}, when $P_z$ is finite, the range of $|x|$ is not bounded by unity.
But as the boosted momentum increases, the distribution sharpens
and narrows, decreasing the contribution coming from the $|x|>1$ regions, just what we would expect 
in the lightcone distribution.
This is not hard to understand (as we discussed in our earlier work~\cite{Lin:2014zya}): in the infinite-momentum frame, no constituents of the nucleon can carry more momentum than the nucleon as a whole. However, since the momentum in our calculation is finite, the PDF does not have to vanish at $x=1$. 
The peak location for the
density and helicity distributions remains roughly the same for $P_z=$ 2 and 3, but in the case of the
transversity, the peak shifts toward $x=0$ for $P_z=3$.
Note that there is a substantial difference in magnitude between $P_z$ = 2 and 3, and an even more severe difference in shape between $P_z=1$ and the others. 
We note that since $x$ is defined as $k/P_z$ and $k$ is arbitrary, we can make $k$ as small as desired to obtain small-$x$ physics.
However, the small-$x$ region corresponds to long-distance physics, which requires longer physical links to probe. This is similar to the finite-volume effect commonly seen in LQCD calculations, except the large-$z$ links are essential to obtain a reasonable description of the physics in this region.

\begin{figure}
\includegraphics[width=0.32\textwidth]{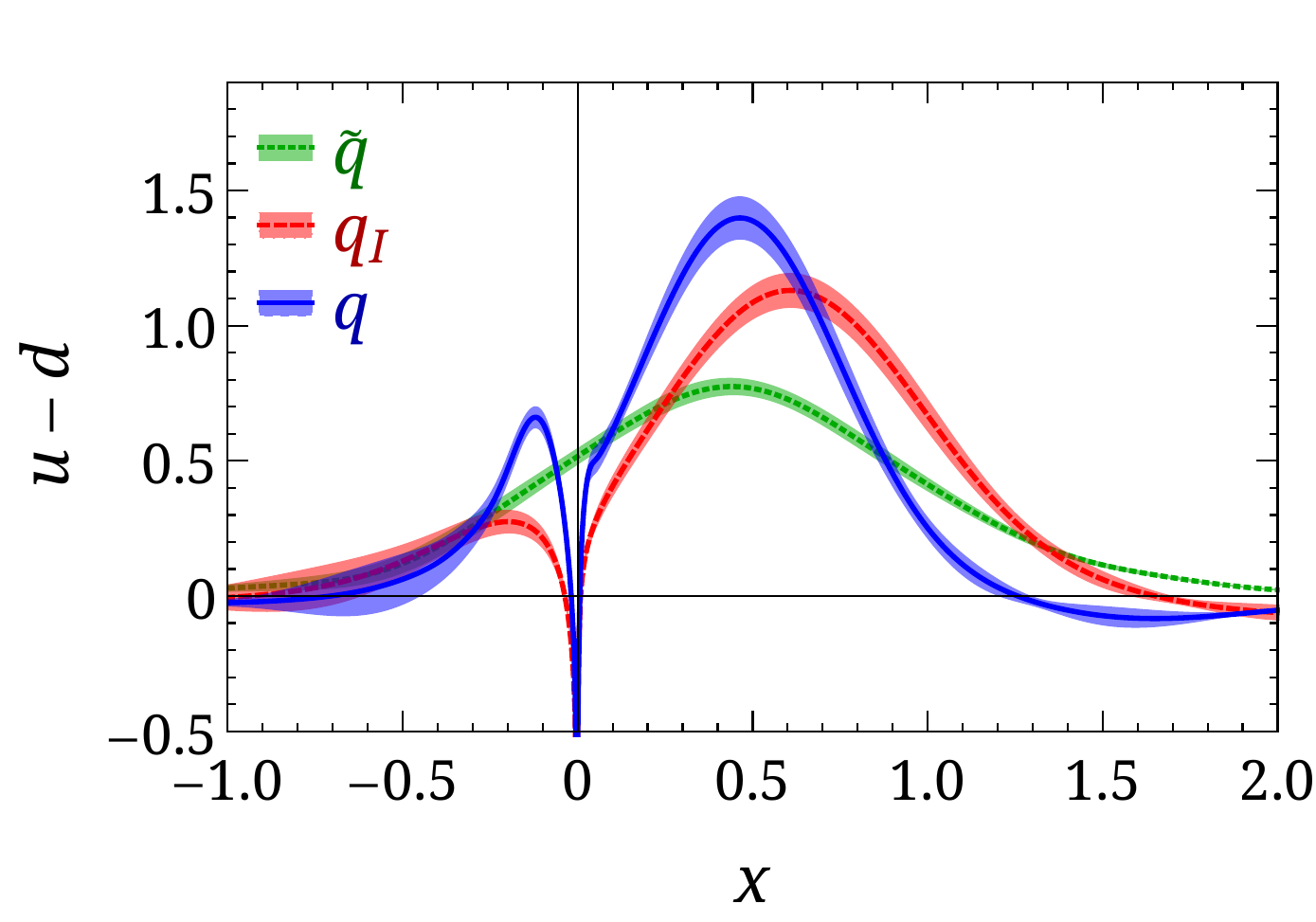}
\includegraphics[width=0.32\textwidth]{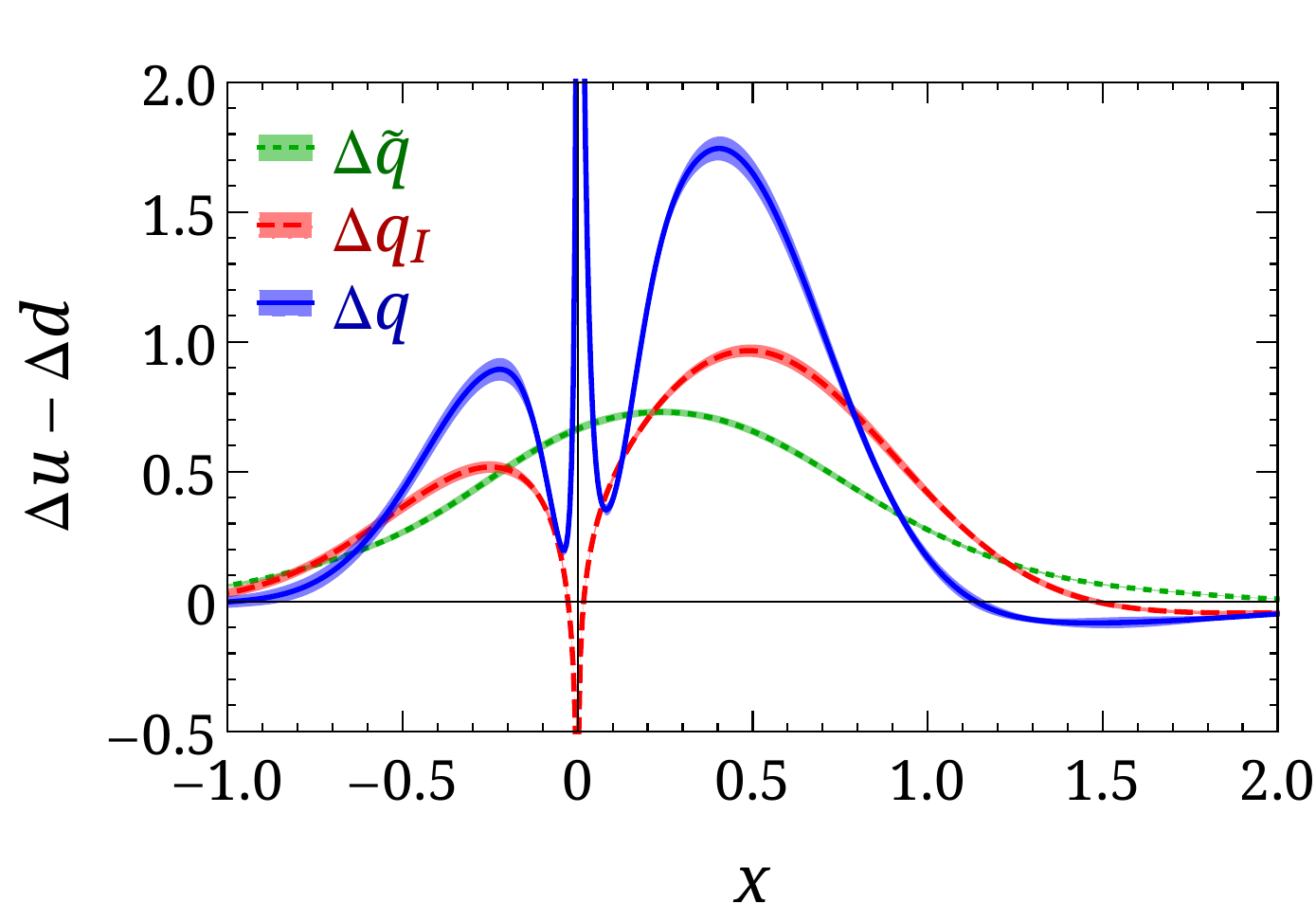}
\includegraphics[width=0.32\textwidth]{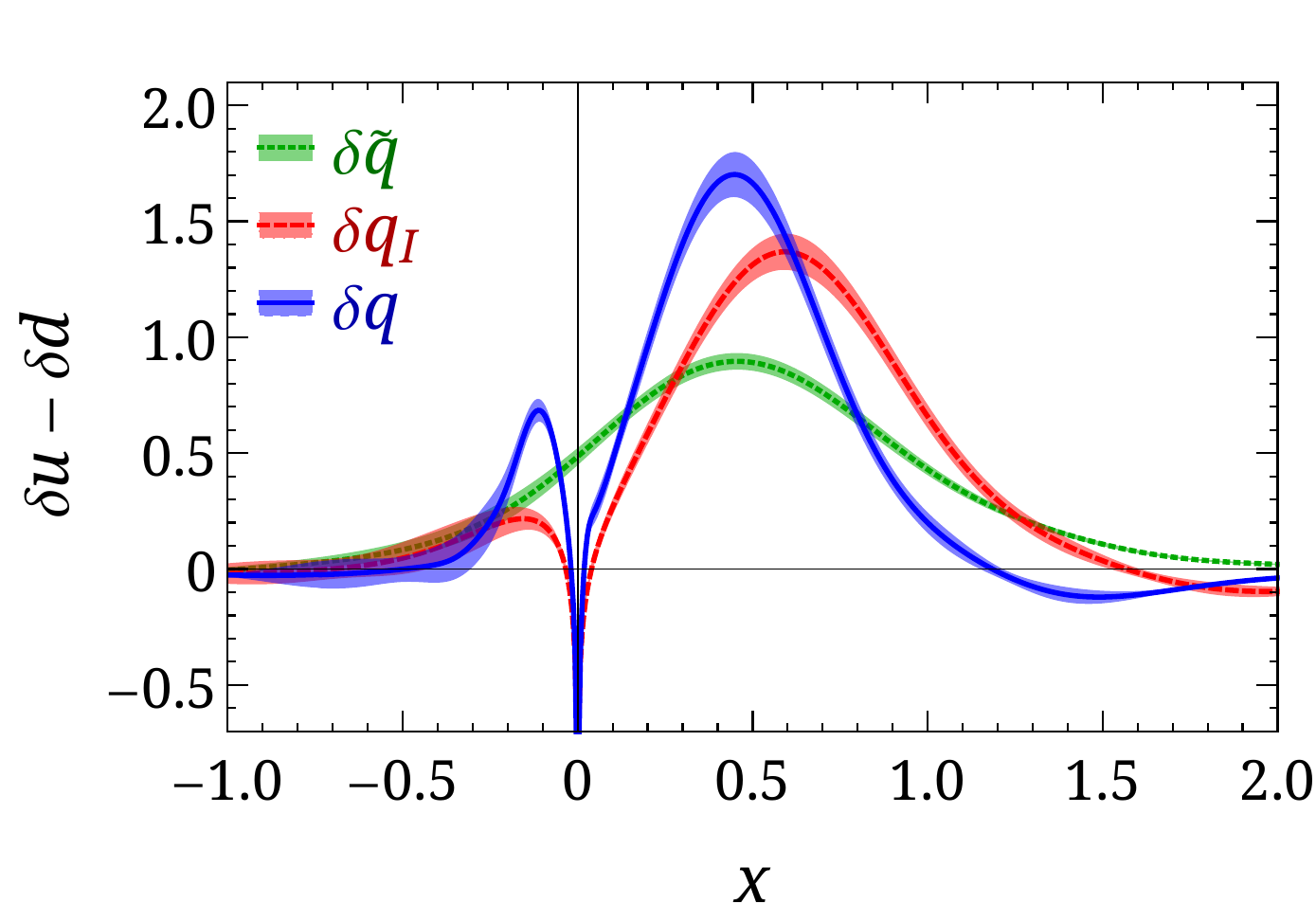}

\includegraphics[width=0.32\textwidth]{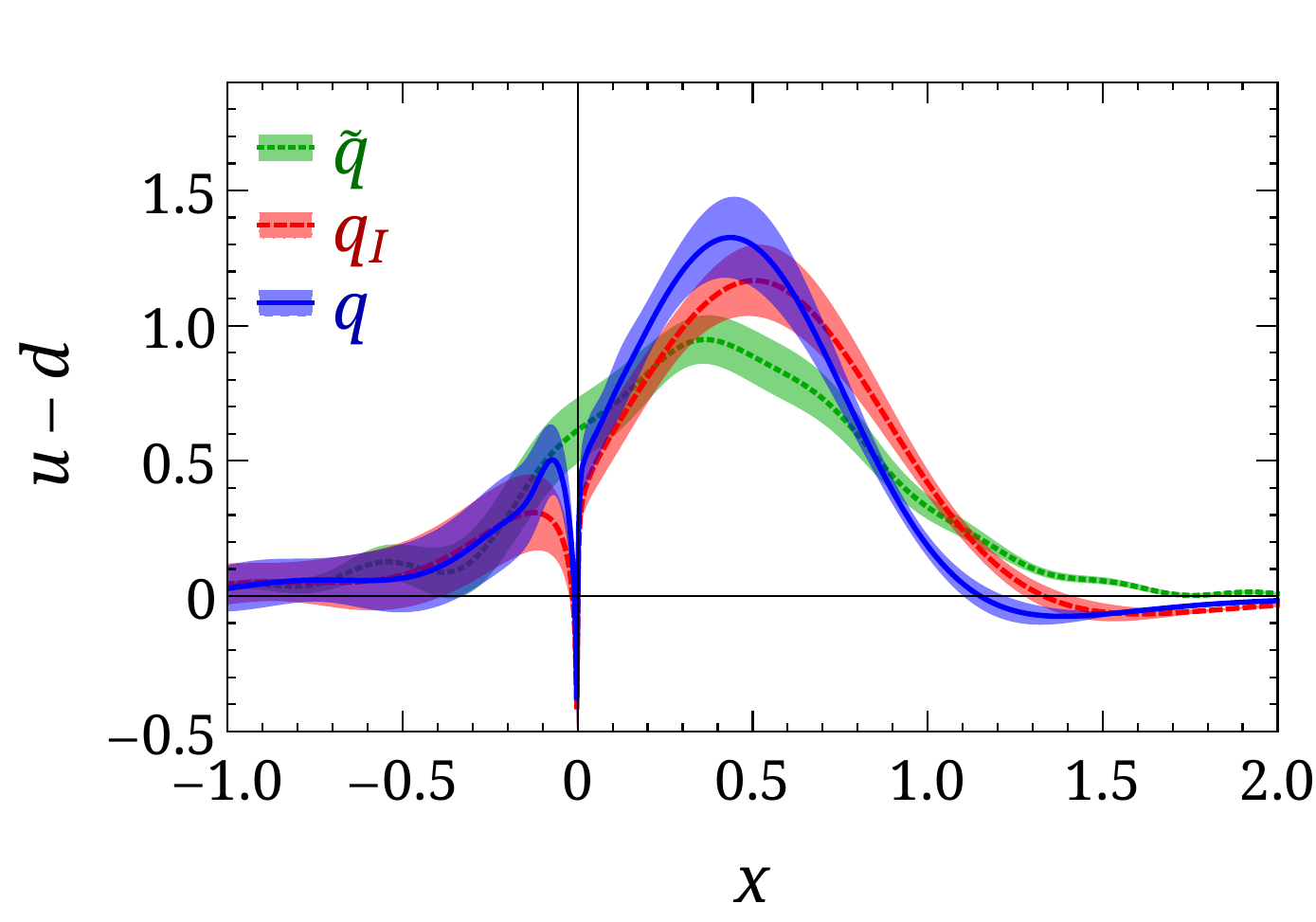}
\includegraphics[width=0.32\textwidth]{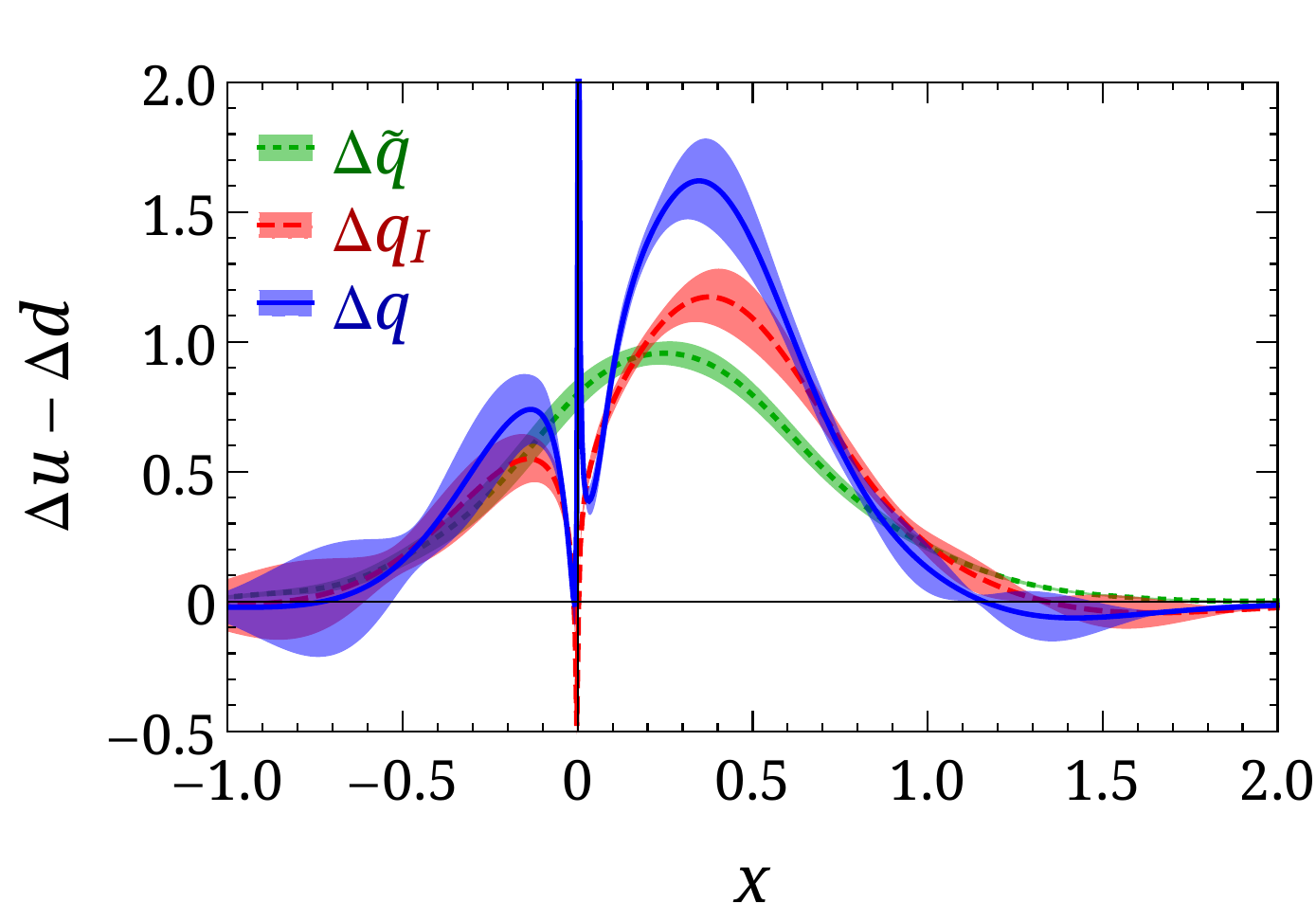}
\includegraphics[width=0.32\textwidth]{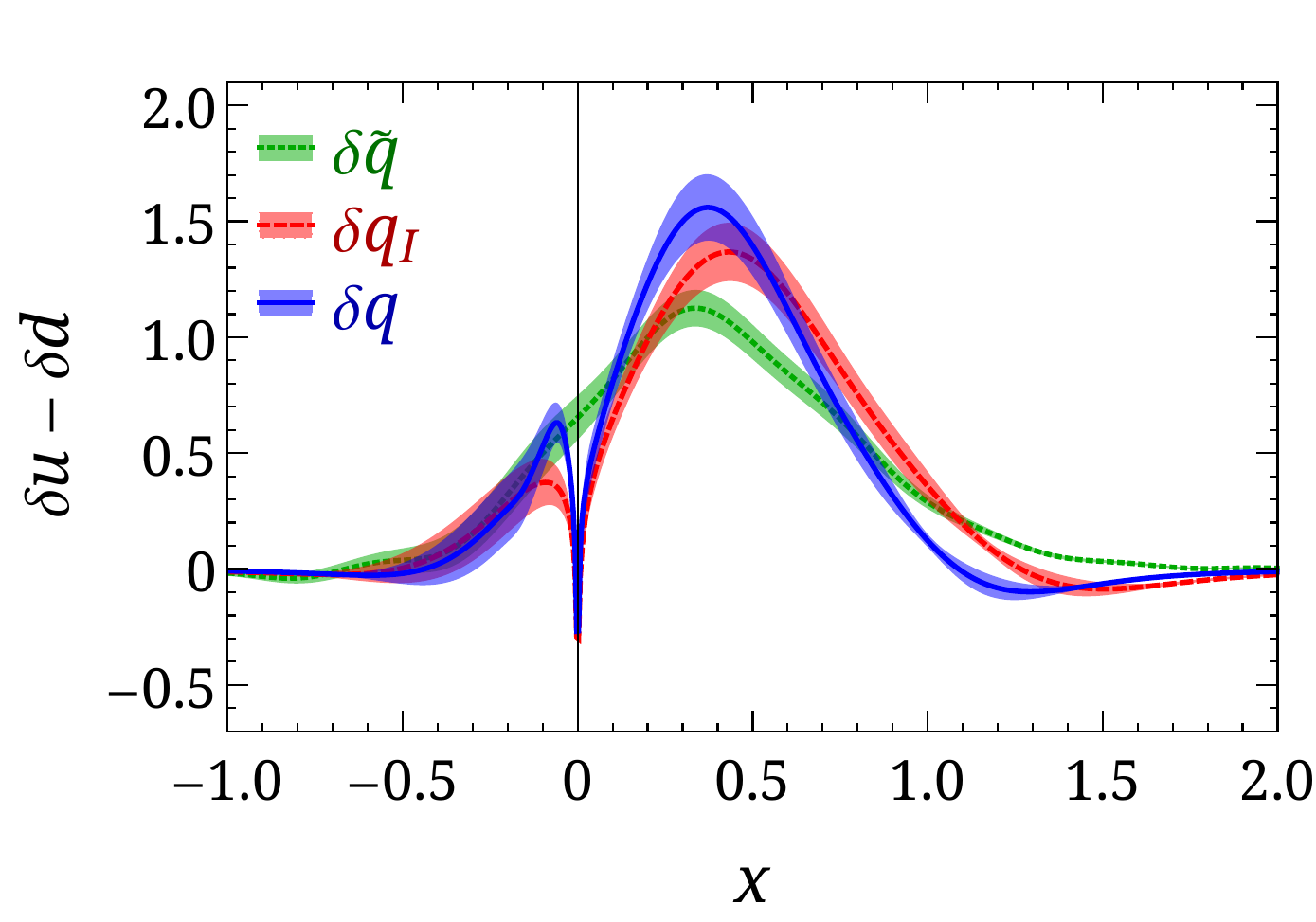}

\caption{The nucleon isovector quasi-PDF (green), with one-loop correction (red), 
and with after one-loop and mass correction (i.e. $q_{II}$).
(blue) for the
 quark density (left), helicity (middle) and transversity (right) as functions of $x$ for the higher two boosted 
momenta $P_z = 2$ (top row) and $3$ (bottom row) in units of $2\pi/L$. 
}
\label{fig:corrections}
\end{figure}

To improve the quasi-distribution closer to the infinite-momentum frame (IMF) proton distribution functions, we follow the recipes described in
Sec.~\ref{sec:corrections} for the one-loop and mass corrections. 
The effects of the one-loop (with $\alpha_s$ set to 0.2) and the final quark distribution (one-loop first, 
followed by mass correction) and original quasi-distribution are shown in Fig.~\ref{fig:corrections} for $P_z=2$ and $P_z=3$. We found that corrections for $P_z=1$ distributions are poorly behaved due to the smallness of the boosted momentum; the results are ignored here. 
First, we compare the quasi- (green band) and one-loop--corrected (red band) distributions. 
For quark density, helicity and transversity distributions, 
we find a significant dip caused by the one-loop correction near $x=0$.
The depth of this dip 
increases as we increase the resolution in $x$, $dx$, used in the integral; 
this artifact may disappear with proper one-loop renormalization in the future calculations.
We also observe a clear evidence of higher values of the peak in the positive-$x$ area and pushing outward of the peak location of the distribution. In the large-$x$ region, the distribution is pulling back, making it rarer for quarks to carry a large
fraction of momentum as one approaches the IMF, which is what we expect. 
For the $P_z=3$ distribution, the magnitude of the changes due to the one-loop correction decrease, as expected.
As we expand the reach of the lattice calculation to larger values of $P_z$, the corrections will be even smaller. 
The pushing outward in the large-$x$ region
may be caused
by the validity of the one-loop correction requiring larger momentum. Future calculations should be
designed to study this further with larger momentum and higher statistics. 

We then apply the mass-correction formula to the one-loop--corrected distribution, shown as blue bands in 
Fig.~\ref{fig:corrections} for all distributions and both $P_z\in\{2,3\}$. 
The peaks are shifted toward $x=0$, the distribution sharpens, and the large-$x$ region distribution is suppressed further, as expected. 
In both the quark density and transversity distributions,
the mass correction also reduces the depth of the dip caused by the one-loop correction formula, 
and the effect of the mass correction also diminishes for the $P_z=3$ case. 
However, for the helicity, the mass-correction
causes a significant unphysical spike rising near $x=0$ due to the singularity in the double-integral terms. 
We note that the peak significantly decreases between $P_z=2$ and $P_z=3$,
and this should be reduced with larger $P_z$ data in the future. 
The height of the peak depends on the resolution of the integral, but has very small effect on the zeroth moment.  
In addition, the mass-correction formulae used in this paper differ from
what we used in our earlier publication, Ref.~\cite{Lin:2014zya}.
This change shifts the central value of the unpolarized and longitudinally polarized up-down quark
asymmetry and increases the estimated errors. However, the results remain consistent within the given errors.

\begin{figure}
\includegraphics[width=0.32\textwidth]{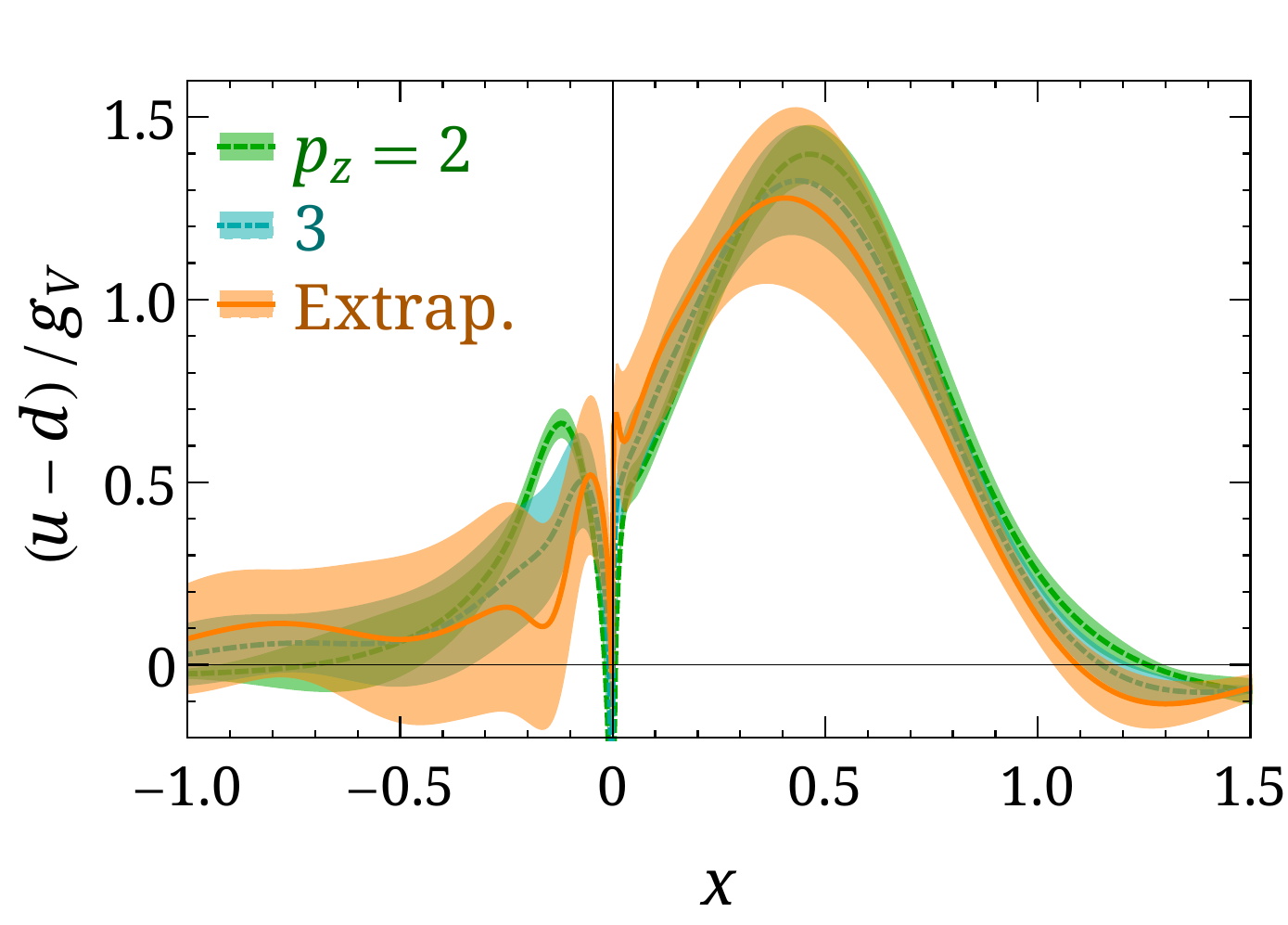}
\includegraphics[width=0.32\textwidth]{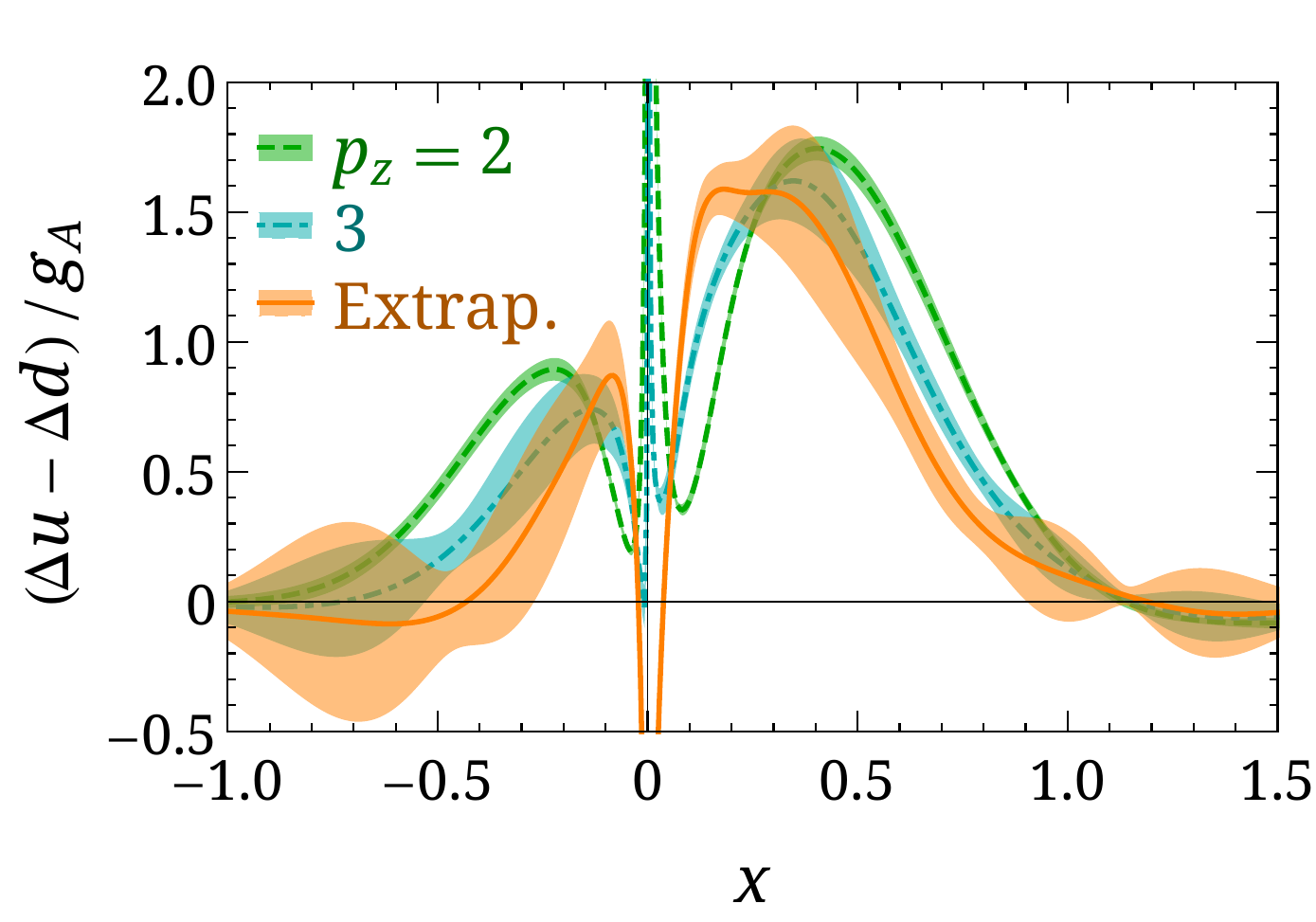}
\includegraphics[width=0.32\textwidth]{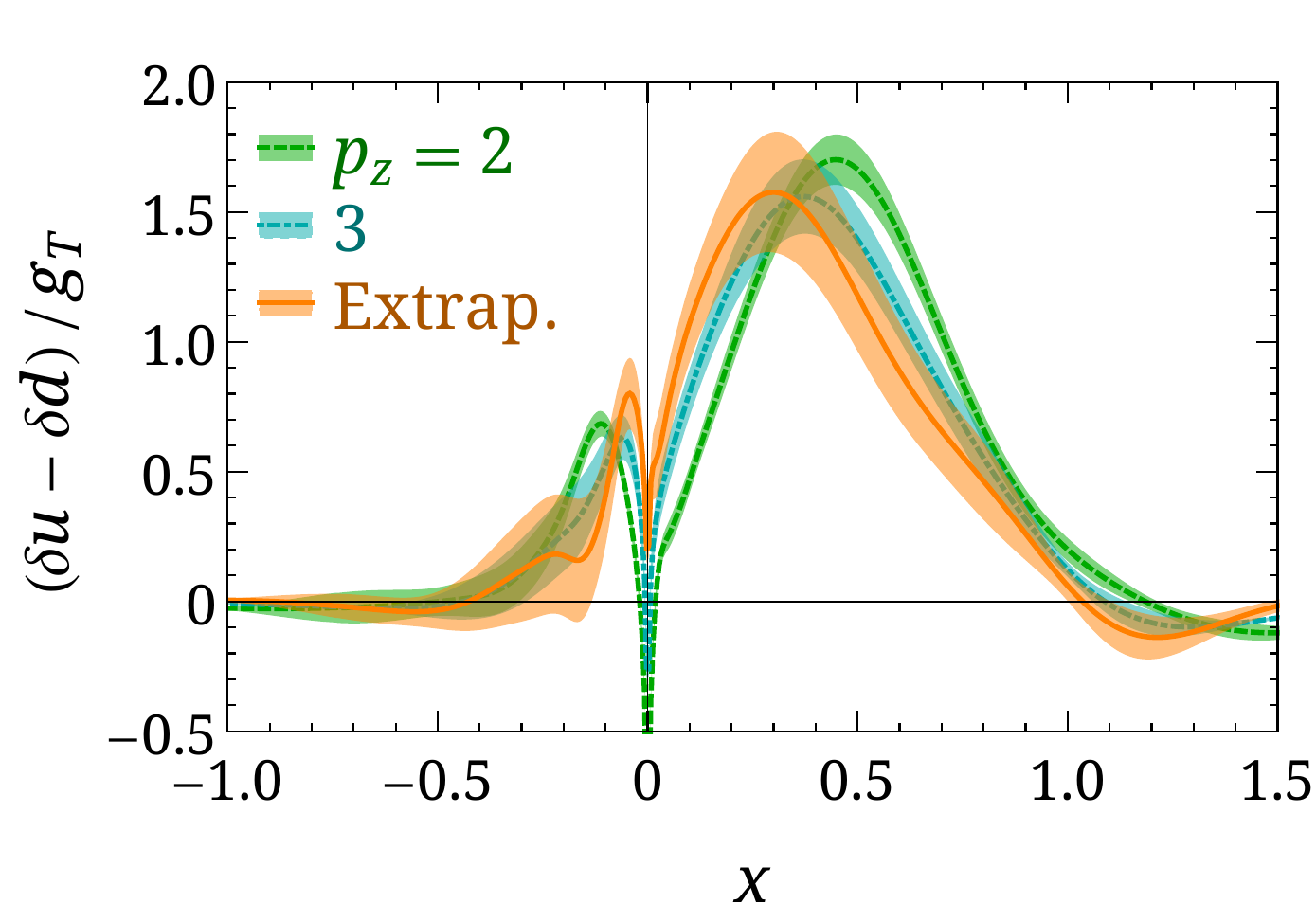}
\vspace*{8pt}
\caption{The momentum-dependence of the nucleon isovector distributions after one-loop and mass correction (i.e. $q_{II}$) for 
 quark density (left), helicity (middle) and transversity (right) as functions of $x$. 
The orange band shows the momentum extrapolation using the higher two momenta.
}
\label{fig:IpM-distribution}
\end{figure}

To further reduce the remaining
$\mathcal{O}(\Lambda_\text{QCD}^2/P_z^2)$ correction due to higher-twist operators, 
we extrapolate to infinite momentum using the form $a + b /P_z^2$ at each $x$ point.
The resulting distribution, shown in Fig.~\ref{fig:IpM-distribution}, has $|x| > 1$ region within 2 sigma of zero;
thus, we recover the correct support for the physical distribution within error.
Note that the smallest reliable region of $x$ is related to the largest
momentum on available on the lattice $\mathcal{O}(1/a)$, which is roughly the
inverse of length of the lattice volume in the link direction; therefore, 
we expect large systematic uncertainty in the region $x \in [-0.08,0.08]$.
In the case of quark density,
there are also indications of momentum convergence within 2 sigma from $P_z=2$ and 3 data. 
In addition, the final extrapolated distribution (orange band) is consistent with the largest momentum 
distribution. However, for the polarized distributions, even larger $P_z$ calculations are needed to improve the convergence rate and reduce the uncertainty due to extrapolation, especially for the helicity.
 
There are many aspects that need to be improved to get the systematics under control, as indicated at various points in the earlier sections. The operator renormalization also needs to be determined to one-loop level or better in the future calculations. We intend in this work mainly to demonstrate that one can achieve light-cone quantities with reasonable accuracy using currently available computational resources, and it opens the door for many more lattice-QCD calculations on parton physics.

\section{Discussion}\label{sec:discussion}

\begin{figure*}
\includegraphics[width=0.47\textwidth]{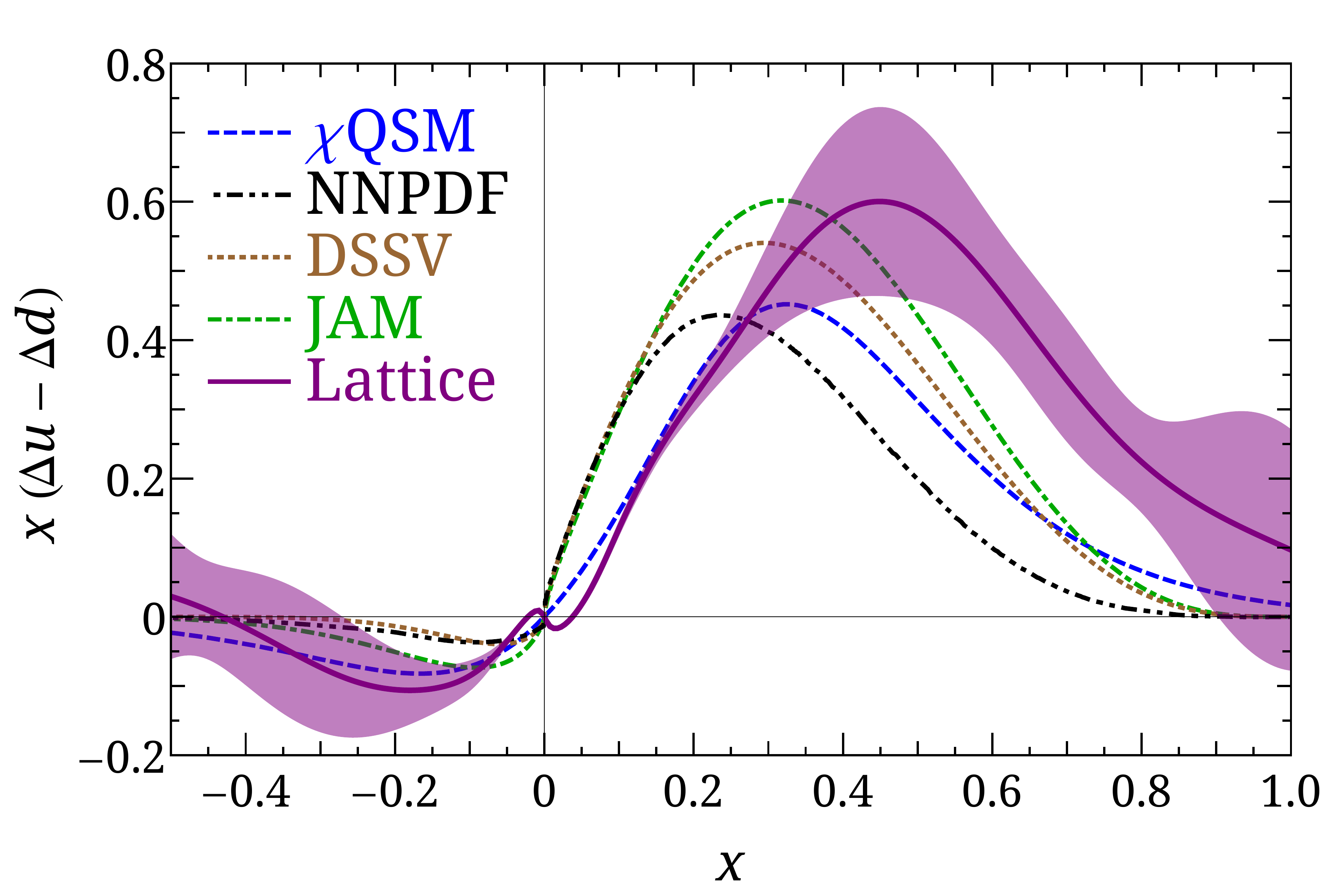}
\includegraphics[width=0.47\textwidth]{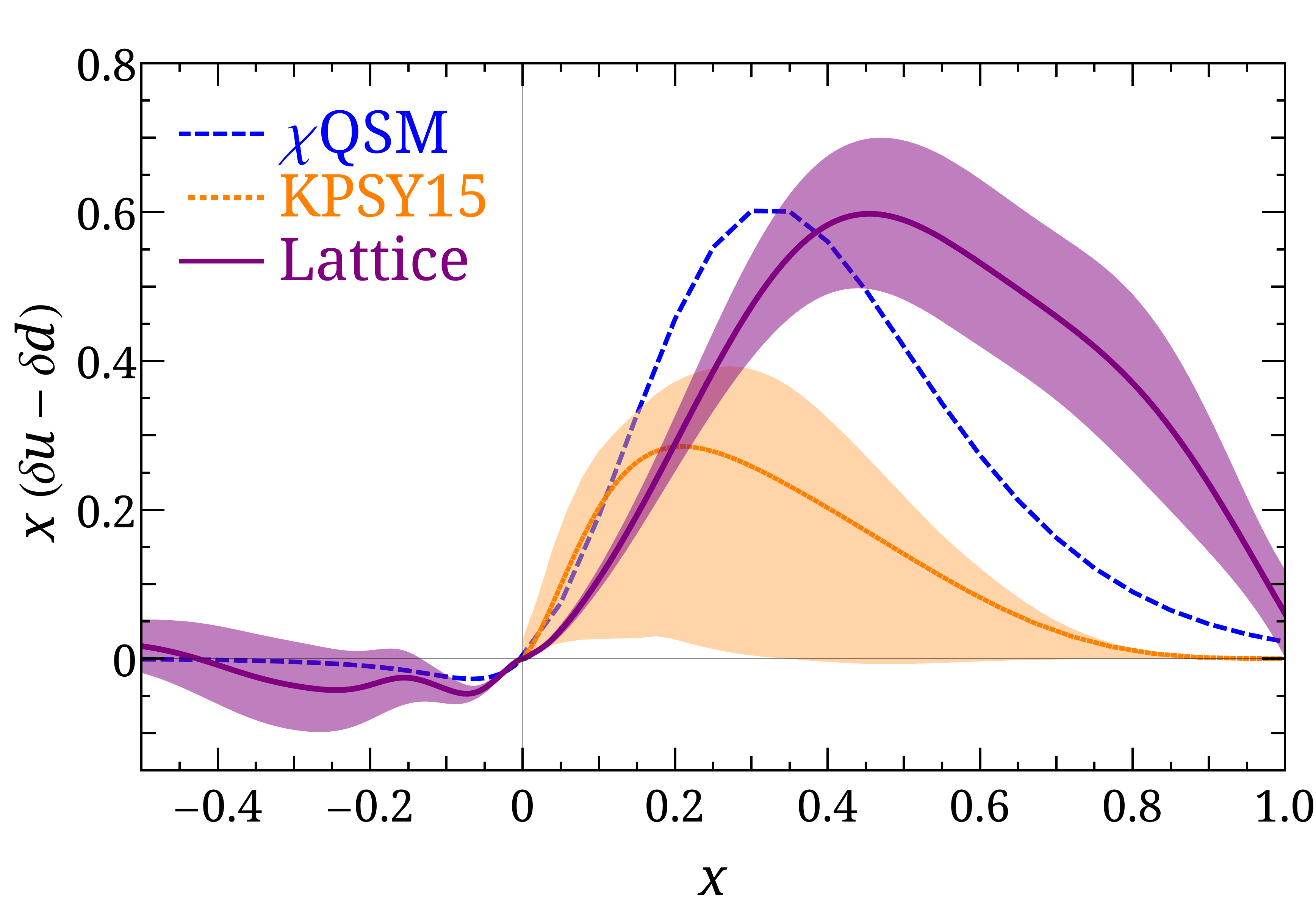}
\caption{(left)
The isovector helicity distribution $x (\Delta u(x)-\Delta d(x))$
(purple band) computed on the lattice, along with selected
global polarized analyses by
NNPDFpol1.1~\cite{Nocera:2014gqa},
JAM~\cite{Jimenez-Delgado:2013boa} (green dot-dashed) and
DSSV09~\cite{deFlorian:2009vb} (brown dotted line),
and a model calculation $\chi$QSM~\cite{Schweitzer:2001sr} (blue dashed line).
Note that the uncertainties in the global analyses are omitted here for visibility reasons.
(right)
The isovector transversity distribution $x(\delta u(x)-\delta d(x))$ computed on the lattice in this work,
along with $\chi$QSM~\cite{Schweitzer:2001sr}
(blue dashed line) and latest phenomenological analysis from Ref.~\cite{Kang:2015msa} (labeled as KPSY15, Orange band).
The corresponding sea-quark distributions are 
$\Delta \overline{q}(x) = \Delta q(-x)$ 
and
$\delta \overline{q}(x) = -\delta q(-x)$.}
\label{fig:helicity-transversity}
\end{figure*}

In this section, we take the distributions from the previous section after all corrections, shown in Fig.~\ref{fig:helicity-transversity}, and discuss
the physics implications. We will focus on the results new to this paper, mainly the
isovector helicity and nucleon transversity distribution. 
We strongly believe that it is worth more lattice-QCD effort to improve our knowledge of the polarized
PDFs, which still lack precision experimental data over most $x$ regions, especially
the antiquark distribution, which we emphasize in this section. 

The left panel of Fig.~\ref{fig:helicity-transversity} shows the helicity distribution
$x (\Delta u(x)-\Delta d(x))$, along with selected recent global analyses
by JAM~\cite{Jimenez-Delgado:2013boa}, CCVS09~\cite{deFlorian:2009vb}, and
NNPDFpol1.1~\cite{Nocera:2014gqa}, whose nucleon isovector distribution uncertainties have been ignored.
Also note that the plots now show the distribution multiplied by $x$,
since this form is used in global-analysis parametrizations.
We see more weight distributed in the large-$x$ region, which could shift toward smaller $x$ as we lower 
the quark masses. This is because lower quark mass increases the long-range correlations in ${\Delta h}_\text{lat}(z)$,
which in turn increases the small-$x$ contribution in the Fourier transformation. 
Since the increasing small-$x$ distribution will decrease the large-$x$ distribution due to charge conservation,
we expect this inconsistency to reduce as we go to smaller quark mass. 
We see there are noticeable differences between the extracted polarized PDFs
depending on the experimental cuts, theory inputs, parametrization, and so on.
For example, JAM excludes SIDIS data, leaving the sign of the light antiquark
determined by the valence and the magnitude determined from sum rules. 
DSSV also relies on assumptions such as SU(3) symmetry to constrain the 
analysis and adds a very small symmetry-breaking term. A direct lattice study
of hyperon axial couplings~\cite{Lin:2007ap} suggested that SU(3) breaking is
roughly 20\% at the physical point, bigger than these assumptions. 
Similar assumptions also made in the NNPDFpol1.1~\cite{Nocera:2014gqa}.
These assumptions are unavoidable due to the difficulties of getting constraint data from polarized experiments.
Future experiments with neutral- and
charged-current DIS (such as at EIC) will provide useful measurements
to constrain our understanding of the antiquark helicity distribution.

Our result for antiquark helicity favor more polarized up quark than down
flavor\footnote{Note that one should ignore the $x\in\{-0.08,0.08\}$ regions
since there is large uncertainty associated with the distribution in this region.},
which is consistent with the current PDF analysis and model
calculations, such as chiral quark soliton model
($\chi$QSM)~\cite{Schweitzer:2001sr}. This was first pointed out in
our earlier paper~\cite{Lin:2014zya} (using a different mass-correction formulation), which concentrated on
the sea flavor asymmetry in the unpolarized distribution; it was also noted in
preliminary studies in conference proceedings~\cite{Lin:2014gaa,Lin:2014yra,Lin:2015vxw}. The sea flavor asymmetry was
confirmed in the full analysis of the Run-9 data by both
STAR~\cite{Adamczyk:2014xyw} and PHENIX~\cite{Adare:2015gsd} collaborations.
RHIC experiments on longitudinal single-spin asymmetry and
parity-violating $W$ production at RHIC might shed more light on the
polarized sea distribution~\cite{Aschenauer:2013woa}.

We see a moderate polarized total sea asymmetry,
$\int_{0.08}^1 \Delta \overline{u}(x)-\Delta \overline{d}(x)=0.14(9)$, which is smaller than the previous determination~\cite{Lin:2014zya} but still consistent within errors. The update is due to the application of the mass-correction formula of Eq.~\ref{h_mass} instead of Eq.~\ref{Knbar}. The latter requires transforming back and forth between the PDF and moments, which introduces oscillatory artifacts.
Most QCD models predict smaller polarized sea asymmetry;
for example, see the recent review article by Chang and Peng,
in particular Table~5 of Ref.~\cite{Chang:2014jba}. 
$\chi$QSM, a large-$N_c$ model,
gives rather different results by predicting a large polarized sea asymmetry: 0.31.
Unfortunately, our current statistical error does not   
help rule out many models yet based on the total sea asymmetry.
On the experimental global analysis side, the total polarized sea asymmetry estimated by DSSV09 is
consistent with zero within 2 sigma, and the central value is also smaller ($\approx 0.07$)
than the unpolarized case. Current results for STAR~\cite{Adamczyk:2014xyw}
and PHENIX~\cite{Adare:2015gsd} in the middle-$x$ range do not clarify what the total
asymmetry would be.
The upcoming RHIC data from Run-13 with significantly improved statistics may
shed some light on this matter.
The upcoming Fermilab Drell-Yan experiments (E1027/E1039) can also provide precise
experimental input on the polarized sea asymmetry magnitude.

The transversity distribution is the least known PDF among the three
PDF structures studied;
there is much less information available due to the difficulties in experiments.
There have been a few attempts to extract the transversity distribution, but
they suffer from fundamental defects.
Ref.~\cite{Anselmino:2008jk} makes various assumptions such as the evolution form and that there is no antiquark contribution.
Ref.~\cite{Radici:2013vra} uses dihadron fragmentation functions using data from HERMES and COMPASS analysis of pion-pair production in DIS off a transversely polarized target for two combinations of ``valence'' ($q+\bar{q}$)
helicity distribution. They have a proper $Q^2$ evolution but still rely on
assumptions such as the Soffer inequality.
Kang et~al.~\cite{Kang:2015msa} has improved evolutions implemented in their analysis, but they also make the assumption
that the sea asymmetry is zero. The distribution for the positive $x$ goes quickly to zero,
likely due to lack of data.

Our transversity result is shown in the
the right panel of Fig.~\ref{fig:helicity-transversity}, along with an estimate from a QCD model,
$\chi$QSM~\cite{Schweitzer:2001sr} and the latest transversity fit from Ref.~\cite{Kang:2015msa}\footnote{Note that the error band of this isovector structure from Kang et~al. has been added up linearly from the up and down components due to the asymmetric errorband reported in the components; the error given here might be larger than if derived using the correlations of the original analysis. Also, note that the scale is set at $10\mbox{ GeV}^2$; however, there is only a small difference in their central values between lower-$Q^2$ scale and $10\mbox{ GeV}^2$.}.
Surprisingly, our result is rather similar to $\chi$QSM within 90\% confidence, but with slower descent to zero in 
the $x\approx 1$ region, similar to the quark distribution. This can be, again, due to the heavier pion mass used in the 
calculation, as well as the need to push for even larger momenta.
In contrast, the phenomenological results from Ref.~\cite{Kang:2015msa} fall faster as $x$ approaches near 1.

Our result favors $\delta \overline{d}(x) > \delta \overline{u}(x)$ with total sea asymmetry
0.10(8), whose central value is still larger than most model predictions (for example, $\chi$QSM estimates $0.082$ asymmetry)
and in contradiction to the assumption that the antiquark is consistent with zero in some
transversity extractions using experimental data~\cite{Anselmino:2008jk,Radici:2013vra,Kang:2014zza,Kang:2015msa}. 
One interesting thing to note is that the central values of the lattice determination of the tensor charge $g_T$ (that is, $\int_{-1}^{+1} \delta u(x) -\delta d(x)$) extrapolated to the continuum limit from various groups are consistently higher than the phenomenological ones who assume zero total sea asymmetry in transversity; see the summary plot Fig.~10 in Ref.~\cite{Bhattacharya:2015wna}. This may indicate nonzero sea contribution with the same sign as our prediction here, or missing larger-$x$ data in containing their fit. 
It would be interesting to see whether such a nonzero sea asymmetry remains in the future high-statistics physical quark mass ensemble; it is certainly contrary to traditional expectation. Improved phenomenological analysis with new experimental data would also help to narrow the phenomenological uncertainties and explore the discrepancy.

The cleanest measurement of the transversity would have both a polarized beam and polarized target, but given the limited setups available, once again, more data are needed.
PHENIX and STAR will be able to help give more insight into this quantity.
Planned experiments, such as SoLID at Jefferson Lab, can provide good transversity measurements for a wide range of positive $x$.
The Drell-Yan experiment at FNAL (E1027+E1039) can in principle extract
sea-asymmetry information in the near future to settle the size of the total transversely polarized sea.

\section*{Acknowledgments}

The LQCD calculations were performed using the Chroma software
suite~\cite{Edwards:2004sx}.
Computations for this work were carried out in part on facilities of the
USQCD Collaboration, which are funded by the Office of Science of the
U.S. Department of Energy, and on Hyak clusters at the University of Washington
managed by UW Information Technology, using hardware awarded by NSF grant
PHY-09227700.
We thank MILC Collaboration for sharing the lattices used to perform this study.
This work was partially supported by the U.S. Department of Energy via grants DE-FG02-93ER-40762, a grant (No.~11DZ2260700) from the Office of Science and Technology in Shanghai Municipal Government, grants from National Science Foundation of China (No.~11175114, No.~11405104), a DFG grant SCHA~458/20-1, the Ministry of Science and Technology, Taiwan under Grant Nos.~102-2112-M-002-013-MY3 and 105-2918-I-002 -003 and the CASTS of NTU. The work of HWL and SDC were supported by the DOE grant DE-FG02-97ER4014 and DE-FG02-00ER41132.
The work of HWL is currently supported in part by the M. Hildred Blewett Fellowship of the American Physical Society, www.aps.org.
HWL thanks the Institute for Nuclear Theory at the University of Washington for its hospitality during the completion of this work. JWC would like to thank the hospitality of the Rudolph Peierls Centre for Theoretical Physics of the University of Oxford and Oxford Holography group, DAMTP of University of Cambridge, and 
Helmholtz-Institut f\"{u}r Strahlen- und Kernphysik and Bethe Center for Theoretical Physics, Universit\"{a}t Bonn.

\section*{Appendix A: One-Loop Matching}\label{sec:oneloop_appendix}

In this Appendix, we list the one-loop matching factors used throughout this paper. As the UV cutoff in a practical calculation is finite, we use the results of Eqs.~5, 6, 21 and 24 in Ref.~\cite{Xiong:2013bka}. 

In the unpolarized case, the matching factors are given as follows:
\begin{align}\label{unpolvertexnoexp}
Z^{(1)}(x)/C_F&=\left\{ \begin{array} {ll} \frac{1+x^2}{1-x}\ln \frac{x(\Lambda(x)-x P^z)}{(x-1)(\Lambda(1-x)+P^z(1-x))}+1-\frac{x P^z}{\Lambda(x)}+\frac{x\Lambda(1-x)+(1-x)\Lambda(x)}{(1-x)^2 P^z}\ , & x>1\ , \\
\ \\
\frac{1+x^2}{1-x}\ln\frac{(P^z)^2}{\mu^2}+\frac{1+x^2}{1-x}\ln \frac{4x(1-x)(\Lambda(x)-x P^z)}{\Lambda(1-x)+(1-x)P^z}-\frac{2x}{1-x}+1-\frac{x P^z}{\Lambda(x)}\ & \\
+\frac{x\Lambda(1-x)+(1-x)\Lambda(x)}{(1-x)^2 P^z}\ , & 0<x<1\ , \\
\ \\
\frac{1+x^2}{1-x}\ln \frac{(x-1)(\Lambda(x)-x P^z)}{x(\Lambda(1-x)+(1-x)P^z)}-1-\frac{x P^z}{\Lambda(x)}+\frac{x\Lambda(1-x)+(1-x)\Lambda(x)}{(1-x)^2P^z}\ , & x<0 \ ,\end{array} \right.
\end{align}
where $\Lambda(x) = \sqrt{\Lambda^2+x^2 P_z^2}$. We have not taken the $\Lambda \gg x P_z$ limit, because they could be the same order on the lattice.

Near $x=1$, one has an extra contribution from the self-energy correction
\begin{align}\label{selfnoexp}
\delta Z^{(1)}_F/C_F&=\int dy \left\{ \begin{array} {ll} -\frac{1+y^2}{1-y}\ln \frac{y(\Lambda(y)-y P^z)}{(y-1)(\Lambda(1-y)+P^z(1-y))}-1-\frac{y\Lambda(1-y)+(1-y)\Lambda(y)}{(1-y)^2 P^z}\ & \\
+\frac{y^2 P^z}{\Lambda(y)}+\frac{y(1-y)P^z}{\Lambda(1-y)}+\frac{\Lambda(y)-\Lambda(1-y)}{P^z}\ , & y>1\ , \\
\ \\
-\frac{1+y^2}{1-y}\ln\frac{(P^z)^2}{\mu^2}-\frac{1+y^2}{1-y}\ln \frac{4y(1-y)(\Lambda(y)-y P^z)}{\Lambda(1-y)+(1-y)P^z}+\frac{2y(2y-1)}{1-y}+1\ & \\
-\frac{y\Lambda(1-y)+(1-y)\Lambda(y)}{(1-y)^2P^z}+\frac{y^2 P^z}{\Lambda(y)}+\frac{y(1-y)P^z}{\Lambda(1-y)}+\frac{\Lambda(y)-\Lambda(1-y)}{P^z}\ , & 0<y<1\ , \\
\ \\
-\frac{1+y^2}{1-y}\ln \frac{(y-1)(\Lambda(y)-y P^z)}{y(\Lambda(1-y)+(1-y)P^z)}+1-\frac{y\Lambda(1-y)+(1-y)\Lambda(y)}{(1-y)^2P^z}\ & \\
+\frac{y(1-y)P^z}{\Lambda(1-y)}+\frac{y^2 P^z}{\Lambda(y)}+\frac{\Lambda(y)-\Lambda(1-y)}{P^z}\ , & y<0\ . \end{array} \right.
\end{align}
For the helicity distribution, we have
\begin{align}\label{helivertexnoexp}
\Delta  Z^{(1)}(x)/C_F&=\left\{ \begin{array} {ll} \frac{1+x^2}{1-x}\ln \frac{x(\Lambda(x)-x P^z)}{(x-1)(\Lambda(1-x)+P^z(1-x))}+1-\frac{x P^z}{\Lambda(x)}+\frac{x\Lambda(1-x)+(1-x)\Lambda(x)}{(1-x)^2 P^z}\ , & x>1\ , \\
\ \\
\frac{1+x^2}{1-x}\ln\frac{(P^z)^2}{\mu^2}+\frac{1+x^2}{1-x}\ln \frac{4x(1-x)(\Lambda(x)-x P^z)}{\Lambda(1-x)+(1-x)P^z}-\frac{2}{1-x}+3\ & \\
-\frac{x P^z}{\Lambda(x)}+\frac{x\Lambda(1-x)+(1-x)\Lambda(x)}{(1-x)^2 P^z}\ , & 0<x<1\ . \\
\ \\
\frac{1+x^2}{1-x}\ln \frac{(x-1)(\Lambda(x)-x P^z)}{x(\Lambda(1-x)+(1-x)P^z)}-1-\frac{x P^z}{\Lambda(x)}+\frac{x\Lambda(1-x)+(1-x)\Lambda(x)}{(1-x)^2P^z}\ , & x<0\ , \end{array} \right.
\end{align}
and for the transversity distribution, we have
\begin{align}\label{transvertexnoexp}
\delta  Z^{(1)}(x)/C_F&=\left\{ \begin{array} {ll} \frac{2x}{1-x}\ln \frac{x(\Lambda(x)-xP^z)}{(x-1)(\Lambda(1-x)+P^z(1-x))}+\frac{x\Lambda(1-x)+(1-x)\Lambda(x)}{(1-x)^2 P^z}\ , & x>1\ , \\
\ \\
\frac{2x}{1-x}\ln\frac{(P^z)^2}{\mu^2}+\frac{2x}{1-x}\ln \frac{4x(1-x)(\Lambda(x)-x P^z)}{\Lambda(1-x)+(1-x)P^z}-\frac{2x}{1-x}+\frac{x\Lambda(1-x)+(1-x)\Lambda(x)}{(1-x)^2 P^z}\ , & 0<x<1\ , \\
\ \\
\frac{2x}{1-x}\ln \frac{(x-1)(\Lambda(x)-x P^z)}{x(\Lambda(1-x)+(1-x)P^z)}+\frac{x\Lambda(1-x)+(1-x)\Lambda(x)}{(1-x)^2P^z}\ , & x<0\ , \end{array} \right.
\end{align}
where the quark self-energy contribution is the same as Eq.~\ref{selfnoexp}.

\section*{Appendix B: Target-Mass Correction for Quasi-Distributions}
\label{sec:tmc_appendix}

In this Appendix, we derive the target-mass corrections to the unpolarized and helicity parton distributions. 

\subsection{Unpolarized distribution}

For the unpolarized parton distribution, the series sum in Eq.~\ref{Kn} can be explicitly performed, and the result for an even $n$ ($=2k$) is
\begin{equation}\label{evensum}
\sum_{j=0}^k  C_{n-j}^j c^j =\frac{1}{\sqrt{1+4c}}\Big[\left(\frac{f_-}{2}\right)^{2k+1}+\left(\frac{f_+}{2}\right)^{2k+1}\Big],
\end{equation}
and for an odd $n$ ($=2k+1$) is
\begin{equation}\label{oddsum}
\sum_{j=0}^k C_{n-j}^j c^j =\frac{1}{\sqrt{1+4c}}\Big[-\left(\frac{f_-}{2}\right)^{2k+2}+\left(\frac{f_+}{2}\right)^{2k+2}\Big].
\end{equation}
where $f_{\pm}=\sqrt{1+4c}\pm 1$. 

With Eqs.~\ref{evensum} and \ref{oddsum}, we perform an inverse Mellin transform to the moment relation of Eq.~\ref{Kn}
\begin{equation}
\frac{1}{2\pi i}\int_{-i\infty}^{i\infty} dn \, s^{-n} \langle x^{n-1} \rangle .
\end{equation}
For $n=2k$, we obtain
\begin{equation}\label{pdfrel}
\tilde q(x)-\tilde q(-x)=\frac{1}{\sqrt{1+4c}}\sum_{i=\pm}\frac{f_i}{2}\Big[q\Big(\frac{2x}{f_i}\Big)-q\Big(\frac{-2x}{f_i}\Big)\Big],
\end{equation}
and for $n=2k+1$, we have
\begin{equation}\label{oddpdfrel}
\tilde q(x)+\tilde q(-x)=\frac{1}{\sqrt{1+4c}}\Big\{\frac{f_+}{2}\Big[q\Big(\frac{2x}{f_+}\Big)+q\Big(\frac{-2x}{f_+}\Big)\Big]-\frac{f_-}{2}\Big[q\Big(\frac{2x}{f_-}\Big)+q\Big(\frac{-2x}{f_-}\Big)\Big]\Big\},
\end{equation}
where we have used the following representation for the Dirac $\delta$-function~\cite{Steffens:2012jx}
\begin{equation}
\delta(\ln u)=\frac{1}{2\pi i}\int_{-i\infty}^{i\infty}dn\, u^n.
\end{equation}
From the above two equations, we have
\begin{equation}\label{qinq0}
\tilde q(x)=\frac{1}{\sqrt{1+4c}}\Big[\frac{f_+}{2}q\Big(\frac{2x}{f_+}\Big)-\frac{f_-}{2}q\Big(\frac{-2x}{f_-}\Big)\Big].
\end{equation}
Note that this is different from the result obtained in Ref.~\cite{Alexandrou:2015rja}. Their result does not conserve quark number. In our case it is easy to check the quark-number conservation
\begin{equation}
\int_{-\infty}^{\infty} dx\, \tilde q(x)=\frac{1}{\sqrt{1+4c}}\int dx \big[\frac{f_+^2}{4}-\frac{f_-^2}{4}\big]q(x)=\int_{-\infty}^{\infty} dx\, q(x),
\end{equation}
or $\langle x^0\rangle_{\tilde q}=\langle x^0\rangle_q$.

Eq.~\ref{qinq0} writes $\tilde q(x)$ in terms of $q(x)$, where the former is the quantity that can be directly computed on the lattice and the latter is the usual parton distribution. In practice, we would like to extract $q(x)$ from $\tilde q(x)$. To see how this can be done, let us rewrite Eq.~\ref{Kn} for an even $n=2k$ as
\begin{equation}\label{momentinverserel}
\langle x^{2k-1}\rangle_q=\langle x^{2k-1}\rangle_{\tilde q}\frac{\sqrt{1+4c}}{\left(\frac{f_-}{2}\right)^{2k+1}+\left(\frac{f_+}{2}\right)^{2k+1}}=\langle x^{2k-1}\rangle_{\tilde q}\frac{\sqrt{1+4c}}{\left(\frac{f_+}{2}\right)^{2k+1}}\sum_{n=0}^\infty (-1)^n\left(\frac{f_-}{f_+}\right)^{(2k+1)n}.
\end{equation}
The inverse Mellin transform then leads to
\begin{align}
q(x)-q(-x)=\sqrt{1+4c}\sum_{n=0}^\infty \frac{2(-f_-)^n}{f_+^{n+1}}\Big[\tilde q\Big(\frac{f_+^{n+1}x}{2f_-^n}\Big)-\tilde q\Big(\frac{-f_+^{n+1}x}{2f_-^n}\Big)\Big].
\end{align}
Similarly, we have
\begin{equation}
q(x)+q(-x)=\sqrt{1+4c}\sum_{n=0}^\infty \frac{2 f_-^n}{f_+^{n+1}}\Big[\tilde q\Big(\frac{f_+^{n+1}x}{2f_-^n}\Big)+\tilde q\Big(\frac{-f_+^{n+1}x}{2f_-^n}\Big)\Big].
\end{equation}
Therefore,
\begin{align}
q(x)&=\sqrt{1+4c}\sum_{n=0}^\infty \frac{f_-^n}{f_+^{n+1}}\Big[(1+(-1)^n)\tilde q\Big(\frac{f_+^{n+1}x}{2f_-^n}\Big)+(1-(-1)^n)\tilde q\Big(\frac{-f_+^{n+1}x}{2f_-^n}\Big)\Big]\non\\
&=\sqrt{1+4c}\sum_{n=0}^\infty \frac{(4c)^n}{f_+^{2n+1}}\Big[(1+(-1)^n)\tilde q\Big(\frac{f_+^{2n+1}x}{2(4c)^n}\Big)+(1-(-1)^n)\tilde q\Big(\frac{-f_+^{2n+1}x}{2(4c)^n}\Big)\Big],
\end{align}
where in the last line we have used $f_+ f_-=4c$. Since $f_+\gg f_-$ or $c$ and the quasi-distribution $\tilde q(x)$ vanishes asymptotically for large $x$, the above sum is dominated by the first term with $n=0$. In practical calculations, we can reach a reasonable accuracy by taking only the first few terms in the sum.

\subsection {Helicity distribution}

Now let us look at the quark helicity distribution. In this case, Eq.~\ref{Knbar} gives the following relation between moments
\begin{equation}
\langle x^{n-1}\rangle_{\Delta q} \frac{1}{n}\sum_{j=0}^{\infty}c^j\frac{(n-j)!}{j!(n-2j-1)!}=\langle x^{n-1}\rangle_{\Delta \tilde q} .
\end{equation}
The result of the series sum for an even $n=2k$ is
\begin{equation}
\sum_{j=0}^{k-1}c^j\frac{(n-j)!}{j!(n-2j-1)!}=\frac{1}{1+4c}\Big[\Big(2k+1 -\frac{1}{\sqrt{1+4c}}\Big)\left(\frac{f_+}{2}\right)^{2k+1}-\Big(2k+1 +\frac{1}{\sqrt{1+4c}}\Big)\left(\frac{f_-}{2}\right)^{2k+1}\Big],
\end{equation}
and for an odd $n=2k+1$ is
\begin{equation}
\sum_{j=0}^{k}c^j\frac{(n-j)!}{j!(n-2j-1)!}=\frac{1}{1+4c}\Big[\Big(2k+2 -\frac{1}{\sqrt{1+4c}}\Big)\left(\frac{f_+}{2}\right)^{2k+2}+\Big(2k+2 +\frac{1}{\sqrt{1+4c}}\Big)\left(\frac{f_-}{2}\right)^{2k+2}\Big].
\end{equation}
From the above equations, we have
\begin{align}
2k \Big[\langle x^{2k-1}\rangle_{\Delta \tilde q}-\frac{\langle x^{2k-1}\rangle_{\Delta q}}{1+4c}\Big(\left(\frac{f_+}{2}\right)^{2k+1}-\left(\frac{f_-}{2}\right)^{2k+1}\Big)\Big]&=\frac{2c \langle x^{2k-1}\rangle_{\Delta q}}{(1+4c)^\frac{3}{2}}\Big[\left(\frac{f_+}{2}\right)^{2k}-\left(\frac{f_-}{2}\right)^{2k}\Big], \non\\
(2k+1) \Big[\langle x^{2k}\rangle_{\Delta \tilde q}-\frac{\langle x^{2k}\rangle_{\Delta q}}{1+4c}\Big(\left(\frac{f_+}{2}\right)^{2k+2}+\left(\frac{f_-}{2}\right)^{2k+2}\Big)\Big]&=\frac{2c \langle x^{2k}\rangle_{\Delta q}}{(1+4c)^\frac{3}{2}}\Big[\left(\frac{f_+}{2}\right)^{2k+1}+\left(\frac{f_-}{2}\right)^{2k+1}\Big].
\end{align}
Their inverse Mellin transform then leads to
\begin{align}
&-\frac{\partial}{\partial s}\Big\{\frac{|s|}{s}\Big[\Delta\tilde q(s)-\Delta\tilde q(-s)-\frac{1}{1+4c}\left(\frac{f_+}{2}\Big(\Delta q\Big(\frac{2s}{f_+}\Big)-\Delta q\Big(\frac{-2s}{f_+}\Big)\Big)-\frac{f_-}{2}\Big(\Delta q\Big(\frac{2s}{f_-}\Big)-\Delta q\Big(\frac{-2s}{f_-}\Big)\Big)\right)\Big]\Big\}\non\\
&=\frac{2c}{(1+4c)^{\frac{3}{2}}|s|}\Big\{\Big[\Delta q\Big(\frac{2s}{f_+}\Big)-\Delta q\Big(\frac{-2s}{f_+}\Big)\Big]-\Big[\Delta q\Big(\frac{2s}{f_-}\Big)-\Delta q\Big(\frac{-2s}{f_-}\Big)\Big]\Big\},\non\\
&-\frac{\partial}{\partial s}\Big\{\frac{|s|}{s}\Big[\Delta \tilde q(s)+\Delta \tilde q(-s)-\frac{1}{1+4c}\left(\frac{f_+}{2}\Big(\Delta q\Big(\frac{2s}{f_+}\Big)+\Delta q\Big(\frac{-2s}{f_+}\Big)\Big)+\frac{f_-}{2}\Big(\Delta q\Big(\frac{2s}{f_-}\Big)+\Delta q\Big(\frac{-2s}{f_-}\Big)\Big)\Big)\right)\Big]\Big\}\non\\
&=\frac{2c}{(1+4c)^{\frac{3}{2}}|s|}\Big\{\Big[\Delta q\Big(\frac{2s}{f_+}\Big)+\Delta q\Big(\frac{-2s}{f_+}\Big)\Big]+\Big[\Delta q\Big(\frac{2s}{f_-}\Big)+\Delta q\Big(\frac{-2s}{f_-}\Big)\Big]\Big\}.
\end{align}
From these two equations, we obtain
\begin{equation}
-\frac{\partial}{\partial x}\Big[\Delta \tilde q(x)-\frac{1}{1+4c}\Big(\frac{f_+}{2}\Delta q\Big(\frac{2x}{f_+}\Big)+\frac{f_-}{2}\Delta q\Big(\frac{-2x}{f_-}\Big)\Big)\Big]=\frac{2c}{(1+4c)^{\frac{3}{2}}x}\Big[\Delta q\Big(\frac{2x}{f_+}\Big)+\Delta q\Big(\frac{-2x}{f_-}\Big)\Big].
\end{equation}
Its solution is given by
\begin{equation}
\Delta \tilde q(x)=\frac{1}{1+4c}\Big(\frac{f_+}{2}\Delta q\Big(\frac{2x}{f_+}\Big)+\frac{f_-}{2}\Delta q\Big(\frac{-2x}{f_-}\Big)\Big)-\int_{\pm\infty}^x \frac{dy}{y}\frac{2c}{(1+4c)^{\frac{3}{2}}}\Big[\Delta q\Big(\frac{2y}{f_+}\Big)+\Delta q\Big(\frac{-2y}{f_-}\Big)\Big],
\end{equation}
where we have used the fact that the quasi-distribution vanishes sufficiently fast for $x\to\pm\infty$. It is irrelevant to choose $\infty$ or $-\infty$ in the lower limit of the second integral. To facilitate numerical implementation, we choose it as $\infty$ ($-\infty$) for $x>0$ ($x<0$).

Since $f_+\gg f_-$ or $c$, we can recursively solve the above equation for $\Delta q(x)$ by making a change of variable $x\to f_+ x/2, y\to f_+ y/2$. 
For simplicity, let us denote
\begin{equation}
a=1+4c, \hspace{2em} b=\frac{f_+}{\sqrt{1+4c}}, \hspace{2em} r=\frac{f_-}{f_+}, \hspace{2em} \Delta(y)=\Delta \tilde q\left(\frac{f_+}{2}y\right)+\Delta \tilde q\left(-\frac{f_+}{2}\frac{y}{r}\right).
\end{equation}
We then have
\begin{align}
\Delta q(x<0)&=\frac{2a}{f_{+}}\Bigg\{\Delta \tilde{q}\left( \frac{f_{+}}{2}%
x\right) -r\left[\Delta \tilde{q}\left( -\frac{f_{+}}{2}\frac{x}{r}\right)
-\int_{-\infty }^{x}\frac{dy}{y}\,b\,\Delta (y)\right]\non \\
& +r^{2}\left[\Delta \tilde{q}\left( \frac{f_{+}}{2}\frac{x}{r^{2}}\right)
-\int_{-\infty }^{x}\frac{dy}{y}\,b\,\Delta \Big(-\frac{y}{r}\Big)%
+\int_{-\infty }^{x}\frac{dy}{y}\int_{-\infty }^{y}\frac{dz}{z}\,b^{2}\Delta
(z)-\int_{-\infty }^{-\frac{x}{r}}\frac{dy}{y}\,b\,\Delta (y)\right.\non \\
& \left.+\int_{-\infty }^{x}\frac{dy}{y}\int_{-\infty }^{-\frac{y}{r}}\frac{dz}{z}%
\,b^{2}\Delta (z)\right]\\
& -r^3 \left[ \Delta \tilde{q}\left( -\frac{%
f_{+}}{2r^{3}}x\right) -b^{3}\left( \int_{-\infty }^{z}\frac{dk}{k}%
\int_{-\infty }^{y}\frac{dz}{z}\int_{-\infty }^{x}\frac{dy}{y}\Delta
(k)+\int_{-\infty }^{z}\frac{dk}{k}\int_{-\infty }^{-\frac{y}{r}}\frac{dz}{z}%
\int_{-\infty }^{x}\frac{dy}{y}\Delta (k)\right. \right.  \\
& +\left. \int_{-\infty }^{-\frac{z}{r}}\frac{dk}{k}\int_{-\infty }^{y}\frac{%
dz}{z}\int_{-\infty }^{x}\frac{dy}{y}\Delta (k)+\int_{-\infty }^{-\frac{z}{r}%
}\frac{dk}{k}\int_{-\infty }^{-\frac{y}{r}}\frac{dz}{z}\int_{-\infty }^{x}%
\frac{dy}{y}\Delta (k)\right)  \\
& +b^{2}\left( \int_{-\infty }^{\frac{y}{r^{2}}}\frac{dk}{k}\int_{-\infty
}^{x}\frac{dy}{y}\Delta (k)+\int_{-\infty }^{-\frac{y}{r}}\frac{dk}{k}%
\int_{-\infty }^{x}\frac{dy}{y}\Delta (k)+\int_{-\infty }^{y}\frac{dk}{k}%
\int_{-\infty }^{-\frac{x}{r}}\frac{dy}{y}\Delta (k)\right.  \\
& \left. +\int_{-\infty }^{-\frac{y}{r}}\frac{dk}{k}\int_{-\infty }^{-\frac{x%
}{r}}\frac{dy}{y}\Delta (k)+\int_{-\infty }^{\frac{y}{r}}\frac{dk}{k}%
\int_{-\infty }^{x}\frac{dy}{y}\Delta (-k)+\int_{-\infty }^{-\frac{y}{r^{2}}}%
\frac{dk}{k}\int_{-\infty }^{x}\frac{dy}{y}\Delta (-k)\right)  \\
& \left.\left. -b\left( 2\int_{-\infty }^{\frac{x}{r^{2}}}\frac{dk}{k}\Delta
(k)+\int_{-\infty }^{-\frac{x}{r^{2}}}\frac{dk}{k}\Delta (-k)\right) \right] \right\}
 +\mathcal{O}(r^{4}),\non\\
\Delta q(x>0)&=\Delta q(x<0)[-\infty\to\infty].
\end{align}
We have derived the mass correction up to $\mathcal O(r^3)$. Although in the present work we did not implement $\mathcal O(r^3)$ correction due to its computational complexity, the above result will be useful for future improvements with more computational resources.

\section*{Appendix C: The $\Lambda_\text{QCD}^2/P_z^2$ Corrections}
\label{sec:appendix}

The symmetric traceless twist-2 operator in Eq.~\ref{O3} is
\begin{equation}
O^{(\mu_1\cdots \mu_n)} =
  \bar{\psi}(0)\gamma^{(\mu_1} iD^{\mu_2} \cdots iD^{\mu_n)} \psi(0).
\end{equation}
Then
\begin{equation}
\lambda_{\mu_1}\lambda_{\mu_2}\cdots \lambda_{\mu_n} O^{(\mu_1\cdots \mu_n)} \simeq
  \left(1-\frac{\lambda^2}{4n} g^{\mu\nu}
    \frac{\partial^2}{\partial \lambda^\mu \partial \lambda^\nu} \right)
  \bar{\psi}(0) \lambda \cdot \gamma \left( \lambda \cdot iD \right)^{n-1} \psi(0)
  + \mathcal{O}\left(\left( \lambda^2\right)^2\right).  \label{trace}
\end{equation}

Under operator expansion, the equal-time quark bilinear has
contributions from the twist-2 part, which is symmetric traceless, and from
the trace part
\begin{equation}
\bar{\psi}(0)\lambda \cdot \gamma \Gamma \left( 0,z \right) \psi (z\lambda )
  \simeq \sum_{n=1}^\infty \frac{\left(-iz\right)^{n-1}}{\left(n-1\right)!}
    \lambda_{\mu_1}\lambda_{\mu_2} \cdots \lambda_{\mu_n} O^{(\mu_1\cdots \mu_n)}
    + \widetilde{\mathcal{O}}_\text{tr}(z),
\end{equation}
where
\begin{eqnarray}
\widetilde{\mathcal{O}}_\text{tr}(z) &=& 
  \int_0^1 \!dt\,\frac{\lambda^2}{4} g^{\mu\nu}
    \frac{\partial^2}{\partial \lambda^\mu \partial \lambda^\nu}
    \bar{\psi}(0) \lambda \cdot \gamma
      \sum_{n=1}^\infty \frac{\left(tz\lambda \cdot D\right)^{n-1}}
                             {\left(n-1\right)!}
    \psi(0)  \notag \\
  &\equiv& \int_0^1 \!dt\, \frac{\lambda^2}{2}\mathcal{O}_\text{tr}(tz).
\end{eqnarray}

The derivatives can be carried out step by step as follows:
\begin{align}
&\phantom{=}
  \frac{1}{2} g^{\mu\nu} \frac{\partial^2}
                              {\partial \lambda^\mu \partial \lambda^\nu}
  \bar{\psi}(0) \lambda \cdot \gamma
    \exp \left(tz\lambda \cdot D\right) \psi(0) \notag \\
&= \frac{1}{2} g^{\mu\nu} \frac{\partial^2}
                               {\partial \lambda^\mu \partial \lambda^\nu}
  \bar{\psi}(0) \lambda \cdot \gamma
  \exp \left(t\Delta z\lambda \cdot D\right) \cdots
    \exp \left( \Delta z\lambda \cdot D\right) \psi(0) \notag \\
&= g_{\mu\nu} \bar{\psi}(0) \left[ \sum_{t_1}\gamma^\nu
  \exp \left(t_1 z\lambda \cdot D\right) \Delta z D^\mu
  \exp \left( (t-t_1) z\lambda \cdot D\right) \right.  \notag \\
&\phantom{=} \left. {}+\sum_{t_1,t_2} \lambda \cdot \gamma
  \exp \left( t_2 z\lambda \cdot D\right) \Delta z D^\mu
  \exp \left( (t_1-t_2) z\lambda \cdot D\right) \Delta z D^\nu
  \exp \left( (t-t_1) z\lambda \cdot D\right) \right]
  \psi(0)  \notag \\
&\rightarrow \int_0^{tz} \!dz_1\, \bar{\psi}(0) \Bigg[
  \gamma^\nu \Gamma\left(0,z_1\right) D_\nu \Gamma\left(z_1,tz\right) \notag \\
&\phantom{= \int_0^{tz} \!dz_1\, \bar{\psi}(0) \Bigg[}
  {} + \int_0^{z_1} \!dz_2\, \lambda \cdot \gamma
  \Gamma\left(0,z_2\right) D^\nu \Gamma\left(z_2,z_1\right)
  D_\nu \Gamma\left(z_1,tz\right) \Bigg] \psi(tz\lambda).
\end{align}

Therefore,
\begin{equation}
\mathcal{O}_\text{tr}(z) =
  \int_0^z \!dz_1\, \bar{\psi}(0) \left[
  \gamma^\nu \Gamma\left(0,z_1\right) D_\nu \Gamma\left(z_1,z\right)
  \int_0^{z_1} \!dz_2\, \lambda \cdot \gamma
  \Gamma\left(0,z_2\right) D^\nu \Gamma\left(z_2,z_1\right) D_\nu
  \Gamma\left(z_1,z\right) \right] \psi(z\lambda) .
\end{equation}

Then we can define the trace term as a twist-4 PDF that needs to be
subtracted:
\begin{align}\label{tildeq-twist4}
\tilde{q}_\text{twist-4}(x,\Lambda ,P_z) &=
  -\int_{-\infty}^\infty \frac{dz}{4\pi} e^{izk^z}
    \left\langle P\left\vert \widetilde{\mathcal{O}}_\text{tr}(z)
    \right\vert P\right\rangle  \notag \\
&= -\frac{\lambda^2}{2} \int_0^1 \!dt\! \int_{-\infty}^\infty
  \frac{dz}{4\pi} e^{izk^z}
  \left\langle P\left\vert \mathcal{O}_\text{tr}(tz)
  \right\vert P\right\rangle  \notag \\
&= \frac{1}{8\pi} \int_{-\infty}^\infty \!dz\,
  \Gamma_0\left(-ixzP_z\right)
  \left\langle P\left\vert \mathcal{O}_\text{tr}(z)\right\vert P\right\rangle ,
\end{align}
where $\Gamma_0$ is the incomplete Gamma function satisfying
\begin{equation}
\int_0^1 \frac{dt}{t} e^{ix/t} = \Gamma_0\left(-ix\right) .
\end{equation}

\ifx\@bibitemShut\undefined\let\@bibitemShut\relax\fi
\makeatother

\begin{thebibliography}{48}
\expandafter\ifx\csname natexlab\endcsname\relax\def\natexlab#1{#1}\fi
\expandafter\ifx\csname bibnamefont\endcsname\relax
  \def\bibnamefont#1{#1}\fi
\expandafter\ifx\csname bibfnamefont\endcsname\relax
  \def\bibfnamefont#1{#1}\fi
\expandafter\ifx\csname citenamefont\endcsname\relax
  \def\citenamefont#1{#1}\fi
\expandafter\ifx\csname url\endcsname\relax
  \def\url#1{\texttt{#1}}\fi
\expandafter\ifx\csname urlprefix\endcsname\relax\def\urlprefix{URL }\fi
\providecommand{\bibinfo}[2]{#2}
\providecommand{\eprint}[2][]{\url{#2}}

\bibitem[{\citenamefont{Alekhin et~al.}(2012)\citenamefont{Alekhin, Blumlein,
  and Moch}}]{Alekhin:2012ig}
\bibinfo{author}{\bibfnamefont{S.}~\bibnamefont{Alekhin}},
  \bibinfo{author}{\bibfnamefont{J.}~\bibnamefont{Blumlein}}, \bibnamefont{and}
  \bibinfo{author}{\bibfnamefont{S.}~\bibnamefont{Moch}},
  \bibinfo{journal}{Phys. Rev.} \textbf{\bibinfo{volume}{D86}},
  \bibinfo{pages}{054009} (\bibinfo{year}{2012}), \eprint{1202.2281}.

\bibitem[{\citenamefont{Gao et~al.}(2014)\citenamefont{Gao, Guzzi, Huston, Lai,
  Li, Nadolsky, Pumplin, Stump, and Yuan}}]{Gao:2013xoa}
\bibinfo{author}{\bibfnamefont{J.}~\bibnamefont{Gao}},
  \bibinfo{author}{\bibfnamefont{M.}~\bibnamefont{Guzzi}},
  \bibinfo{author}{\bibfnamefont{J.}~\bibnamefont{Huston}},
  \bibinfo{author}{\bibfnamefont{H.-L.} \bibnamefont{Lai}},
  \bibinfo{author}{\bibfnamefont{Z.}~\bibnamefont{Li}},
  \bibinfo{author}{\bibfnamefont{P.}~\bibnamefont{Nadolsky}},
  \bibinfo{author}{\bibfnamefont{J.}~\bibnamefont{Pumplin}},
  \bibinfo{author}{\bibfnamefont{D.}~\bibnamefont{Stump}}, \bibnamefont{and}
  \bibinfo{author}{\bibfnamefont{C.~P.} \bibnamefont{Yuan}},
  \bibinfo{journal}{Phys. Rev.} \textbf{\bibinfo{volume}{D89}},
  \bibinfo{pages}{033009} (\bibinfo{year}{2014}), \eprint{1302.6246}.

\bibitem[{\citenamefont{Radescu}(2010)}]{Radescu:2010zz}
\bibinfo{author}{\bibfnamefont{V.}~\bibnamefont{Radescu}}
  (\bibinfo{collaboration}{ZEUS, H1}), \bibinfo{journal}{PoS}
  \textbf{\bibinfo{volume}{ICHEP2010}}, \bibinfo{pages}{168}
  (\bibinfo{year}{2010}).

\bibitem[{\citenamefont{Cooper-Sarkar}(2011)}]{CooperSarkar:2011aa}
\bibinfo{author}{\bibfnamefont{A.~M.} \bibnamefont{Cooper-Sarkar}}
  (\bibinfo{collaboration}{ZEUS, H1}), \bibinfo{journal}{PoS}
  \textbf{\bibinfo{volume}{EPS-HEP2011}}, \bibinfo{pages}{320}
  (\bibinfo{year}{2011}), \eprint{1112.2107}.

\bibitem[{\citenamefont{Martin et~al.}(2009)\citenamefont{Martin, Stirling,
  Thorne, and Watt}}]{Martin:2009iq}
\bibinfo{author}{\bibfnamefont{A.~D.} \bibnamefont{Martin}},
  \bibinfo{author}{\bibfnamefont{W.~J.} \bibnamefont{Stirling}},
  \bibinfo{author}{\bibfnamefont{R.~S.} \bibnamefont{Thorne}},
  \bibnamefont{and} \bibinfo{author}{\bibfnamefont{G.}~\bibnamefont{Watt}},
  \bibinfo{journal}{Eur. Phys. J.} \textbf{\bibinfo{volume}{C63}},
  \bibinfo{pages}{189} (\bibinfo{year}{2009}), \eprint{0901.0002}.

\bibitem[{\citenamefont{Ball et~al.}(2013)}]{Ball:2012cx}
\bibinfo{author}{\bibfnamefont{R.~D.} \bibnamefont{Ball}} \bibnamefont{et~al.},
  \bibinfo{journal}{Nucl. Phys.} \textbf{\bibinfo{volume}{B867}},
  \bibinfo{pages}{244} (\bibinfo{year}{2013}), \eprint{1207.1303}.

\bibitem[{\citenamefont{Chatrchyan et~al.}(2012)}]{CMS:2012nga}
\bibinfo{author}{\bibfnamefont{S.}~\bibnamefont{Chatrchyan}}
  \bibnamefont{et~al.} (\bibinfo{collaboration}{CMS}),
  \bibinfo{journal}{Science} \textbf{\bibinfo{volume}{338}},
  \bibinfo{pages}{1569} (\bibinfo{year}{2012}).

\bibitem[{\citenamefont{Aad et~al.}(2012)}]{ATLAS:2012oga}
\bibinfo{author}{\bibfnamefont{G.}~\bibnamefont{Aad}} \bibnamefont{et~al.}
  (\bibinfo{collaboration}{ATLAS}), \bibinfo{journal}{Science}
  \textbf{\bibinfo{volume}{338}}, \bibinfo{pages}{1576} (\bibinfo{year}{2012}).

\bibitem[{\citenamefont{Aad et~al.}(2014)}]{Aad:2014xca}
\bibinfo{author}{\bibfnamefont{G.}~\bibnamefont{Aad}} \bibnamefont{et~al.}
  (\bibinfo{collaboration}{ATLAS}), \bibinfo{journal}{JHEP}
  \textbf{\bibinfo{volume}{05}}, \bibinfo{pages}{068} (\bibinfo{year}{2014}),
  \eprint{1402.6263}.

\bibitem[{\citenamefont{Chatrchyan et~al.}(2014)}]{Chatrchyan:2013mza}
\bibinfo{author}{\bibfnamefont{S.}~\bibnamefont{Chatrchyan}}
  \bibnamefont{et~al.} (\bibinfo{collaboration}{CMS}), \bibinfo{journal}{Phys.
  Rev.} \textbf{\bibinfo{volume}{D90}}, \bibinfo{pages}{032004}
  (\bibinfo{year}{2014}), \eprint{1312.6283}.

\bibitem[{\citenamefont{Gockeler et~al.}(2005)\citenamefont{Gockeler, Horsley,
  Pleiter, Rakow, and Schierholz}}]{Gockeler:2004wp}
\bibinfo{author}{\bibfnamefont{M.}~\bibnamefont{Gockeler}},
  \bibinfo{author}{\bibfnamefont{R.}~\bibnamefont{Horsley}},
  \bibinfo{author}{\bibfnamefont{D.}~\bibnamefont{Pleiter}},
  \bibinfo{author}{\bibfnamefont{P.~E.~L.} \bibnamefont{Rakow}},
  \bibnamefont{and}
  \bibinfo{author}{\bibfnamefont{G.}~\bibnamefont{Schierholz}}
  (\bibinfo{collaboration}{QCDSF}), \bibinfo{journal}{Phys. Rev.}
  \textbf{\bibinfo{volume}{D71}}, \bibinfo{pages}{114511}
  (\bibinfo{year}{2005}), \eprint{hep-ph/0410187}.

\bibitem[{\citenamefont{Davoudi and Savage}(2012)}]{Davoudi:2012ya}
\bibinfo{author}{\bibfnamefont{Z.}~\bibnamefont{Davoudi}} \bibnamefont{and}
  \bibinfo{author}{\bibfnamefont{M.~J.} \bibnamefont{Savage}},
  \bibinfo{journal}{Phys. Rev.} \textbf{\bibinfo{volume}{D86}},
  \bibinfo{pages}{054505} (\bibinfo{year}{2012}), \eprint{1204.4146}.

\bibitem[{\citenamefont{Detmold and Lin}(2006)}]{Detmold:2005gg}
\bibinfo{author}{\bibfnamefont{W.}~\bibnamefont{Detmold}} \bibnamefont{and}
  \bibinfo{author}{\bibfnamefont{C.~J.~D.} \bibnamefont{Lin}},
  \bibinfo{journal}{Phys. Rev.} \textbf{\bibinfo{volume}{D73}},
  \bibinfo{pages}{014501} (\bibinfo{year}{2006}), \eprint{hep-lat/0507007}.

\bibitem[{\citenamefont{Ji}(2013)}]{Ji:2013dva}
\bibinfo{author}{\bibfnamefont{X.}~\bibnamefont{Ji}}, \bibinfo{journal}{Phys.
  Rev. Lett.} \textbf{\bibinfo{volume}{110}}, \bibinfo{pages}{262002}
  (\bibinfo{year}{2013}), \eprint{1305.1539}.

\bibitem[{\citenamefont{Ji}(2014)}]{Ji:2014gla}
\bibinfo{author}{\bibfnamefont{X.}~\bibnamefont{Ji}}, \bibinfo{journal}{Sci.
  China Phys. Mech. Astron.} \textbf{\bibinfo{volume}{57}},
  \bibinfo{pages}{1407} (\bibinfo{year}{2014}), \eprint{1404.6680}.

\bibitem[{\citenamefont{Xiong et~al.}(2014)\citenamefont{Xiong, Ji, Zhang, and
  Zhao}}]{Xiong:2013bka}
\bibinfo{author}{\bibfnamefont{X.}~\bibnamefont{Xiong}},
  \bibinfo{author}{\bibfnamefont{X.}~\bibnamefont{Ji}},
  \bibinfo{author}{\bibfnamefont{J.-H.} \bibnamefont{Zhang}}, \bibnamefont{and}
  \bibinfo{author}{\bibfnamefont{Y.}~\bibnamefont{Zhao}},
  \bibinfo{journal}{Phys. Rev.} \textbf{\bibinfo{volume}{D90}},
  \bibinfo{pages}{014051} (\bibinfo{year}{2014}), \eprint{1310.7471}.

\bibitem[{\citenamefont{Ji et~al.}(2015)\citenamefont{Ji, Schäfer, Xiong, and
  Zhang}}]{Ji:2015qla}
\bibinfo{author}{\bibfnamefont{X.}~\bibnamefont{Ji}},
  \bibinfo{author}{\bibfnamefont{A.}~\bibnamefont{Schäfer}},
  \bibinfo{author}{\bibfnamefont{X.}~\bibnamefont{Xiong}}, \bibnamefont{and}
  \bibinfo{author}{\bibfnamefont{J.-H.} \bibnamefont{Zhang}},
  \bibinfo{journal}{Phys. Rev.} \textbf{\bibinfo{volume}{D92}},
  \bibinfo{pages}{014039} (\bibinfo{year}{2015}), \eprint{1506.00248}.

\bibitem[{\citenamefont{Xiong and Zhang}(2015)}]{Xiong:2015nua}
\bibinfo{author}{\bibfnamefont{X.}~\bibnamefont{Xiong}} \bibnamefont{and}
  \bibinfo{author}{\bibfnamefont{J.-H.} \bibnamefont{Zhang}},
  \bibinfo{journal}{Phys. Rev.} \textbf{\bibinfo{volume}{D92}},
  \bibinfo{pages}{054037} (\bibinfo{year}{2015}), \eprint{1509.08016}.

\bibitem[{\citenamefont{Ji and Zhang}(2015)}]{Ji:2015jwa}
\bibinfo{author}{\bibfnamefont{X.}~\bibnamefont{Ji}} \bibnamefont{and}
  \bibinfo{author}{\bibfnamefont{J.-H.} \bibnamefont{Zhang}},
  \bibinfo{journal}{Phys. Rev.} \textbf{\bibinfo{volume}{D92}},
  \bibinfo{pages}{034006} (\bibinfo{year}{2015}), \eprint{1505.07699}.

\bibitem[{\citenamefont{Li}(2016)}]{Li:2016amo}
\bibinfo{author}{\bibfnamefont{H.-n.} \bibnamefont{Li}} (\bibinfo{year}{2016}),
  \eprint{1602.07575}.

\bibitem[{\citenamefont{Bali et~al.}(2016)\citenamefont{Bali, Lang, Musch, and
  Schäfer}}]{Bali:2016lva}
\bibinfo{author}{\bibfnamefont{G.~S.} \bibnamefont{Bali}},
  \bibinfo{author}{\bibfnamefont{B.}~\bibnamefont{Lang}},
  \bibinfo{author}{\bibfnamefont{B.~U.} \bibnamefont{Musch}}, \bibnamefont{and}
  \bibinfo{author}{\bibfnamefont{A.}~\bibnamefont{Schäfer}}
  (\bibinfo{year}{2016}), \eprint{1602.05525}.

\bibitem[{\citenamefont{Lin}(2014{\natexlab{a}})}]{Lin:2014gaa}
\bibinfo{author}{\bibfnamefont{H.-W.} \bibnamefont{Lin}},
  \bibinfo{journal}{Int. J. Mod. Phys. Conf. Ser.}
  \textbf{\bibinfo{volume}{25}}, \bibinfo{pages}{1460039}
  (\bibinfo{year}{2014}{\natexlab{a}}).

\bibitem[{\citenamefont{Lin}(2014{\natexlab{b}})}]{Lin:2014yra}
\bibinfo{author}{\bibfnamefont{H.-W.} \bibnamefont{Lin}},
  \bibinfo{journal}{PoS} \textbf{\bibinfo{volume}{LATTICE2013}},
  \bibinfo{pages}{293} (\bibinfo{year}{2014}{\natexlab{b}}).

\bibitem[{\citenamefont{Lin et~al.}(2015)\citenamefont{Lin, Chen, Cohen, and
  Ji}}]{Lin:2014zya}
\bibinfo{author}{\bibfnamefont{H.-W.} \bibnamefont{Lin}},
  \bibinfo{author}{\bibfnamefont{J.-W.} \bibnamefont{Chen}},
  \bibinfo{author}{\bibfnamefont{S.~D.} \bibnamefont{Cohen}}, \bibnamefont{and}
  \bibinfo{author}{\bibfnamefont{X.}~\bibnamefont{Ji}}, \bibinfo{journal}{Phys.
  Rev.} \textbf{\bibinfo{volume}{D91}}, \bibinfo{pages}{054510}
  (\bibinfo{year}{2015}), \eprint{1402.1462}.

\bibitem[{\citenamefont{Alexandrou et~al.}(2015)\citenamefont{Alexandrou,
  Cichy, Drach, Garcia-Ramos, Hadjiyiannakou, Jansen, Steffens, and
  Wiese}}]{Alexandrou:2015rja}
\bibinfo{author}{\bibfnamefont{C.}~\bibnamefont{Alexandrou}},
  \bibinfo{author}{\bibfnamefont{K.}~\bibnamefont{Cichy}},
  \bibinfo{author}{\bibfnamefont{V.}~\bibnamefont{Drach}},
  \bibinfo{author}{\bibfnamefont{E.}~\bibnamefont{Garcia-Ramos}},
  \bibinfo{author}{\bibfnamefont{K.}~\bibnamefont{Hadjiyiannakou}},
  \bibinfo{author}{\bibfnamefont{K.}~\bibnamefont{Jansen}},
  \bibinfo{author}{\bibfnamefont{F.}~\bibnamefont{Steffens}}, \bibnamefont{and}
  \bibinfo{author}{\bibfnamefont{C.}~\bibnamefont{Wiese}},
  \bibinfo{journal}{Phys. Rev.} \textbf{\bibinfo{volume}{D92}},
  \bibinfo{pages}{014502} (\bibinfo{year}{2015}), \eprint{1504.07455}.

\bibitem[{\citenamefont{Adamczyk et~al.}(2014)}]{Adamczyk:2014xyw}
\bibinfo{author}{\bibfnamefont{L.}~\bibnamefont{Adamczyk}} \bibnamefont{et~al.}
  (\bibinfo{collaboration}{STAR}), \bibinfo{journal}{Phys. Rev. Lett.}
  \textbf{\bibinfo{volume}{113}}, \bibinfo{pages}{072301}
  (\bibinfo{year}{2014}), \eprint{1404.6880}.

\bibitem[{\citenamefont{Adare et~al.}(2015)}]{Adare:2015gsd}
\bibinfo{author}{\bibfnamefont{A.}~\bibnamefont{Adare}} \bibnamefont{et~al.}
  (\bibinfo{collaboration}{PHENIX}) (\bibinfo{year}{2015}),
  \eprint{1504.07451}.

\bibitem[{\citenamefont{Ma and Qiu}(2014)}]{Ma:2014jla}
\bibinfo{author}{\bibfnamefont{Y.-Q.} \bibnamefont{Ma}} \bibnamefont{and}
  \bibinfo{author}{\bibfnamefont{J.-W.} \bibnamefont{Qiu}}
  (\bibinfo{year}{2014}), \eprint{1404.6860}.

\bibitem[{\citenamefont{Follana et~al.}(2007)\citenamefont{Follana, Mason,
  Davies, Hornbostel, Lepage, Shigemitsu, Trottier, and Wong}}]{Follana:2006rc}
\bibinfo{author}{\bibfnamefont{E.}~\bibnamefont{Follana}},
  \bibinfo{author}{\bibfnamefont{Q.}~\bibnamefont{Mason}},
  \bibinfo{author}{\bibfnamefont{C.}~\bibnamefont{Davies}},
  \bibinfo{author}{\bibfnamefont{K.}~\bibnamefont{Hornbostel}},
  \bibinfo{author}{\bibfnamefont{G.~P.} \bibnamefont{Lepage}},
  \bibinfo{author}{\bibfnamefont{J.}~\bibnamefont{Shigemitsu}},
  \bibinfo{author}{\bibfnamefont{H.}~\bibnamefont{Trottier}}, \bibnamefont{and}
  \bibinfo{author}{\bibfnamefont{K.}~\bibnamefont{Wong}}
  (\bibinfo{collaboration}{HPQCD, UKQCD}), \bibinfo{journal}{Phys. Rev.}
  \textbf{\bibinfo{volume}{D75}}, \bibinfo{pages}{054502}
  (\bibinfo{year}{2007}), \eprint{hep-lat/0610092}.

\bibitem[{\citenamefont{Bazavov et~al.}(2013)}]{Bazavov:2012xda}
\bibinfo{author}{\bibfnamefont{A.}~\bibnamefont{Bazavov}} \bibnamefont{et~al.}
  (\bibinfo{collaboration}{MILC}), \bibinfo{journal}{Phys. Rev.}
  \textbf{\bibinfo{volume}{D87}}, \bibinfo{pages}{054505}
  (\bibinfo{year}{2013}), \eprint{1212.4768}.

\bibitem[{\citenamefont{Hasenfratz and Knechtli}(2001)}]{Hasenfratz:2001hp}
\bibinfo{author}{\bibfnamefont{A.}~\bibnamefont{Hasenfratz}} \bibnamefont{and}
  \bibinfo{author}{\bibfnamefont{F.}~\bibnamefont{Knechtli}},
  \bibinfo{journal}{Phys. Rev.} \textbf{\bibinfo{volume}{D64}},
  \bibinfo{pages}{034504} (\bibinfo{year}{2001}), \eprint{hep-lat/0103029}.

\bibitem[{\citenamefont{Bhattacharya
  et~al.}(2015{\natexlab{a}})\citenamefont{Bhattacharya, Cirigliano, Cohen,
  Gupta, Joseph, Lin, and Yoon}}]{Bhattacharya:2015wna}
\bibinfo{author}{\bibfnamefont{T.}~\bibnamefont{Bhattacharya}},
  \bibinfo{author}{\bibfnamefont{V.}~\bibnamefont{Cirigliano}},
  \bibinfo{author}{\bibfnamefont{S.}~\bibnamefont{Cohen}},
  \bibinfo{author}{\bibfnamefont{R.}~\bibnamefont{Gupta}},
  \bibinfo{author}{\bibfnamefont{A.}~\bibnamefont{Joseph}},
  \bibinfo{author}{\bibfnamefont{H.-W.} \bibnamefont{Lin}}, \bibnamefont{and}
  \bibinfo{author}{\bibfnamefont{B.}~\bibnamefont{Yoon}}
  (\bibinfo{collaboration}{PNDME}), \bibinfo{journal}{Phys. Rev.}
  \textbf{\bibinfo{volume}{D92}}, \bibinfo{pages}{094511}
  (\bibinfo{year}{2015}{\natexlab{a}}), \eprint{1506.06411}.

\bibitem[{\citenamefont{Bhattacharya
  et~al.}(2015{\natexlab{b}})\citenamefont{Bhattacharya, Cirigliano, Gupta,
  Lin, and Yoon}}]{Bhattacharya:2015esa}
\bibinfo{author}{\bibfnamefont{T.}~\bibnamefont{Bhattacharya}},
  \bibinfo{author}{\bibfnamefont{V.}~\bibnamefont{Cirigliano}},
  \bibinfo{author}{\bibfnamefont{R.}~\bibnamefont{Gupta}},
  \bibinfo{author}{\bibfnamefont{H.-W.} \bibnamefont{Lin}}, \bibnamefont{and}
  \bibinfo{author}{\bibfnamefont{B.}~\bibnamefont{Yoon}},
  \bibinfo{journal}{Phys. Rev. Lett.} \textbf{\bibinfo{volume}{115}},
  \bibinfo{pages}{212002} (\bibinfo{year}{2015}{\natexlab{b}}),
  \eprint{1506.04196}.

\bibitem[{\citenamefont{Bhattacharya et~al.}(2014)\citenamefont{Bhattacharya,
  Cohen, Gupta, Joseph, Lin, and Yoon}}]{Bhattacharya:2013ehc}
\bibinfo{author}{\bibfnamefont{T.}~\bibnamefont{Bhattacharya}},
  \bibinfo{author}{\bibfnamefont{S.~D.} \bibnamefont{Cohen}},
  \bibinfo{author}{\bibfnamefont{R.}~\bibnamefont{Gupta}},
  \bibinfo{author}{\bibfnamefont{A.}~\bibnamefont{Joseph}},
  \bibinfo{author}{\bibfnamefont{H.-W.} \bibnamefont{Lin}}, \bibnamefont{and}
  \bibinfo{author}{\bibfnamefont{B.}~\bibnamefont{Yoon}},
  \bibinfo{journal}{Phys. Rev.} \textbf{\bibinfo{volume}{D89}},
  \bibinfo{pages}{094502} (\bibinfo{year}{2014}), \eprint{1306.5435}.

\bibitem[{\citenamefont{Jimenez-Delgado
  et~al.}(2014)\citenamefont{Jimenez-Delgado, Accardi, and
  Melnitchouk}}]{Jimenez-Delgado:2013boa}
\bibinfo{author}{\bibfnamefont{P.}~\bibnamefont{Jimenez-Delgado}},
  \bibinfo{author}{\bibfnamefont{A.}~\bibnamefont{Accardi}}, \bibnamefont{and}
  \bibinfo{author}{\bibfnamefont{W.}~\bibnamefont{Melnitchouk}},
  \bibinfo{journal}{Phys. Rev.} \textbf{\bibinfo{volume}{D89}},
  \bibinfo{pages}{034025} (\bibinfo{year}{2014}), \eprint{1310.3734}.

\bibitem[{\citenamefont{de~Florian et~al.}(2009)\citenamefont{de~Florian,
  Sassot, Stratmann, and Vogelsang}}]{deFlorian:2009vb}
\bibinfo{author}{\bibfnamefont{D.}~\bibnamefont{de~Florian}},
  \bibinfo{author}{\bibfnamefont{R.}~\bibnamefont{Sassot}},
  \bibinfo{author}{\bibfnamefont{M.}~\bibnamefont{Stratmann}},
  \bibnamefont{and}
  \bibinfo{author}{\bibfnamefont{W.}~\bibnamefont{Vogelsang}},
  \bibinfo{journal}{Phys. Rev.} \textbf{\bibinfo{volume}{D80}},
  \bibinfo{pages}{034030} (\bibinfo{year}{2009}), \eprint{0904.3821}.

\bibitem[{\citenamefont{Nocera et~al.}(2014)\citenamefont{Nocera, Ball, Forte,
  Ridolfi, and Rojo}}]{Nocera:2014gqa}
\bibinfo{author}{\bibfnamefont{E.~R.} \bibnamefont{Nocera}},
  \bibinfo{author}{\bibfnamefont{R.~D.} \bibnamefont{Ball}},
  \bibinfo{author}{\bibfnamefont{S.}~\bibnamefont{Forte}},
  \bibinfo{author}{\bibfnamefont{G.}~\bibnamefont{Ridolfi}}, \bibnamefont{and}
  \bibinfo{author}{\bibfnamefont{J.}~\bibnamefont{Rojo}}
  (\bibinfo{collaboration}{NNPDF}), \bibinfo{journal}{Nucl. Phys.}
  \textbf{\bibinfo{volume}{B887}}, \bibinfo{pages}{276} (\bibinfo{year}{2014}),
  \eprint{1406.5539}.

\bibitem[{\citenamefont{Lin and Orginos}(2009)}]{Lin:2007ap}
\bibinfo{author}{\bibfnamefont{H.-W.} \bibnamefont{Lin}} \bibnamefont{and}
  \bibinfo{author}{\bibfnamefont{K.}~\bibnamefont{Orginos}},
  \bibinfo{journal}{Phys. Rev.} \textbf{\bibinfo{volume}{D79}},
  \bibinfo{pages}{034507} (\bibinfo{year}{2009}), \eprint{0712.1214}.

\bibitem[{\citenamefont{Schweitzer et~al.}(2001)\citenamefont{Schweitzer,
  Urbano, Polyakov, Weiss, Pobylitsa, and Goeke}}]{Schweitzer:2001sr}
\bibinfo{author}{\bibfnamefont{P.}~\bibnamefont{Schweitzer}},
  \bibinfo{author}{\bibfnamefont{D.}~\bibnamefont{Urbano}},
  \bibinfo{author}{\bibfnamefont{M.~V.} \bibnamefont{Polyakov}},
  \bibinfo{author}{\bibfnamefont{C.}~\bibnamefont{Weiss}},
  \bibinfo{author}{\bibfnamefont{P.~V.} \bibnamefont{Pobylitsa}},
  \bibnamefont{and} \bibinfo{author}{\bibfnamefont{K.}~\bibnamefont{Goeke}},
  \bibinfo{journal}{Phys. Rev.} \textbf{\bibinfo{volume}{D64}},
  \bibinfo{pages}{034013} (\bibinfo{year}{2001}), \eprint{hep-ph/0101300}.

\bibitem[{\citenamefont{Lin}(2015)}]{Lin:2015vxw}
\bibinfo{author}{\bibfnamefont{H.-W.} \bibnamefont{Lin}}, \bibinfo{journal}{Few
  Body Syst.} \textbf{\bibinfo{volume}{56}}, \bibinfo{pages}{455}
  (\bibinfo{year}{2015}).

\bibitem[{\citenamefont{Aschenauer et~al.}(2013)}]{Aschenauer:2013woa}
\bibinfo{author}{\bibfnamefont{E.~C.} \bibnamefont{Aschenauer}}
  \bibnamefont{et~al.} (\bibinfo{year}{2013}), \eprint{1304.0079}.

\bibitem[{\citenamefont{Chang and Peng}(2014)}]{Chang:2014jba}
\bibinfo{author}{\bibfnamefont{W.-C.} \bibnamefont{Chang}} \bibnamefont{and}
  \bibinfo{author}{\bibfnamefont{J.-C.} \bibnamefont{Peng}},
  \bibinfo{journal}{Prog. Part. Nucl. Phys.} \textbf{\bibinfo{volume}{79}},
  \bibinfo{pages}{95} (\bibinfo{year}{2014}), \eprint{1406.1260}.

\bibitem[{\citenamefont{Anselmino et~al.}(2009)\citenamefont{Anselmino,
  Boglione, D'Alesio, Kotzinian, Murgia, Prokudin, and
  Melis}}]{Anselmino:2008jk}
\bibinfo{author}{\bibfnamefont{M.}~\bibnamefont{Anselmino}},
  \bibinfo{author}{\bibfnamefont{M.}~\bibnamefont{Boglione}},
  \bibinfo{author}{\bibfnamefont{U.}~\bibnamefont{D'Alesio}},
  \bibinfo{author}{\bibfnamefont{A.}~\bibnamefont{Kotzinian}},
  \bibinfo{author}{\bibfnamefont{F.}~\bibnamefont{Murgia}},
  \bibinfo{author}{\bibfnamefont{A.}~\bibnamefont{Prokudin}}, \bibnamefont{and}
  \bibinfo{author}{\bibfnamefont{S.}~\bibnamefont{Melis}},
  \bibinfo{journal}{Nucl. Phys. Proc. Suppl.} \textbf{\bibinfo{volume}{191}},
  \bibinfo{pages}{98} (\bibinfo{year}{2009}), \eprint{0812.4366}.

\bibitem[{\citenamefont{Radici et~al.}(2014)\citenamefont{Radici, Bacchetta,
  and Courtoy}}]{Radici:2013vra}
\bibinfo{author}{\bibfnamefont{M.}~\bibnamefont{Radici}},
  \bibinfo{author}{\bibfnamefont{A.}~\bibnamefont{Bacchetta}},
  \bibnamefont{and} \bibinfo{author}{\bibfnamefont{A.}~\bibnamefont{Courtoy}},
  \bibinfo{journal}{Int. J. Mod. Phys. Conf. Ser.}
  \textbf{\bibinfo{volume}{25}}, \bibinfo{pages}{1460045}
  (\bibinfo{year}{2014}), \eprint{1308.5928}.

\bibitem[{\citenamefont{Kang et~al.}(2016)\citenamefont{Kang, Prokudin, Sun,
  and Yuan}}]{Kang:2015msa}
\bibinfo{author}{\bibfnamefont{Z.-B.} \bibnamefont{Kang}},
  \bibinfo{author}{\bibfnamefont{A.}~\bibnamefont{Prokudin}},
  \bibinfo{author}{\bibfnamefont{P.}~\bibnamefont{Sun}}, \bibnamefont{and}
  \bibinfo{author}{\bibfnamefont{F.}~\bibnamefont{Yuan}},
  \bibinfo{journal}{Phys. Rev.} \textbf{\bibinfo{volume}{D93}},
  \bibinfo{pages}{014009} (\bibinfo{year}{2016}), \eprint{1505.05589}.

\bibitem[{\citenamefont{Kang et~al.}(2015)\citenamefont{Kang, Prokudin, Sun,
  and Yuan}}]{Kang:2014zza}
\bibinfo{author}{\bibfnamefont{Z.-B.} \bibnamefont{Kang}},
  \bibinfo{author}{\bibfnamefont{A.}~\bibnamefont{Prokudin}},
  \bibinfo{author}{\bibfnamefont{P.}~\bibnamefont{Sun}}, \bibnamefont{and}
  \bibinfo{author}{\bibfnamefont{F.}~\bibnamefont{Yuan}},
  \bibinfo{journal}{Phys. Rev.} \textbf{\bibinfo{volume}{D91}},
  \bibinfo{pages}{071501} (\bibinfo{year}{2015}), \eprint{1410.4877}.

\bibitem[{\citenamefont{Edwards and Joo}(2005)}]{Edwards:2004sx}
\bibinfo{author}{\bibfnamefont{R.~G.} \bibnamefont{Edwards}} \bibnamefont{and}
  \bibinfo{author}{\bibfnamefont{B.}~\bibnamefont{Joo}}
  (\bibinfo{collaboration}{SciDAC, LHPC, UKQCD}), \bibinfo{journal}{Nucl. Phys.
  Proc. Suppl.} \textbf{\bibinfo{volume}{140}}, \bibinfo{pages}{832}
  (\bibinfo{year}{2005}), \bibinfo{note}{[,832(2004)]},
  \eprint{hep-lat/0409003}.

\bibitem[{\citenamefont{Steffens et~al.}(2012)\citenamefont{Steffens, Brown,
  Melnitchouk, and Sanches}}]{Steffens:2012jx}
\bibinfo{author}{\bibfnamefont{F.~M.} \bibnamefont{Steffens}},
  \bibinfo{author}{\bibfnamefont{M.~D.} \bibnamefont{Brown}},
  \bibinfo{author}{\bibfnamefont{W.}~\bibnamefont{Melnitchouk}},
  \bibnamefont{and} \bibinfo{author}{\bibfnamefont{S.}~\bibnamefont{Sanches}},
  \bibinfo{journal}{Phys. Rev.} \textbf{\bibinfo{volume}{C86}},
  \bibinfo{pages}{065208} (\bibinfo{year}{2012}), \eprint{1210.4398}.

\end{thebibliography}

\end{document}